\documentclass[fleqn,usenatbib]{mnras}

\usepackage{newtxtext,newtxmath}

\usepackage[T1]{fontenc}

\DeclareRobustCommand{\VAN}[3]{#2}
\let\VANthebibliography\thebibliography
\def\thebibliography{\DeclareRobustCommand{\VAN}[3]{##3}\VANthebibliography}

\usepackage{graphicx}	
\usepackage{amsmath}	
\usepackage{xcolor}
\usepackage{supertabular}
\usepackage[normalem]{ulem}
\usepackage{pdflscape}
\usepackage{lscape}
\usepackage{caption, subcaption}
\usepackage{siunitx}
\usepackage{hyperref}
\graphicspath{{/nvme1/hwang/data/Plots/}}

\newcommand{\clearmaybe}{}

\title[Complex sub-pulse variability in 9 pulsars]{The Thousand-Pulsar-Array programme on MeerKAT -- XVIII. Complex sub-pulse modulation patterns, bi-drifting and mode changing of nine radio pulsars}

\author[H. Wang et al.]{
H. Wang,$^{1}$\thanks{E-mail: haoyue.wang-5@postgrad.manchester.ac.uk}
M. J. Keith,$^{1}$
P. Weltevrede,$^{1}$
G. Wright,$^{1}$
B. Bhattacharyya,$^{2}$
J.~A. Hsu$^{1}$
and X. Song$^{3}$
\\
$^{1}$Jodrell Bank Centre for Astrophysics, Department of Physics and Astronomy, University of Manchester, Manchester M13 9PL, UK\\
$^{2}$National Centre for Radio Astrophysics, Tata Institute of Fundamental Research, Pune 411 007, India\\
$^{3}$ASTRON, the Netherlands Institute for Radio Astronomy, Oude Hoogeveensedijk 4,7991 PD Dwingeloo, The Netherlands \\
}

\date{Accepted XXX. Received YYY; in original form ZZZ}

\pubyear{2025}

\begin{document}
\label{firstpage}
\pagerange{\pageref{firstpage}--\pageref{lastpage}}
\maketitle

\begin{abstract}
We present a detailed analysis of sub-pulse modulations in nine pulsars which 
show evidence of changes in sub-pulse drift direction as a function of pulse longitude in the Thousand Pulsar Array single pulse survey with MeerKAT.
We confirm that all of these are consistent with persistent drift direction changes. These `bi-drifting’ pulsars present a challenge to the classical carousel model for sub-pulse drifting.
In general, bi-drifting in this expanded sample is less clear than some of the previously published cases, which we attribute to narrower profile widths (<20\textdegree) or smaller $P_3$ values (close to 2$P$).
However, given the broad variety of pulse shapes and drift behaviours across the pulsar population, 
it is unsurprising that the phenomenon is not limited to only those where it can most easily be detected.
Four of our samples show at least two emission modes with different profile shapes and drift properties, which seems to be a relatively common feature of bi-drifting pulsars.
We also find jumps in sub-pulse phase between adjacent components in two pulsars.
In addition to our MeerKAT L-band data, we used GMRT observations for four, and MeerKAT UHF observations for two of these pulsars 
to investigate the frequency dependence of sub-pulse drift.
We find subtle changes in the drift as a function of frequency, but no clear overall pattern.
Looking at the distribution of bi-drifting pulsars over $P$, $\dot{P}$ and $P_3$ suggests they are consistent with the underlying population of all drifting pulsars.

\end{abstract}
\clearmaybe


\section{Introduction}
\label{sec:Intro}
The emission from radio pulsars is characterised by a sequence of pulses with period, $P$, the rotational period of the pulsar.
Although the time-averaged shape of the pulse (termed the pulse profile) is usually very stable, individual pulses are often composed of narrower sub-pulses which can exhibit changes in intensity, shape and phase from pulse to pulse.
In addition to random fluctuations, many pulsars show systematic variations in the intensity or phase of these sub-pulses.
A common feature is drifting, where sub-pulses `march' through the pulse profile window over time, and the single pulses form inclined, almost uniformly spaced drift bands \citep{Drake1968, Sutton1970} in a pulse stack, which is a 2-D map of intensity as a function of the pulse longitude and the pulse number. This sub-pulse drifting phenomenon is common, and recent studies (e.g. by \citealt{Song2023}) suggest it occurs in about 60\% of the pulsar population. This feature is more prevalent among the older pulsar population, a bias previously identified by e.g. \citet{Wolszczan1980A,Rankin1986,Weltevrede2006}.
The drift rate is characterised by two quantities.
These are the separation between sub-pulses in pulse longitude, $P_2$, and the time between the repeating drift bands, $P_3$. The drift rate is $P_2$/$P_3$, which is the slope of the drift band in the pulse stack.

For the majority of pulsars with drifting sub-pulses, the drift is consistent with a carousel of radio emitting `sparks' which drift in a circular pattern around the magnetic pole \citep{Ruderman1975}. 
The observed drift occurs as each time the line-of-sight passes over the emission region the sparks have rotated to a different phase.
In this model $P_3$ is determined by the carousel rotation period (often termed $P_4$) divided by the number of sparks, both of which could in principle change over time.
In some cases, the drift may be fast enough that we observe an aliasing effect, e.g. PSR~B0943+10 \citep{Deshpande1999,Gil2003}. The true drift rate may differ significantly from that observed and even appear in reverse direction if there are fluctuations in $P_3$ close to the critical aliasing frequency, something argued to occur in PSRs~B2303+30 \citep{Redman2005}, B0826$-$34 \citep{Gupta2004,Bhattacharyya2008} and J1750$-$3503 \citep{Szary2022}.

The single carousel picture has proved to be inadequate to explain the emission of multi-component pulsars and, on observational grounds, \citet{Rankin1993} introduced the concept of a double carousel plus a central core emission. Depending on the observer's line-of-sight, this results in profiles with between one and five components. In this picture both carousels rotate at a common $P_4$ and have the same number of sparks with alternate spacing, as observed for example for PSR~B0818$-$41 \citep{Bhattacharyya2009}.

A further common feature of pulsar emission is that of emission `mode changing', where the average pulse profile changes between two or more stable shapes \citep{Backer1970,Lyne1971}. 
This can take two forms. In `drift-mode changing' the emission experiences a usually sudden change in $P_3$ (though not in the drift-band spacing $P_2$). Two or even three separate modes may appear, clear examples of which are  PSRs~B0031$-$07 (\citealp{Huguenin1970}, \citealp{Ilie2020}) and B2319$+$60  \citep{Wright1981}. These might be explained by a sudden acceleration of the sparks or a change in their number \citep{Wright2022}. The second form is more dramatic and results in a complete collapse of orderly patterns, resulting in the addition or absence of components. Depending on the observer's sight-line, a central core component may appear in the disordered mode \citep{Backer1976,Rankin1992,Rankin1990}. Thus, this second kind of mode-changing clearly suggests an interdependence between the central core and the surrounding cones.
Mode changing has been observed on a wide range of timescales, from a handful of pulses to several years (e.g. \citealp{L10,S22}).
Although $P_3$ can vary over time and between modes, these changes seem global, with all components being modulated with the same $P_3$ and such that the relative sub-pulse phase between components being constant.
This is observed in pulsars with changing drift rates (PSR~B0818$-$41; \citealp{Bhattacharyya2007}), drift directions (PSR~B0826$-$34; \citealp{Gupta2004}) and synchronised between main-pulse and inter-pulse (PSR~B1702$-$19; \citealp{WeltevredeMI2007}).

Some pulsars also exhibit emission `nulling', during which the flux density of the pulsar goes to zero, or near zero, for a few to several hundreds of pulses \citep{Taylor1971,Wang2007} before switching on again. Sustained nulling has often been argued to be a yet another form of mode-changing while occasional single nulls (`pseudo-nulls') may be due to gaps in the carousel and have a different nature \citep{Redman2005}. 

Typically, single-pulse phenomena, including drifting, can change in different modes, which is often interpreted as a rearrangement of sparks on the polar cap.
However, the classical carousel model cannot explain all observed drifting phenomena and hence cannot be a complete model for drifting in pulsar emission.
In particular, it cannot explain the phenomenon of `bi-drifting', where the drift direction is different for different profile components \citep{champion2005}.  
It is important to distinguish this from different drift directions at different times or in different modes, as this can be explained by changes in spark patterns, especially when combined with the effect of aliasing \citep{Szary2022,Wright2022}.
 
Bi-drifting has been reported for five pulsars in the literature \citep{champion2005,Basu2018,patrick2016,bpm19,Shang2024}.
These observations of bi-drifting have led to the development of new models.
For example, \citet{Qiao2004} propose an additional acceleration region where the electric field is oriented in the opposite direction compared to the conventional acceleration region. This allows carousels to rotate in opposite senses.
Alternatively, if all sparks rotate in a single direction, bi-drifting can still be explained if the carousel is not circular.
Seeking to preserve the original idea of two corotating carousels and exploiting a line-of-sight close to the magnetic pole, \citet{Geoff2017} propose a purely geometric model for bi-drifting, where both inner and outer carousels follow a {\it tilted} elliptical path and use this to model drifting in PSRs~J1842$-$0359 and J1842$-$0359 (B1839$-$04).
Such pattern could also arise from sparks that do not rotate around the magnetic axis, but around the point of maximum potential at the polar cap \citep{vanLeeuwen2012}. \citet{Szary2017} demonstrate that such a model can replicate the observed bi-drifting feature of PSR~J0815+0939.
Other models of sub-pulse drifting do not make use of a circular carousel. For example,
\citet{BasuMitra2020} demonstrate a model with a non-dipolar magnetic field close to the neutron star surface, based on the partially screened gap model suggested by \citet{Gilb2003}. This model is able to reproduce a range of drifting phenomena, including bi-drifting, although only for wide pulse profiles.

In this paper we study in detail the drift patterns observed in nine pulsars, all flagged as potential bi-drifters in the analysis of the MeerTime Thousand-Pulsar-Array (TPA) single pulse dataset by \citet{Song2023}. All of these show complex drifting and/or mode changing behaviour. For one of these, PSR~J1921$+$1948, bi-drifting has been discussed before \citep{Shang2024}.
Full details of the observations used are given in Section~\ref{sec:obs}, and the sub-pulse features of individual pulsars are presented in Section~\ref{sec:results}. Discussions of results are presented in Section~\ref{sec:Dis}, followed by conclusions (Section~\ref{sec:conclusion}). 
\clearmaybe

\section{Observations and Data Analysis}
\label{sec:obs}
\begin{table*}
\caption{List of observations used in this paper. Observations with MeerKAT L-band (MK-L), the UHF band (MK-UHF), and with the GMRT are listed. The observation times, observation duration and the number of bins per pulse period are presented. }
\centering
\begin{tabular}{lrrrrrr}\\
\hline
     Name &  Telescope& Obs Time & Nr of Pulses & Obs Duration &No. of Bins \\
     PSR &  & &   &(hr)& \\
\hline
J1418$-$3921  & MK-L & 2020-01-11-02:43:07 
&  1026  & 0.56 &1024  \\
&GMRT & 2023-05-17-19:12:30& 3281& 1.00&2048\\
J1534$-$4428 & MK-L& 2020-01-03-01:03:02
& 1029& 0.35& 1024 \\
&MK-UHF&2024-01-18-04:04:28 &2946&1.00&4096\\
&GMRT & 2023-05-17-20:23:30&2948&1.00 &2048\\
J1537$-$4912 & MK-L& 2019-07-25-19:22:27
& 2991&  0.25 &1024 \\
J1734$-$0212 (B1732-02)  & MK-L& 2020-05-24-21:08:41 
&710 & 0.17 & 1024\\ 
&MK-L&2020-05-06-00:13:16&710&0.17&1024\\
J1803$-$3329 & MK-L& 2020-05-06-01:17:06
&946&0.17  &4096  \\
& MK-UHF&2024-01-20-07:53:51 &5678&1.00&4096\\
&GMRT& 2023-05-17-21:42:30&6116&1.08 &1600\\
J1834$-$1202 & MK-L& 2020-01-03-08:11:07
&1038 & 0.18  &1024\\
J1843$-$0211 &MK-L& 2020-01-20-12:13:12
& 1030 & 0.58 &1024 \\
J1921+1948 (B1918+19)&MK-L&2019-12-16-14:44:43
&1033&0.24&1024\\
J1921+2003 (B1919+20) &MK-L& 2020-08-20-21:11:27
&1030 & 0.22&1024\\
& GMRT & 2023-04-21-03:24:30& 4738&1.00&2048\\
\hline
\label{table: observation}
\end{tabular}
\end{table*}
The pulsars and observations used are listed in Table~\ref{table: observation}.  
We primarily use the MeerTime TPA dataset as described in \citet{Johnston2020} and \citet{Song2023}. These are recorded with the MeerKAT L-band (MK-L) receiver and the PTUSE instrument \citep{Bailes2020}.
These data use the full MeerKAT array ($\sim$64 dishes, depending on availability) and typically have 1024 channels over a bandwidth of 856 MHz, centred at $\sim$1283.58~MHz, and originally sampled at $\sim$38.28$\,\mu$s. A small number of observations (those between 2019-05 and 2020-02) kept only the central 768 channels, giving 642~MHz bandwidth.

We obtained additional data for this study with the MeerKAT UHF receiver (MK-UHF) for PSRs~J1803$-$3329 and J1534$-$4428. These have a centre frequency of $\sim$815.93~MHz and 4096 channels over a bandwidth of 544~MHz, and  the sampling rate was 120.5$\,\mu$s.
We also made additional observations of PSRs~J1418$-$3921, J1534$-$4428, J1803$-$3329 and J1921+2003 with the Giant Metrewave Radio Telescope (GMRT) at Band 4, with a centre frequency of 750~MHz. The recorded bandwidth is 400 MHz split into 4096 channels and sampled every 327.68$\,\mu$s. However, typically only data below 800 MHz was used in the analysis as this yielded higher signal-to-noise ratios.

\subsection{Data Preparation}

For the MeerKAT data the time-series are processed using the standard MeerTime single pulse pipeline.
Single-pulse datasets are produced using \textsc{dspsr} \citep{vanStraten2011}, having been corrected for dispersion and Faraday rotation.
A channel mask is applied to remove known sources of interference, and an automated process also removes additional frequency channels based on the off-pulse statistics.
The same channels are removed from each single pulse, so that the pulse-to-pulse statistics should not be affected by the interference removal.
The MeerKAT L-band pipeline produces single pulse datasets with 1024 samples per period by standard; however, this was increased to 4096 for PSR~J1803$-$3329 to better resolve the narrow trailing component. The data are averaged in frequency. 
 
The UHF data have 4096 longitude bins, and 32 frequency channels are formed, each being the average of 128 original channels. 
These frequency channels were further averaged to three equal parts, after discarding the very first and very last channels.  
The lower part of the band overlaps with that of the GMRT observations, and the upper part overlaps with the MK-L.

The GMRT data are also processed using \textsc{dspsr}. The removal of frequency channels with a low signal-to-noise ratio was carried out manually by direct inspection of the folded profiles. The frequency band was then averaged for further analysis.
The GMRT data were produced with 2048 samples per period by default, except for PSR~J1803$-$3329, where we used 1600 samples per period due to constraints of the underlying sampling rate.

The MeerKAT L-band observation for each pulsar includes a long observation with at least 1000 pulses. A subset of the pulsars is regularly monitored, recording $\sim\!200$ pulses per month \citep{Keith2024,Avishek2024}. 
In our sample, the long TPA `census' observations of PSRs~J1537$-$4912, J1843$-$0211 and J1921+1948 are supplemented with additional short observations, as detailed in Table~\ref{table: observation long}.
These short observations are processed identically to long observations.
The single pulse data are not flux calibrated, and changes in the digital instrumentation mean that the absolute scale of the data varies significantly from observation to observation.
Therefore, in order to combine the individual short observations and the long observation, we scale the data by assuming the off-pulse rms (after baseline subtraction) is dominated by the system temperature (including contributions from the sky at the position of the pulsar) and accounting for the number of antennas used in a given observation.
The latter is particularly important, since the short observations are made with a subset of the full MeerKAT array ($\sim 32$ dishes), because the monitoring campaign splits the dishes into two sub-arrays.
In practice, this means that we normalise the data so that after scaling, multiplied by the number of antennas used, the off-pulse rms are constant across all observations.

\subsection{Analysis Methods}
In order to study the pulse drift in our sample, the prepared single pulse data are analysed with \textsc{psrsalsa} \citep{patrick2016}, largely following methods similar to \citet{Song2023}. What follows is a summary of the methods used. 

\subsubsection{Baseline removal}

To analyse sub-pulse modulations, we must first remove the off-pulse baseline, which typically varies over the duration of the observation.
For most of the observations, this is done by fitting a linear function to the off-pulse region of each pulse and subtracting this. For MK-L observations of PSRs~J1734$-$0212 and J1921$+$1948, we fit and subtract up to three harmonically related sinusoids with the fundamental frequency of $1/P$.

The GMRT observation of PSR~J1921$+$2003 shows periodic radio frequency interference (RFI)
throughout the observation, which leads to spurious detections of sub-pulse drifting.
Fourier analysis of the off-pulse data showed a primary RFI frequency of around 0.0186 cycles per period (cpp).
To mitigate this, we applied a filter to the data in the Fourier domain. 
Each single pulse is Fourier transformed, the RFI frequency (and up to 10 harmonics) was masked and then inverse transformed.

\subsubsection{Longitude Resolved Fluctuation Spectra}

A longitude resolved fluctuation spectrum (LRFS; \citealt{Backer1970, Ben2003}) is computed by Fourier transforming each column of constant pulse longitude in the pulse stack.
This is done separately for blocks of a fixed number of pulses, termed the FFT length, and the resulting power spectra of each block are then averaged.

Because some pulsars exhibit multiple emission modes and their modulations are not stable with precise periodicities, the FFT length should be large enough to have a sufficient frequency resolution to resolve the spectral features, while not being excessive.
Typically, for our full pulse stack analyses, the FFT length is set to around 550 pulses. For MK-L observations, this corresponds to about half the observation length, while for GMRT and MK-UHF observations, it represents roughly one-tenth of the full observation length.

The LRFS reveals the fluctuation frequency ($1/P_3$) of repeating drift bands, as well as longitude stationary intensity modulations.   
If the pulsar has mode changes with different periodic sub-pulse modulations, different $P_3$ values will appear in the LRFS. To analyse the modes in detail, sections of the pulse stack associated with the different modes are identified by eye as well as possible.

For short observations of the three pulsars in Table~\ref{table: observation long}, we combine the LRFS of each observation. 
We achieve this by extracting pulses from each observation in integer multiples of the FFT length.
This avoids any artefacts caused by combining multiple incoherent pulse sequences.

\subsubsection{Subpulse Phase Analysis}

While the power of the complex spectra can be used to measure $P_3$, the phase information can be used to constrain $P_2$. To determine the phase relationship of the modulation cycle in different profile components, the sub-pulse phase as function of pulse longitude are computed. 
The longitude stationary modulation corresponds to the case where the sub-pulse phases are independent of pulse-longitude: the drift bands are flat in the pulse stack. Phase modulation corresponds to sub-pulse drifting such that the sub-pulse phase has a gradient: on average drift bands in the pulse stack are inclined.
Thus, the sub-pulse phase plot reveals the average shape of the drift bands, as the `sub-pulse phase track'. In this paper, a positive gradient implies a positive drift \citep{patrick2016}, which shows drift towards later pulse longitudes. 
The FFT length used for phase analysis is ideally short enough to allow most power of a spectral feature to fall into a single Fourier bin.

\subsubsection{$P_3$ Folding}
\label{sec: P3Folding}

Another method to visualise time-averaged sub-pulse drift properties is to `fold' the pulses over multiple $P_3$ cycles. Since $P_3$ can fluctuate over time, this needs to be compensated for by correcting for phase offsets. 
Initially, the pulse stack is divided into `folding blocks' of 3$P_3$ pulses (or 5$P_3$ pulses, see below) and is folded with a fixed period of $P_3$. If the actual $P_3$ is slightly different from the specified $P_3$ for folding, or the actual $P_3$ is not an integer number of pulses, the drift band in the second block will appear slightly delayed. This phase offset is determined and used to align the modulation patterns of each individual folding block, allowing the 3$P_3$ long blocks to be coherently added together.  
 
Increasing the length of the folding blocks can smear out the drift band obtained. 
For most of our analysis, 3$P_3$ is a reasonable choice, although for PSRs~J1537$-$4912 and J1734$-$0212, the value is increased to 5$P_3$ to reduce the influence of brighter pulses and obtain clearer average drift bands.
To avoid sporadic bright pulses in the MK-L pulse stack of PSR~J1537$-$4912 from dominating the fitted phase offsets during the folding process, the phase offsets were first determined using the pulse stack with the brightest pulses suppressed by applying clipping. 
These fitted offsets were then applied to the full pulse stack without clipping applied. 

For most observations, the $P_3$ cycle in the $P_3$ fold is divided into 50 equally spaced bins; for PSR~J1537$-$4912 the number of bins in each $P_3$ cycle is 120. 
This leads to oversampling,  since the intrinsic resolution of the $P_3$ fold is 1$P$.
To resolve the oversampling, a `Gaussian smoothing' process is performed, effectively convolving each observed pulse with a Gaussian with width chosen to be the number of output phase bins divided by $P_3$ \citep{patrick2016}.
We refer to this width as the `smoothing factor' in the $P_3$ fold.
For combined pulse stacks with short observations, excess pulses beyond the multiples of 3$P_3$ are neglected at the observation boundaries, to ensure that each block contains a continuous sequence of pulses.

We then divide the folded data by the mean profile.
This emphasises the overall shape of the drift band, including the edges of the profile where the emission is weaker.
Therefore, the resulting $P_3$ fold shows the emission variability relative to the mean intensity at each pulse longitude.
\clearmaybe

\section{Results}
\begin{table}
\caption{The $P_3$ derived from the centroid value of the features in the LRFS in units of the rotation period for each pulsar. Note that the error in the centroid $P_3$ is typically much smaller than the width of the observed $P_3$ distribution.  
PSRs~J1418$-$3921, J1803$-$3329, J1843$-$0211 and J1921$+$1948 have multiple $P_3$s, corresponding to different emission modes. For reference, we also include the rotation period, period derivative $\dot{P}$, the characteristic age $\tau_\mathrm{c}$, the spin down energy loss rate $\dot{E}$. These values are taken from the ATNF Pulsar Catalogue version 2.30 (\citealp{Manchester2005}, \url{https://www.atnf.csiro.au/research/pulsar/psrcat}). }

\centering
\begin{tabular}{lrrrrrr}\\
\hline
Name& $P_3/P$ &Period&$\log_{10}{\!\dot{P}}$&$\log_{10}{\!\tau_\mathrm{c}}$&$\log_{10}{\!\dot{E}}$

\\
PSR&(Centroid)&(s)&&(yrs)&(ergs/s)& 
\\
J1418$-$3921&2.50$\pm$0.02&1.097&$-$15.051&7.290&31.431
\\
&2.53$\pm$0.03&&&&\\

J1534$-$4428&10.6$\pm$0.1&1.221&$-$15.745&8.033&30.591
\\

J1537$-$4912&53$\pm$4&0.301&$-$14.714&6.393&33.447

\\
J1734$-$0212&2.17$\pm$0.03&0.839&$-$15.376&7.500&31.447
\\
J1803$-$3329&12.2$\pm$0.3&0.633&$-$15.472&7.474&31.716
\\
&$2.3\pm 0.1$&&&&\\
J1834$-$1202&24$\pm$2&0.610&$-$17.174&9.155&30.079
\\
J1843$-$0211&2.27$\pm$0.02&2.028&$-$13.840&6.346&31.833
\\
&$3.1\pm 0.1$&&&&\\
J1921$+$1948&3.7$\pm$0.2&0.821&$-$15.048&7.161&31.806
\\
&5.5$\pm$0.2&\\
J1921$+$2003&8.30$\pm$0.07&0.761&$-$16.301&8.382&30.653
\\
\hline
\label{table: observed p3}
\end{tabular}
\end{table}

\label{sec:results}
\begin{figure*}
    \centering
    \includegraphics[width=1\textwidth,height=0.9\textheight]{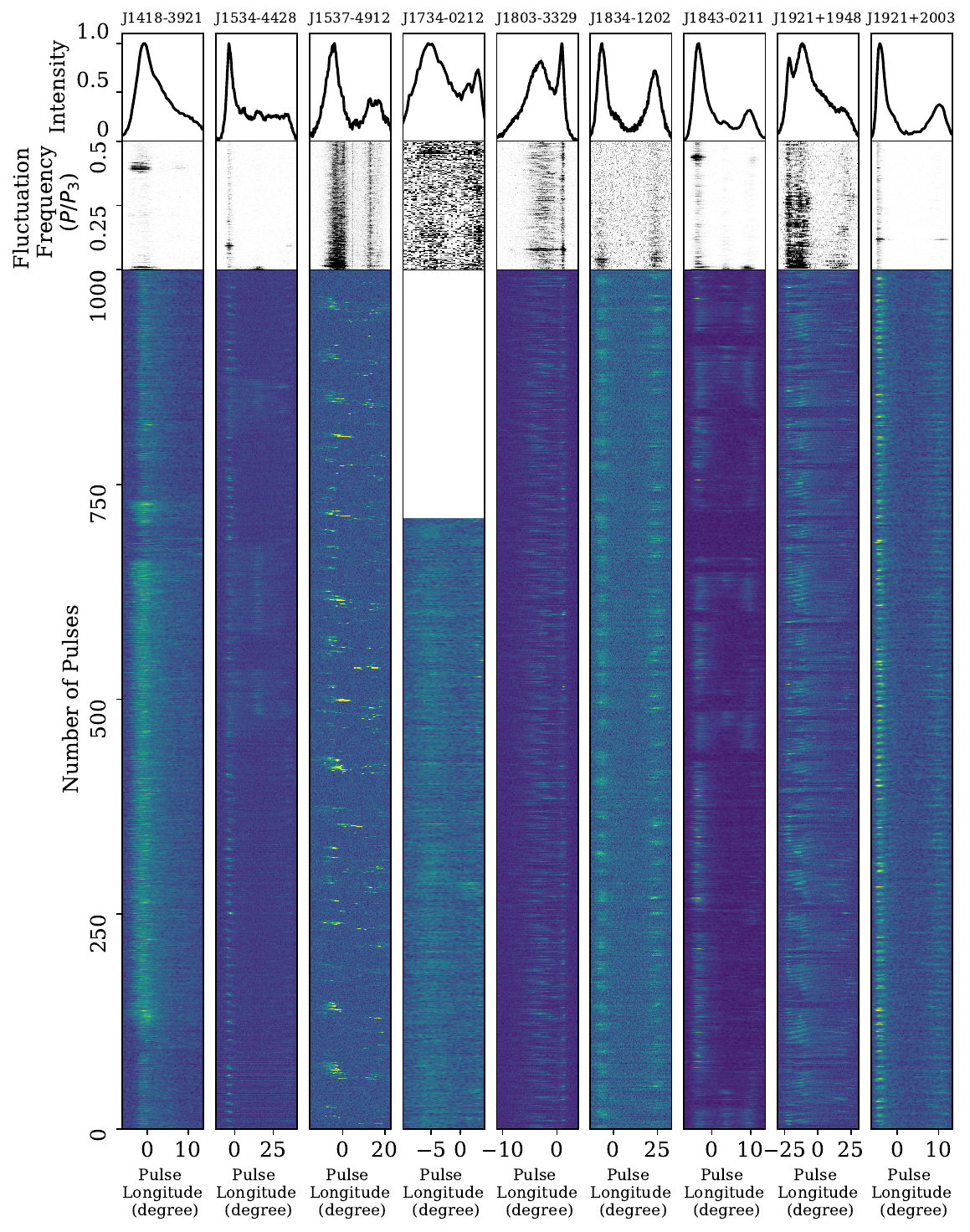}
    \caption{MK-L ($\sim\!1284$\,MHz) pulse stacks, LRFS and mean profiles of our sample of pulsars that show bi-drifting. The bottom panels show the pulse stacks of up to 1000 pulses from these observations of each pulsar. 
    For PSRs~J1537$-$4921 and J1921+2003, the power is clipped at 1/5 the maximum power to bring out the weak single pulse features.
    The top and middle panels show the mean profiles and LRFS of the full pulse stack respectively, with the power clipped at 0.1 of the maximum power to show weaker spectral features clearer.}
\label{fig:PulseStacksTPA}
\end{figure*}

\begin{figure}
    \centering
    \includegraphics[width=0.45\textwidth]{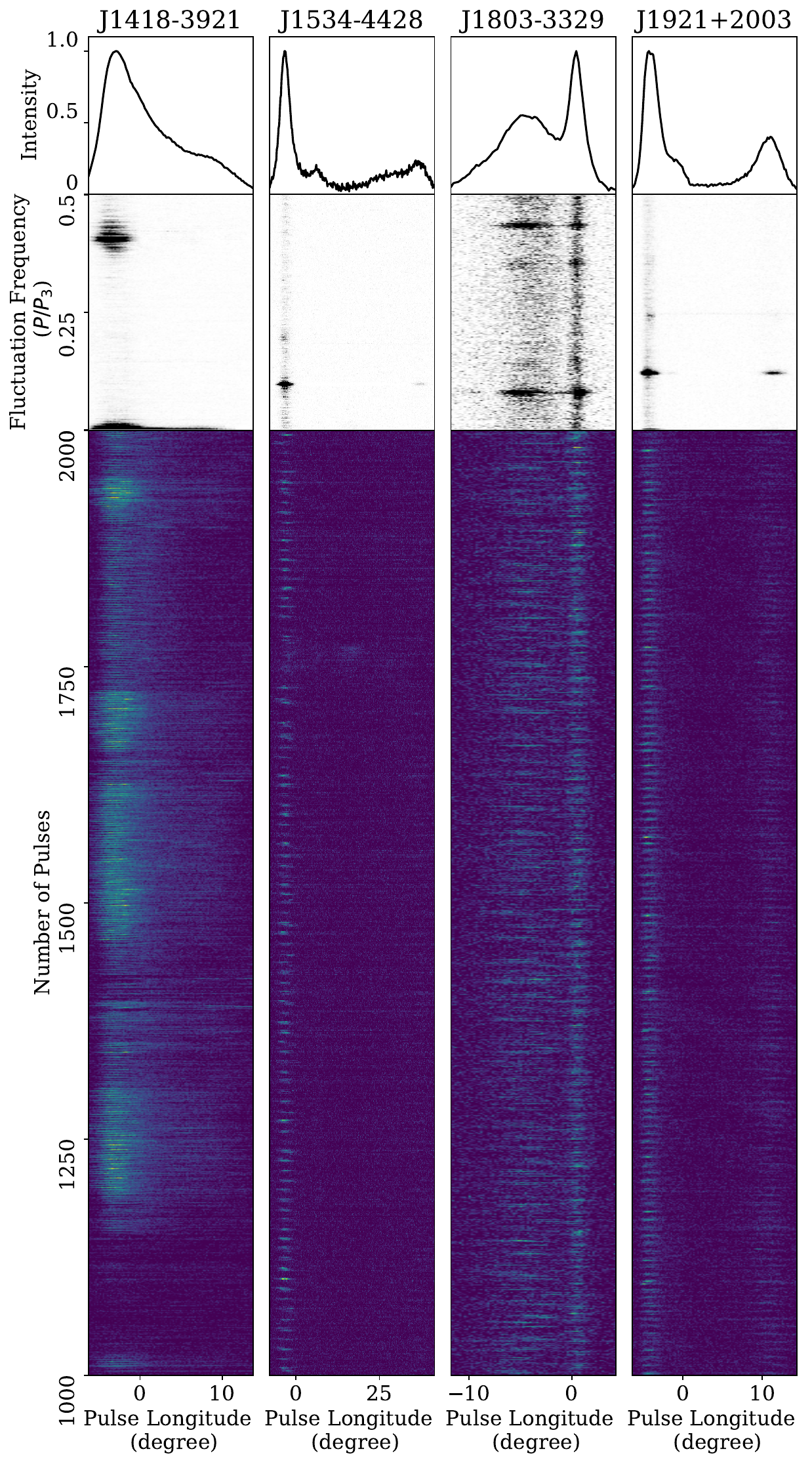}
    \caption{As Fig.~\ref{fig:PulseStacksTPA}, but showing pulse stacks of the GMRT observations ($\sim650$\,MHz). 
    The pulse stacks show 1000 pulses, starting from the 1000\textsuperscript{th} pulse of the observations to show  mode changes.}
    \label{fig:PulseStacksGMRT}
\end{figure}

Fig.~\ref{fig:PulseStacksTPA} shows the pulse stacks of the MK-L data (see also Table~\ref{table: observation}). 
This figure gives a view of the general features of pulse-to-pulse modulation, including the mode changing in pulsars J1418$-$3921, J1534$-$4428, J1843$-$0211 and J1921$+$1948 (see more details in Section. \ref{sec:J1912+1948}).
The sub-pulse modulation of the individual pulsars will be presented in the following sub-sections, with a focus on the potential bi-drifting.
The differences of the emission modes will be addressed. Additionally, 
the GMRT observations at lower frequencies
(Fig.~\ref{fig:PulseStacksGMRT}) and the MK-UHF band will also be discussed.

The $P_3$ values are measured as the centroid values of the spectral features in the LRFS (with \textsc{psrsalsa}, see \citealp{patrick2016}) and are given in Table~\ref{table: observed p3}. 
The associated errors do not reflect the magnitude of the actual fluctuations in $P_3$; instead, they indicate the uncertainties in determining the centroid values of the selected region around the features. The errors are based on multiple selections of regions around features in two halves of the pulse stacks. This method will take into account the subjective choices of the region and the slight variation of $P_3$ in the pulse stack.
As will be discussed in the following subsections, some pulsars have multiple modes with different $P_3$, or multiple $P_3$ that correspond to different kinds of modulation.
The fluctuation of $P_3$ is associated with the width of the fluctuation frequency peak that can be read from the side panels of the LRFS plots (e.g. in Fig.~\ref{fig:J1418-3921TPAModesLRFS}). The $P_3$ used for folding is determined by the centroid value of the region confined close to the fluctuation frequency peak. The widths of fluctuation peaks are larger than the $P_3$ centroid errors quoted in the table. 

\clearmaybe
\subsection{PSR J1418--3921}
\label{sec: J1418-3921}
\begin{figure*}
\centering
\begin{tabular}{cc}
\includegraphics[width=0.45\textwidth]{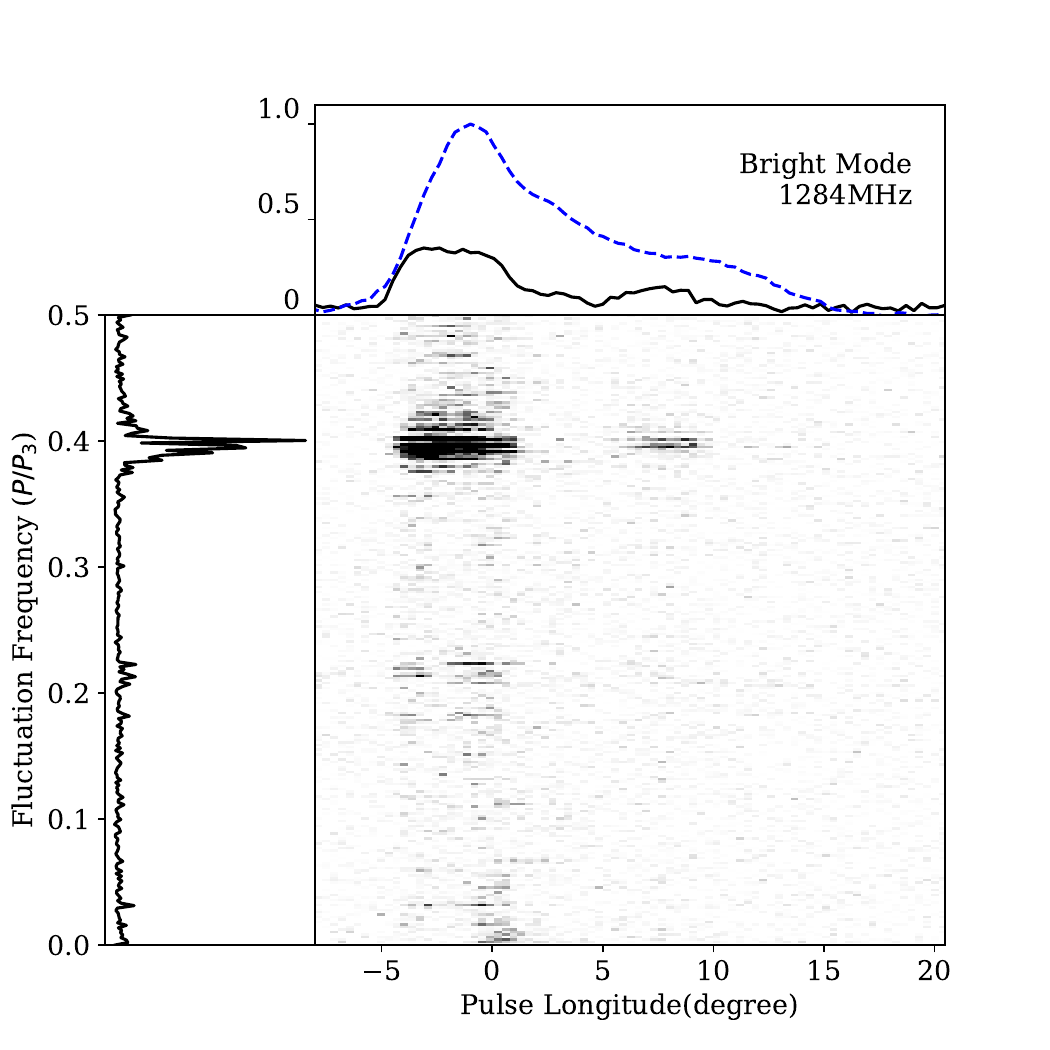}&
\includegraphics[width=0.45\textwidth]{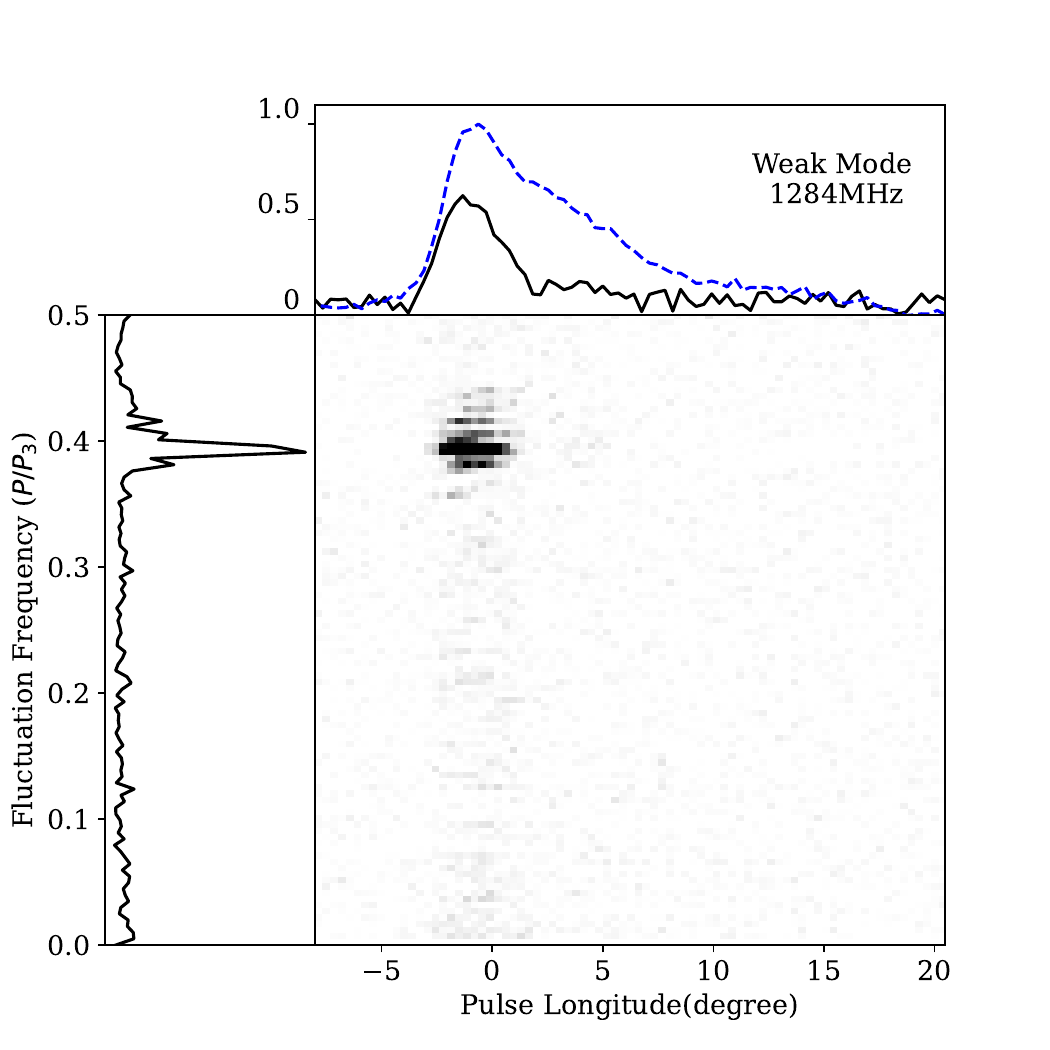}
\end{tabular}
\caption{The figure shows the LRFS of the two modes from the long MK-L observation of PSR~J1418$-$3921. The left plot shows the bright mode and the right plot shows the weak mode. The plots have three panels: the central panel shows the LRFS; the top panel shows the standard deviation of the pulse stack (black solid) which indicates the modulation intensity, and the mean profile of the mode separated pulse stack (blue dashed).
The left side-panel shows the  
integrated fluctuation power, 
which emphasises the strength of the fluctuations as function of fluctuation frequency. The FFT lengths used are 82 for the the weak mode and 512 for the bright mode.}
\label{fig:J1418-3921TPAModesLRFS}
\end{figure*}
\begin{figure*}
    \centering
    \begin{tabular}{cc}
    \includegraphics[width=0.45\textwidth]{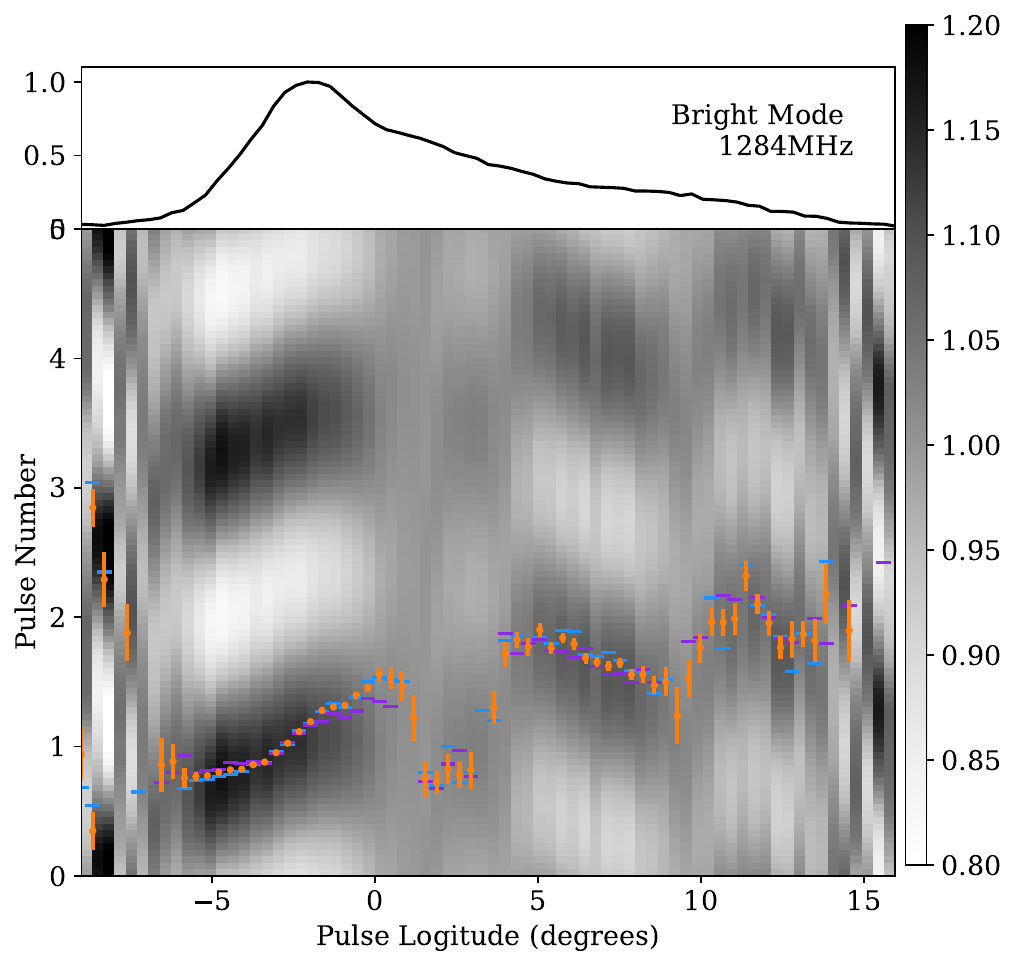}&
    \includegraphics[width=0.45\textwidth]{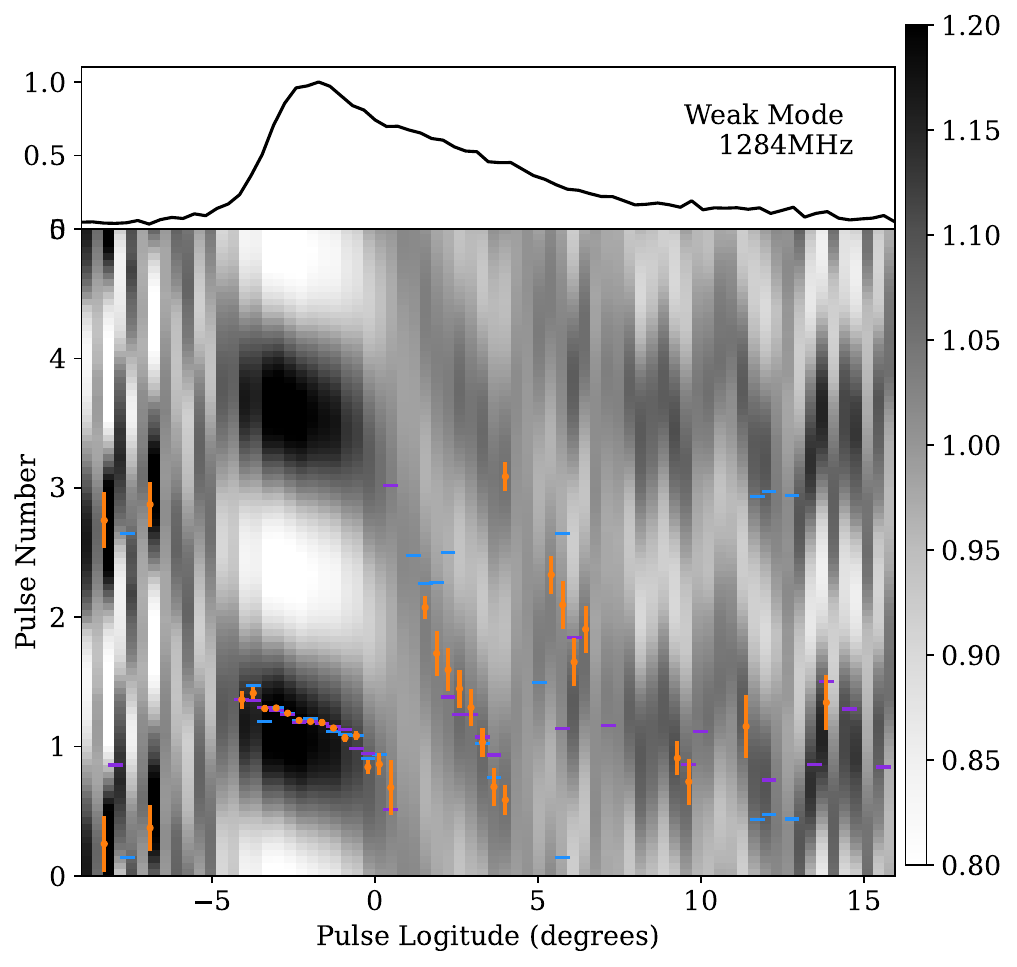}\\
    \includegraphics[width=0.45\textwidth]{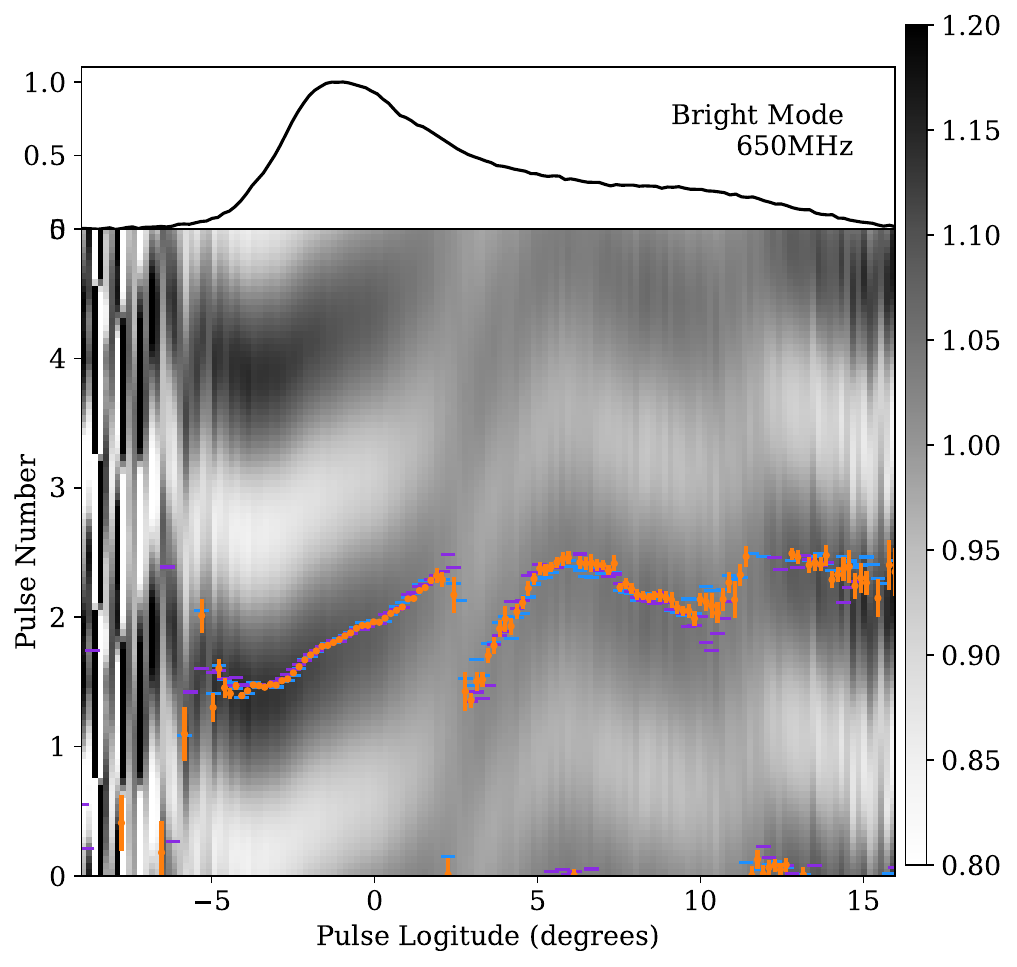}&
    \includegraphics[width=0.45\textwidth]{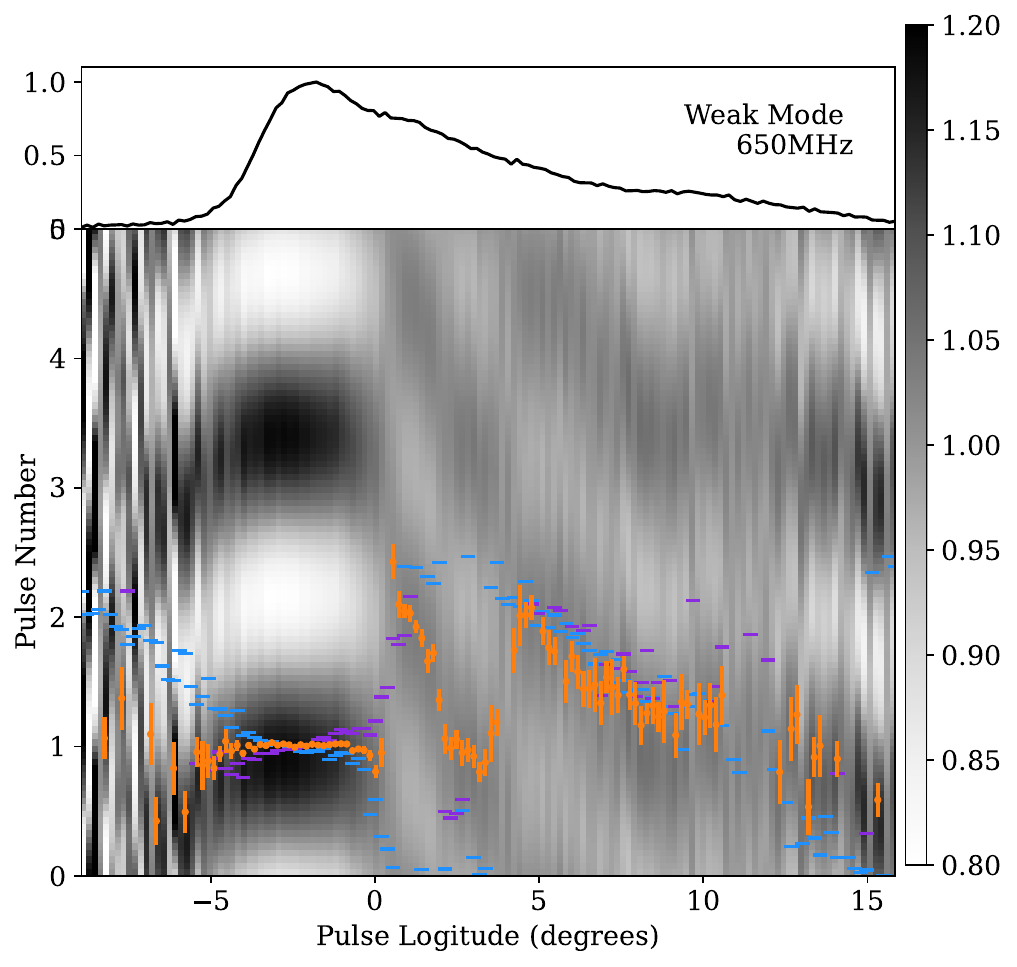}
    \end{tabular}
    \caption{The $P_3$ fold plots of J1418$-$3921 MK-L (top row) and GMRT (bottom row) observations in the bright (left column) and weak (right column) modes. The $P_3$ used for folding is 2.5$P$, the smoothing factor used is 20. The top panel shows the mean pulse profile of the pulse stack. The bottom panel shows the $P_3$ fold together with sub-pulse phases. The sub-pulse phases of the full pulse stack are represented by orange dots with error bars, the phases of the first and second halves of the pulse stack are marked with blue and violet horizontal bars. The plotted phases are limited to those with error bars smaller than 10\% of the $P_3$. 
    The sub-pulse phases are vertically aligned to the highest intensity of the $P_3$ fold features (see the main text for further details). Since the $P_3$-fold is cyclic, two cycles are shown to ensure continuity around the wrap boundary. To make the plot clearer, only one complete phase track is shown.
    The colours represent the intensity fractions relative to the mean. For more details, see Section \ref{sec: P3Folding}.}
    \label{fig:J1418-3921P3Fold}
\end{figure*}

We used two observations of PSR~J1418$-$3921 (Table~\ref{table: observation}), one from MK-L (Fig.~\ref{fig:PulseStacksTPA}) and one from GMRT (Fig.~\ref{fig:PulseStacksGMRT}). 
The average pulse profile is similar in both observations, featuring a bright leading component with a long tail on the trailing edge.  

From the pulse stack shown in Fig.~\ref{fig:PulseStacksTPA}, two modes can be seen, a bright mode and a weak mode.
Across the 4307 pulses in the two observations, the pulsar is in the bright mode for $\sim$52\% of the pulses and in the weak mode for $\sim$34\%. 
In addition, it has very weak pulses that might be nulls for $\sim$10\%.
The transition from the weak mode to the bright mode is accompanied by a section of pulses with a low-intensity leading component (for example, around pulse number 100 in Fig.~\ref{fig:PulseStacksTPA}). 
Although we could not conclude that the mode change is frequency dependent, the mode change appears to occur much more frequently in the GMRT data (Fig.~\ref{fig:PulseStacksGMRT}) compared to the MK-L observation.
We see the null mode only in the first 1500 pulses of the GMRT observation (about 6\% the pulses in this section).

We computed the LRFS of each mode of the pulsar separately. 
The mode separated LRFS plots for the MK-L data are presented in Fig.~\ref{fig:J1418-3921TPAModesLRFS}. The GMRT data (not shown) show similar sub-pulse fluctuation features.
The FFT length used for the spectral analysis was shorter for the weak mode, as the pulsar does not stay in this mode for a long time and there are not many pulses in this mode.

The mean profiles of the two modes can be seen in the upper panels of Fig.~\ref{fig:J1418-3921TPAModesLRFS}. Both modes have very similar shapes, so the main distinguishing character is the average intensity of the pulses.
The bright mode is slightly broader; in particular, the leading shoulder extends about 2 degrees in earlier longitudes than in the weak mode.
The LRFS of the bright mode shows additional power in the trailing part of the profile, which seems to be largely absent in the weak mode.

The two modes share a nearly identical fluctuation frequency, around 0.39~cpp. 
The fluctuation frequency in the weak mode appear to be the same as in the bright mode.
In the bright mode, there seems to be an additional fluctuation feature around 0.22~cpp, which we believe to be the second harmonic at $\sim\!0.78$~cpp that would appear aliased at $\sim\!0.22$~cpp. This harmonic is not seen in the weak mode. 

The $P_3$ fold of the bright and weak modes from the MK-L and GMRT datasets are shown in Fig.~\ref{fig:J1418-3921P3Fold}.
The top panels of the $P_3$ fold plot display the mean profiles, normalised by the peak intensity. 
The $P_3$-folded average drift bands are shown in the main panel. Overlaid are the sub-pulse phases obtained from spectral analysis, which are vertically aligned at the location of the highest intensity of the drift bands.
The MK-L and GMRT observations are each divided into two sections and the sub-pulse phases from each are shown together with the phases obtained from the full pulse stack. The sub-pulse phases from the sections are displayed as violet and blue bars, illustrating the range of variation.
This shows that the patterns obtained are consistent.
The phase tracks for the two halves of the data are aligned in phase with the track of the full pulse stack using a least-squares method.

Both the GMRT and MK-L observations show broadly similar features in the bright mode, including changes in drift slope and direction across the profile, i.e. bi-drifting. There are also discontinuities in the sub-pulse phase tracks.
In the GMRT observation, the leading component (from about $-5$ to 2 degrees) has a positive drift and a trailing component (about 13 to 15 degrees) has a negative drift. There is also another pair of positive and negative drift bands nested between them (about 3 to 11 degrees). The sub-pulse phases are not continuous between the inner and outer pairs of drift bands. It seems to have a phase jump between longitude $2$ and 3 degrees for an approximately 0.4$P_3$ phase difference, although this is less clear in the $P_3$ fold.  
The longitude ranges of the leading and the trailing components are slightly shorter than those observed in the GMRT data. Additionally, there is a step-like section in phase, from about 1 to about 4 degrees pulse longitude. The modulations between the leading and trailing components are not as strong as in the GMRT observation. 

In the MK-L observation, the leading component exhibits a clear negative drift in the weak mode, whereas the trailing component shows some indication of a negative drift, although its significance is low because the phases of the two halves of the data do not agree well. Interestingly, the drift direction of the leading component in the weak mode is opposite to that in the bright mode. This mode changing behaviour is similar to PSR~B2303+30~\citep{Redman2005}, in which the drift direction changes in different modes, while $P_3$ remains the same. 

In contrast, the leading component of the weak mode in the GMRT observation shows a flat drift band in the $P_3$ fold,
and the sub-pulse phase tracks for the two halves suggest the drift direction changes.
The instability of the drift pattern in this mode makes it difficult to draw definitive conclusions about whether the drift direction in the leading component changes at different frequencies.

Looking at a sample of individual pulses shown in Fig.~\ref{fig:J1418-3921Sections}, phase jumps are indeed visible between adjacent profile components in the bright mode (left plot). 
The figure demonstrates that drift features are challenging to identify directly in the pulse stack, especially when $P_3$ is close to 2. In a section of the weak mode (right plot), it appears even more difficult to observe drift features directly. However, statistical analysis strongly suggests the presence of complex bi-drifting in the bright mode of this pulsar.

\clearmaybe
\subsection{PSR J1534--4428}
\label{sec:J1534}
\begin{figure*}
    \centering
    \begin{tabular}{cc}
    \includegraphics[width=0.45\textwidth]{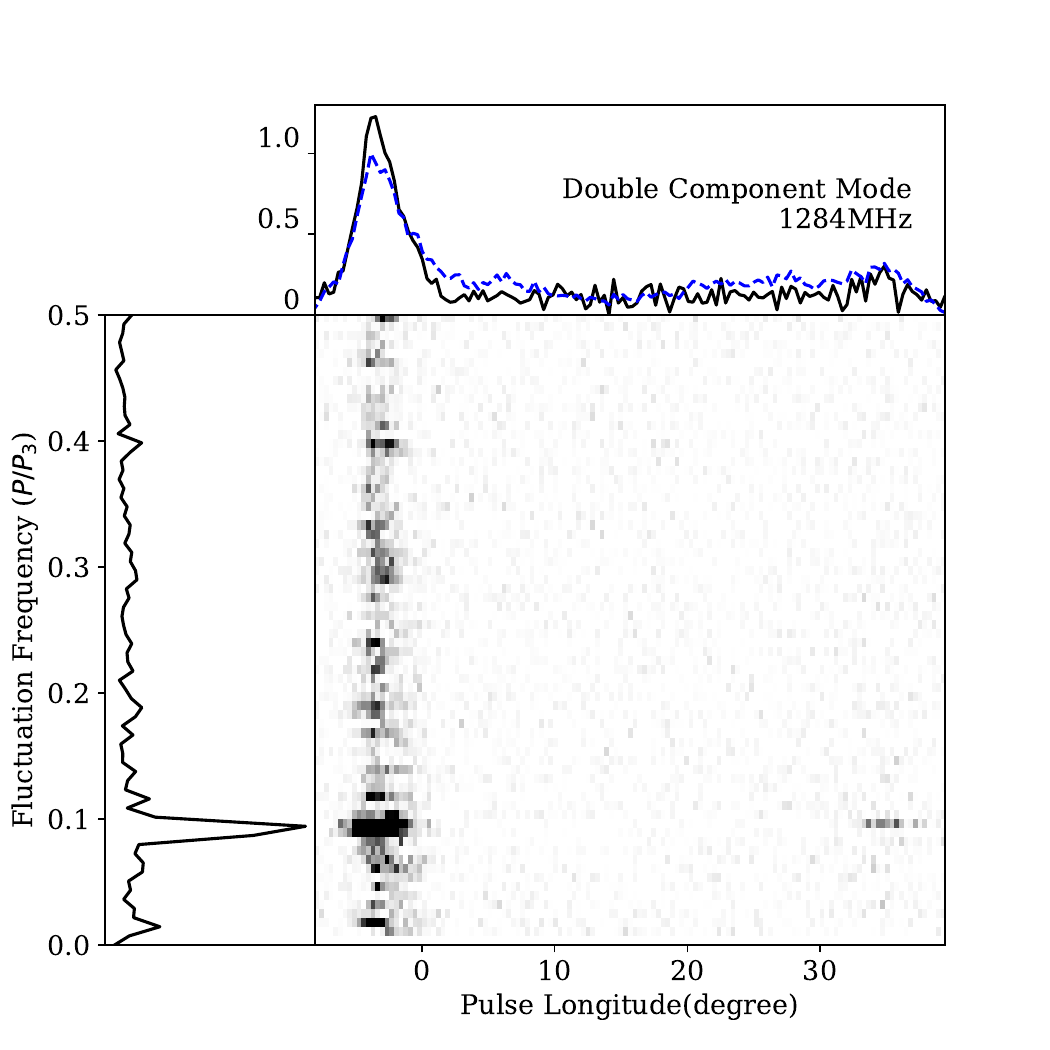}
    &
    \includegraphics[width=0.45\textwidth]{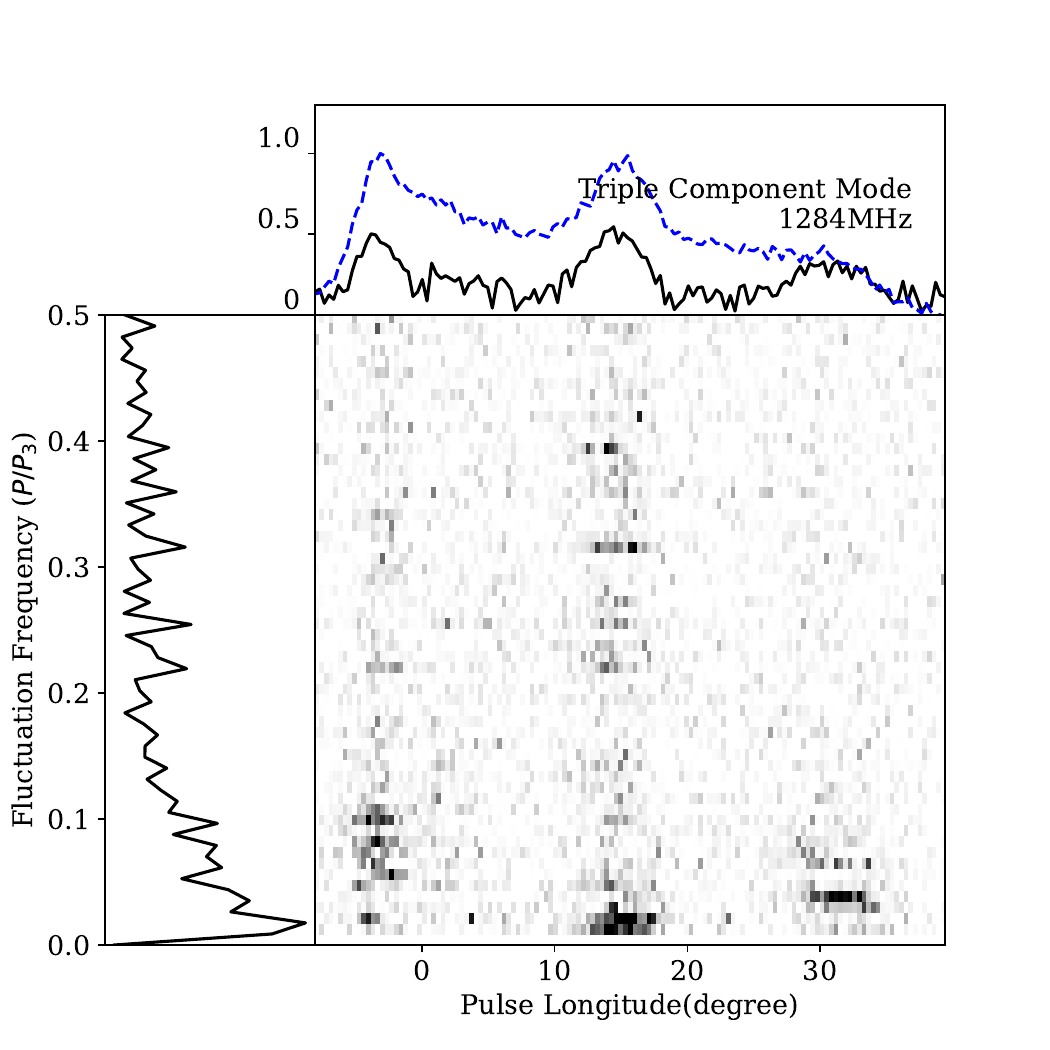}
    \end{tabular}
    \caption{LRFS plots of the different modes of PSR~J1534$-$4428 with the MK-L observation. The panel arrangement is the same as described in Fig.~\ref{fig:J1418-3921TPAModesLRFS}. The FFT lengths used are 138 for the double component and 114 for the triple component modes.}
    \label{fig:J1534-4428LRFS}
\end{figure*}

\begin{figure}
    \centering
     \includegraphics[width=0.45\textwidth]{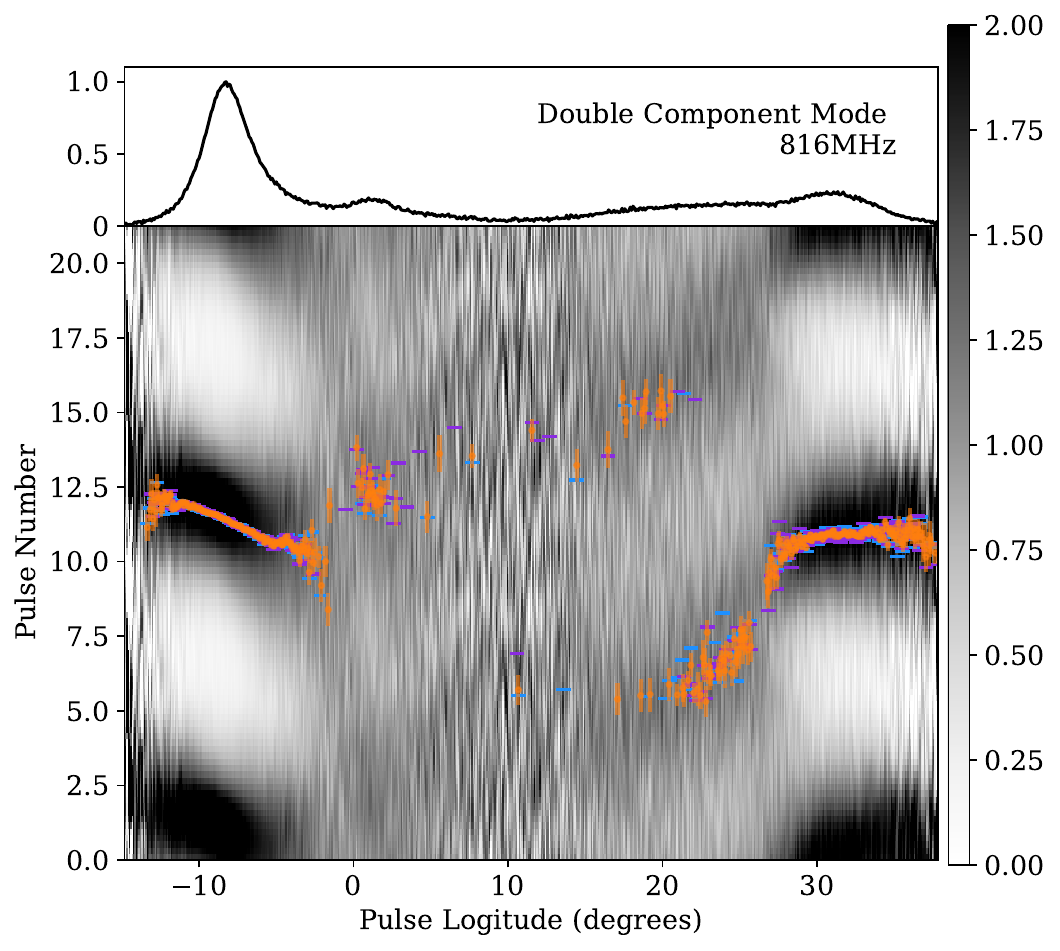}

    \caption{$P_3$ fold plots of double-component mode of PSR~J1534$-$4428 with MK-UHF observation. The $P_3$ used for folding is 10.6248$P$. The smoothing factor used is 4. The detailed description of the plots can be found in Fig.~\ref{fig:J1418-3921P3Fold}. }
    \label{fig:J1534-4428P3Fold}
\end{figure}

We used three observations of PSR~J1534$-$4428 (Table~\ref{table: observation}): one from the MK-L observation (Fig.~\ref{fig:PulseStacksTPA}), one from the GMRT observation (Fig.~\ref{fig:PulseStacksGMRT}) and one from the MK-UHF observation. 
The mean profile of the MK-L observation in Fig.~\ref{fig:PulseStacksTPA} shows multiple components with a width of about 50~degrees. 
The pulsar predominately shows a double-peaked mode with a much stronger leading component. 
Additionally, there is a more symmetric three-component mode with a more prominent central component.
There also seems to be a period of transition from the two component mode to the three component mode where the profile is more symmetric, but central component is much weaker.

The mode-separated LRFS of MK-L observation is shown in Fig.~\ref{fig:J1534-4428LRFS}. 
The double component mode shows fluctuation frequency of about 0.09~cpp, which corresponds to the sub-pulse drift. 
In the three-component mode, the LRFS peaks towards the lowest measurable fluctuation frequency, which is related to the appearance of the middle component during the mode change.

The sub-pulse phase and $P_3$ fold patterns in MK-L, MK-UHF and GMRT observations are similar.
Fig.~\ref{fig:J1534-4428P3Fold} shows the $P_3$ fold of the double component mode from the MK-UHF observation (full band). 
It shows that the leading component has a negative drift, which also can be observed directly in the pulse stack in Fig.~\ref{fig:PulseStacksTPA}.

The drift in the trailing component is hard to see in the $P_3$ fold. 
However, fitting a straight line shows a positive gradient in subpulse phase with a significance of $\sim8\sigma$. 
Consistent positive gradients are found in both halves of the MK-L data ($\sim2\sigma$), as well as the GMRT data ($\sim 1.5\sigma$).

\begin{figure}
    \centering     \includegraphics[width=0.45\textwidth]{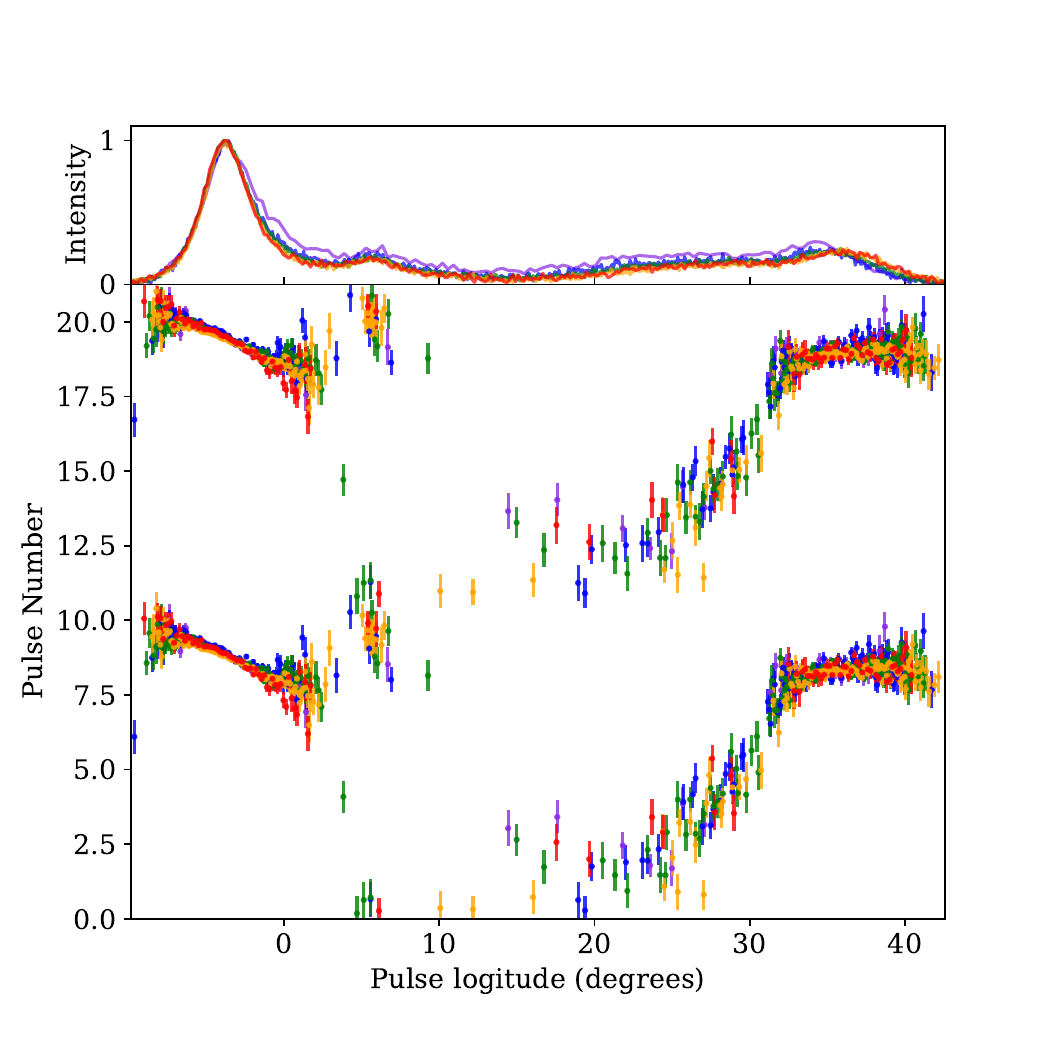}
    \caption{Plot of relative phases of the double-component mode from the MK-L, three frequency bands of MK-UHF and GMRT observations of PSR~J1534$-$4428. The top panel displays the mean profiles of the full pulse stack from the five observations, aligned in longitude. The colours follow the order of light spectrum, representing the frequencies from high (violet for MK-L observation) to low (red for GMRT observation). The bottom panel shows the corresponding phase tracks, with adjusted phases and longitudes. For details on the alignment process, refer to the main text. }
    \label{fig:J1534-4428PhasesTogether}
\end{figure}

To examine the frequency evolution of the double component mode, the profiles and sub-pulse phase tracks from the three MK-UHF sub-bands, as well as the MK-L and GMRT data, are plotted together in Fig.~\ref{fig:J1534-4428PhasesTogether}.
The longitude axis of each profile has been adjusted so that the leading peaks are aligned, as we feel this results in the best alignment of the sub-pulse phase tracks. The sub-pulse phases of the MK-UHF and GMRT data have been adjusted to match the MK-L data.
There are slight changes in profile shape with frequency, especially at the bridge between the first peak and the small bump next to it. 
We do not see any significant evolution of the sub-pulse phase over frequency, and all the observations are consistent.

The overall properties of the single-pulse emission of this pulsar are similar to the known bi-drifting pulsar PSR~B1839$-$04 \citep{patrick2016}.
This pulsar also shows two modes, with an asymmetric double mode showing bi-drifting and a more symmetric mode without clear drift patterns.
The overall profile shape of this pulsar is `mirrored' from PSR~B1839$-$04, which has a more intense trailing component.
\clearmaybe
\subsection{PSR J1537--4912}
\label{sec:J1537-4912}

PSR~J1537$-$4912 has a double-peaked profile, with a brighter leading component and an overall width of around 40 degrees.
The single pulses are characterised by bright clusters of narrow pulses, especially in the trailing component, interspaced with long nulls (up to $\sim20$ per cent).
This erratic single-pulse behaviour leads to a broad spread of power in the the LRFS, although there is a narrow feature around around 0.017~cpp. 

The instability of fluctuation might be related to the young characteristic age (Table~\ref{table: observed p3}) of the pulsar \citep{Rankin1986,Rankin1993}.

\begin{figure}
    \centering
    \includegraphics[width=0.45\textwidth]{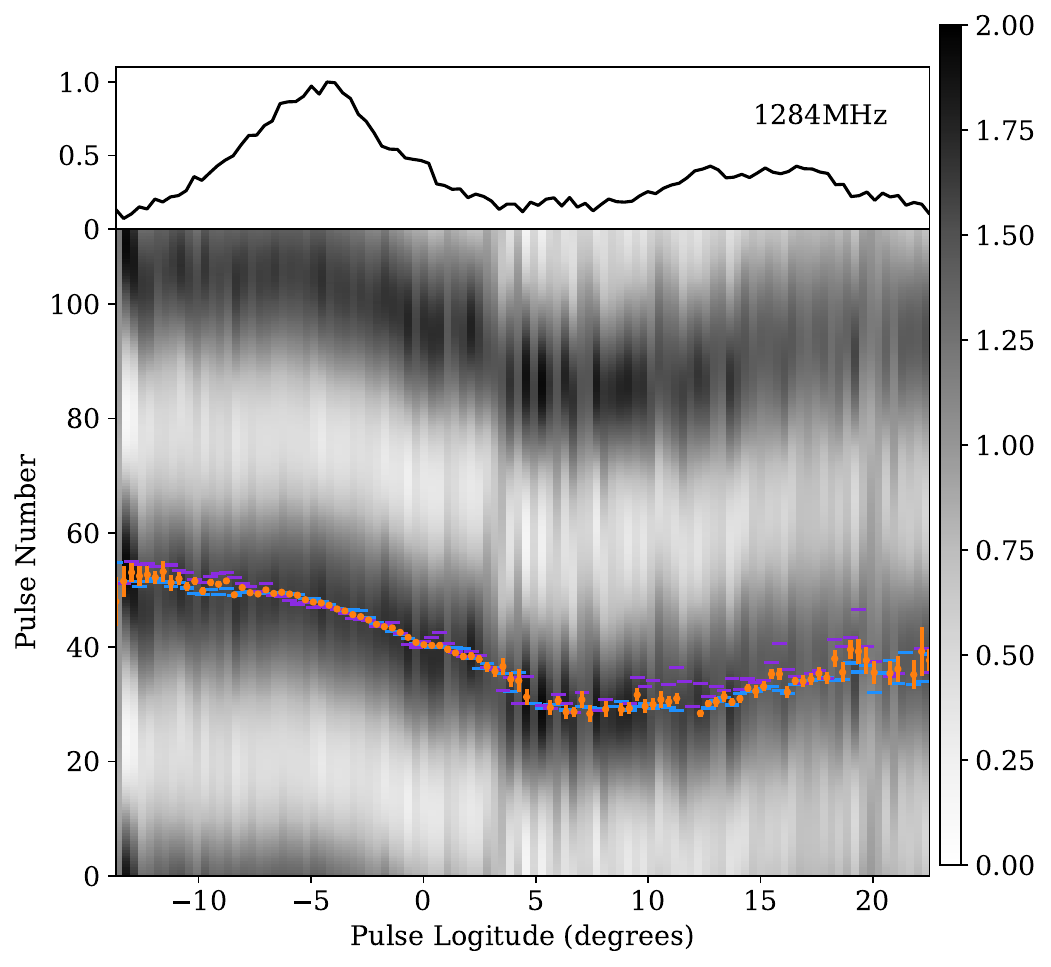}
    \caption{$P_3$ fold of the longest MK-L observation of PSR~J1537$-$4912. The $P_3$ used for folding is 56.577$P$. The smoothing factor is 4. For phase calculations, the FFT length was set to 248. The relative sub-pulse phases are overlaid with the $P_3$ fold with error bars. For a more detailed description of the plot, refer to Fig.~\ref{fig:J1418-3921P3Fold}.}
    \label{fig:J1537-4912TPA}
\end{figure}

Fig.~\ref{fig:J1537-4912TPA} shows the $P_3$ fold plot overlaid with sub-pulse phases as computed from the combination of MK-L observations listed in Table~\ref{table: observation long}.
Clear drift bands are visible with a change in drift direction between the components.

We note that this pulsar is not detected in 22 of the 58 short observations, even though the observations are much longer (typically 300--600 pulses) than the typical nulling interval ($\sim$100 pulses) seen in the long observation.
There are some instances where it remains undetected in consecutive observations over several months, however we do not have sufficient cadence to determine if there really is a pattern or timescale to the longer term nulling.

\clearmaybe
\subsection{PSR J1734--0212}
\label{sec:J1734-0212}
PSR~J1734$-$0212 has a broadly double-peaked profile composed of overlapping components with an overall width of $\sim$13 degrees.
Two MK-L observations, each consisting of 710 pulses, are combined in this analysis.
The sub-pulse modulation of PSR~J1734$-$0212 is not very pronounced, but there is a distinct feature in the LRFS at a fluctuation frequency of 0.46~cpp.

\begin{figure}
    \centering
    \includegraphics[width=0.45\textwidth]{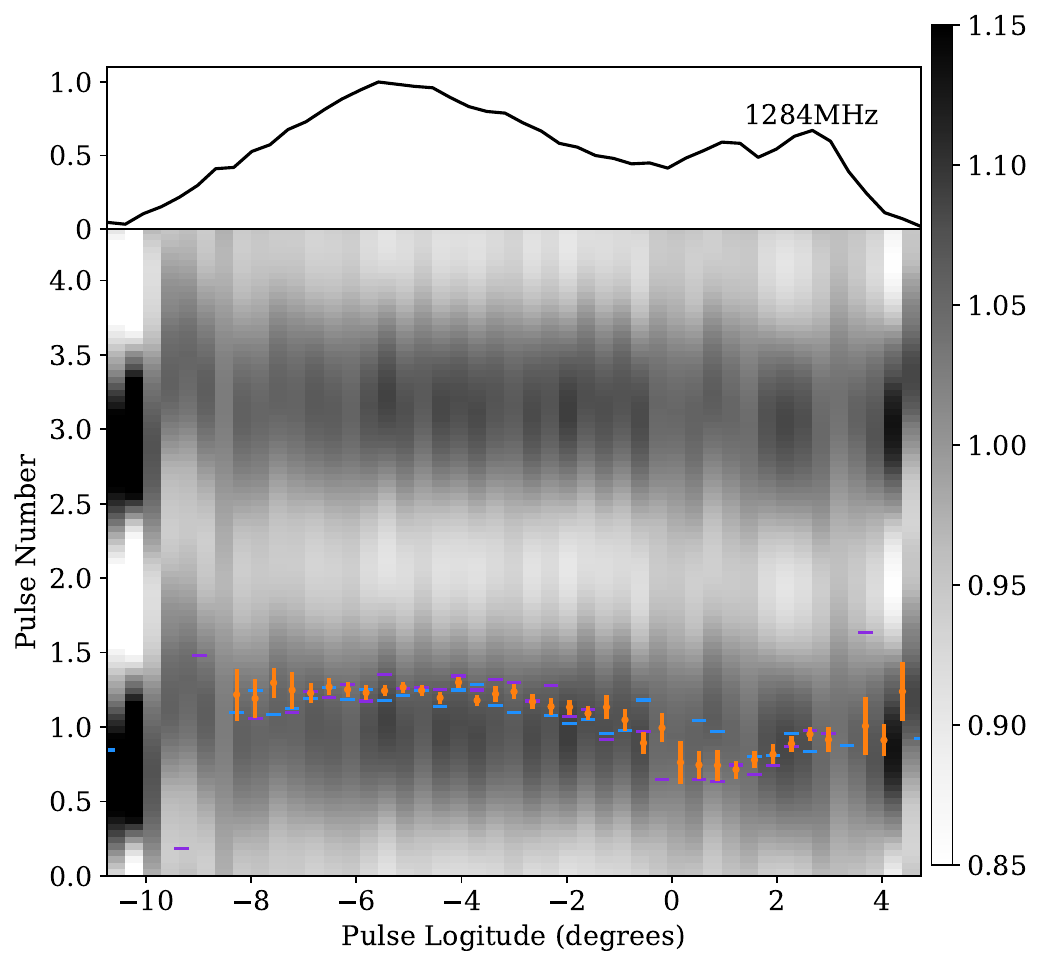}
    \caption{$P_3$ fold plot of PSR~J1734$-$0212 with MK-L observations. The $P_3$ used for folding is 2.173$P$. The smoothing factor is 24. The plot includes some of the off-pulse region. For a more detailed description of the plot, refer to Fig.~\ref{fig:J1418-3921P3Fold}.}
    \label{fig:J1734-0212TPA}
\end{figure}

Fig.~\ref{fig:J1734-0212TPA} shows the $P_3$ fold plot along with the sub-pulse phases. 
The leading component seems to have a slight negative drift between around $-3$ and 0 degrees in longitude and a positive drift in the trailing component.

\clearmaybe

\begin{figure*}
    \centering
    \begin{tabular}{cc}
    \includegraphics[width=0.35\textwidth]{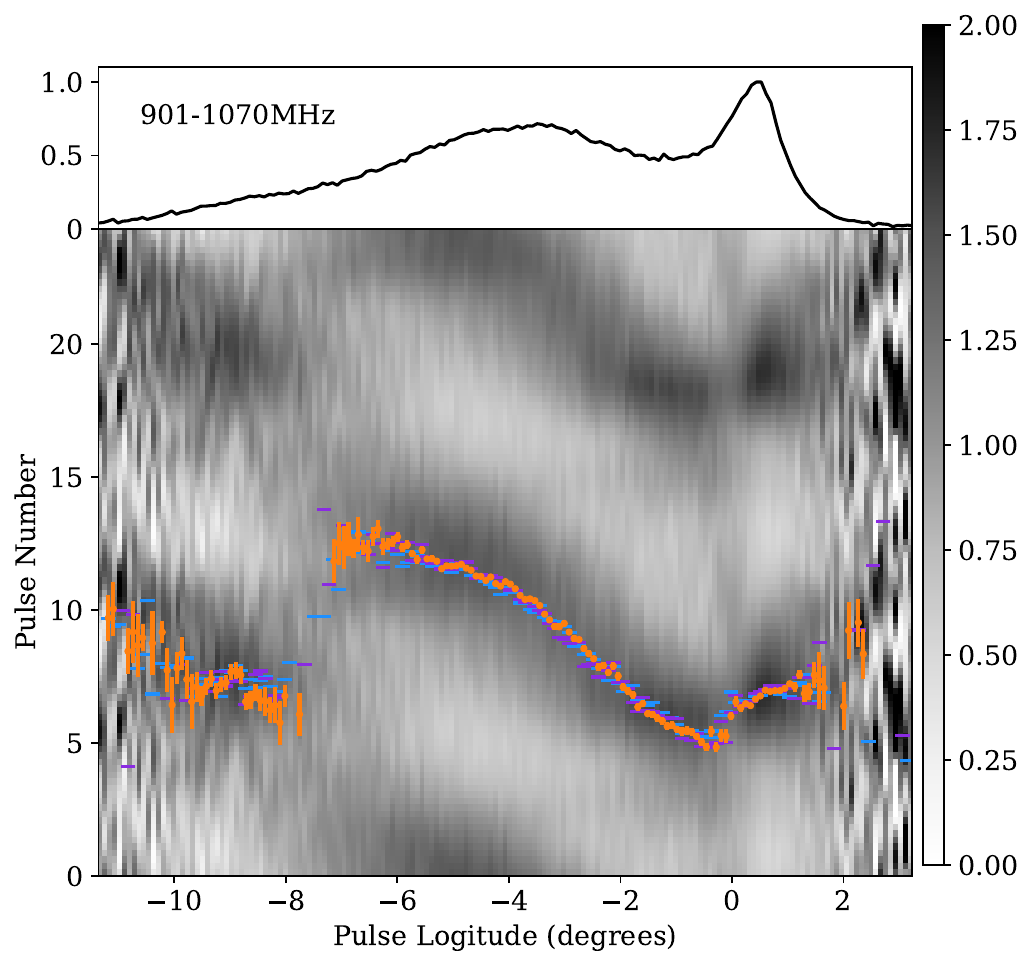}&
     \includegraphics[width=0.35\textwidth]{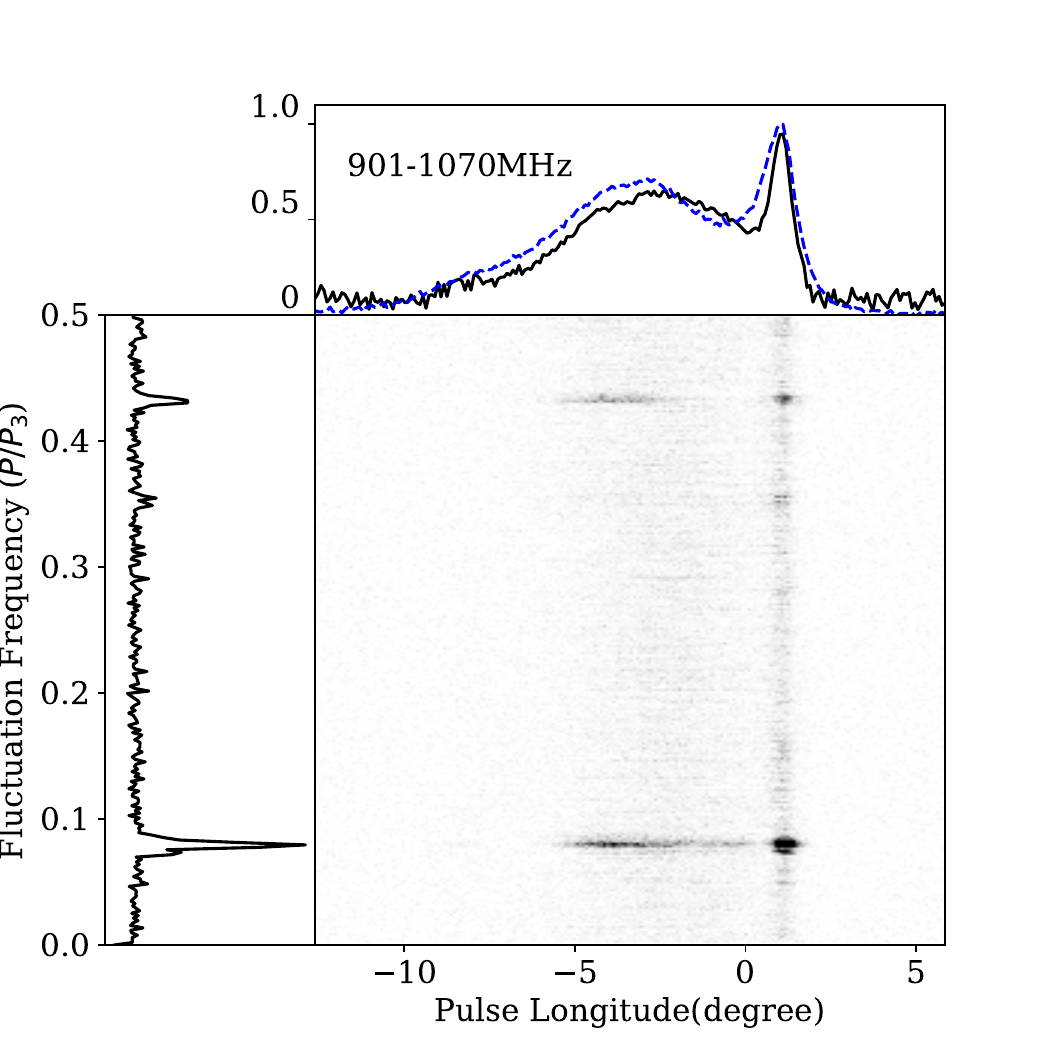}\\
     \includegraphics[width=0.35\textwidth]{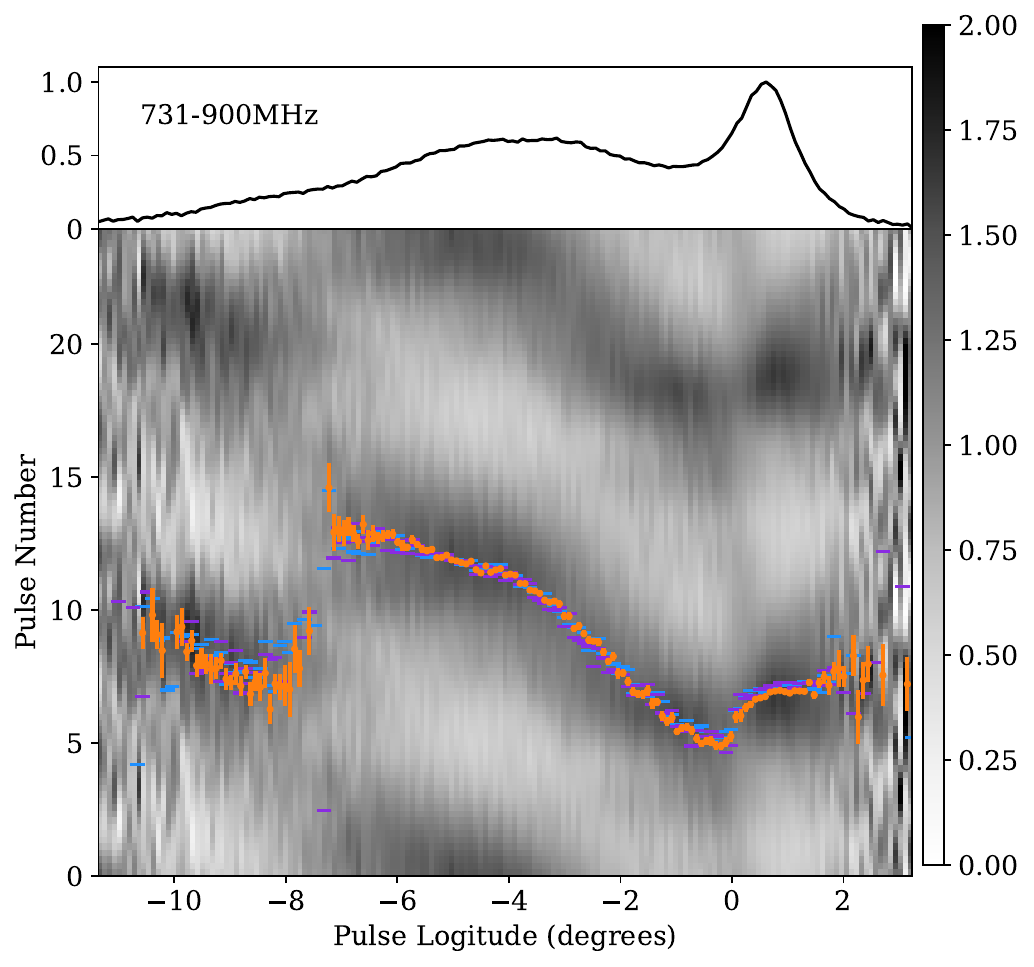}&
     \includegraphics[width=0.35\textwidth]{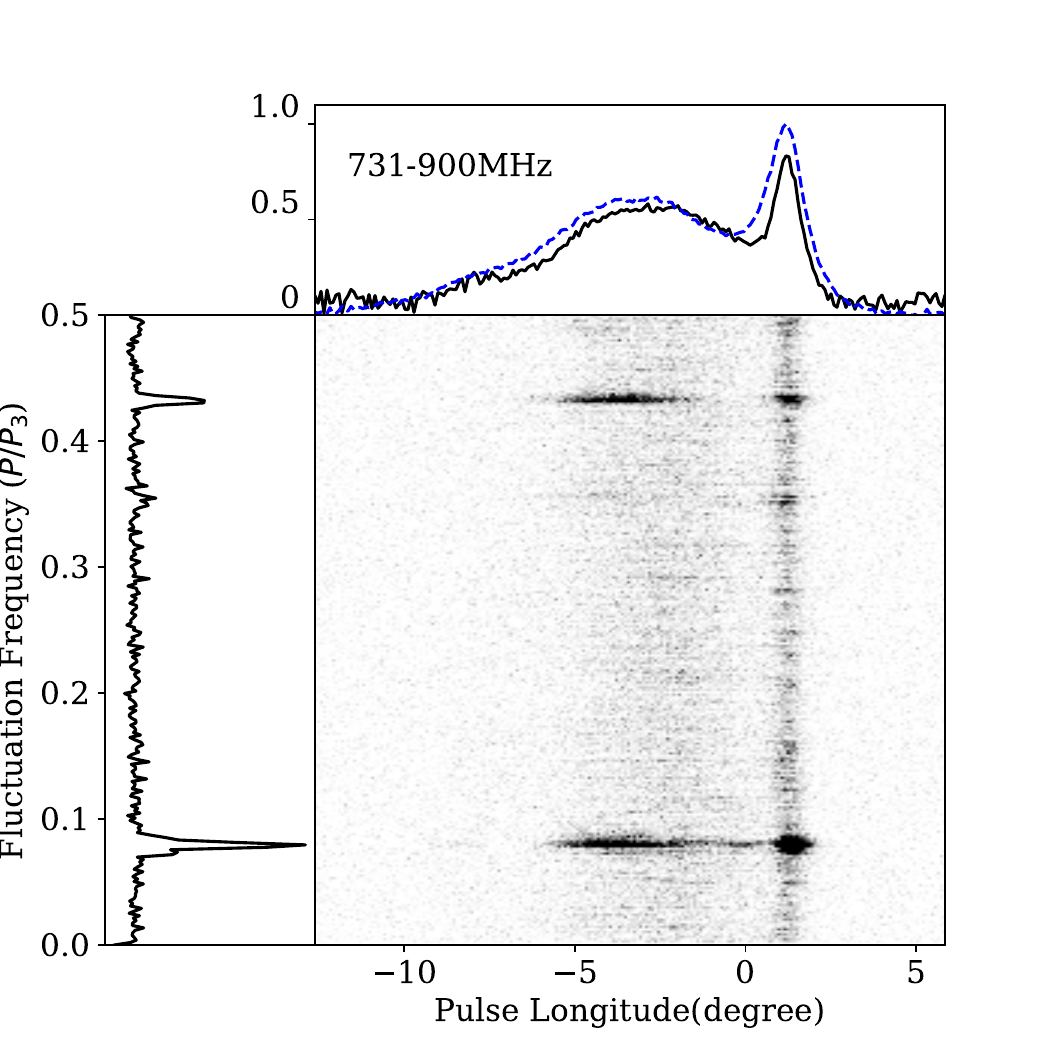}\\
     \includegraphics[width=0.35\textwidth]{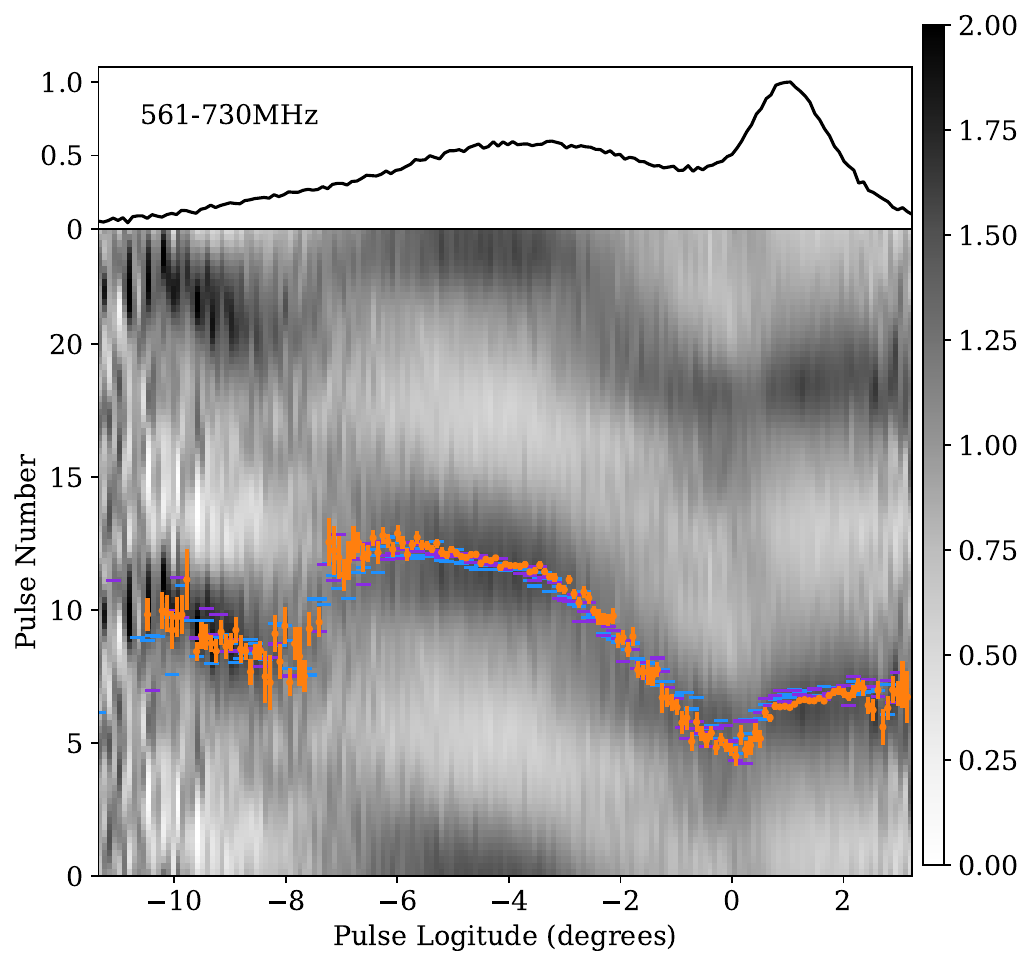}&
     \includegraphics[width=0.35\textwidth]{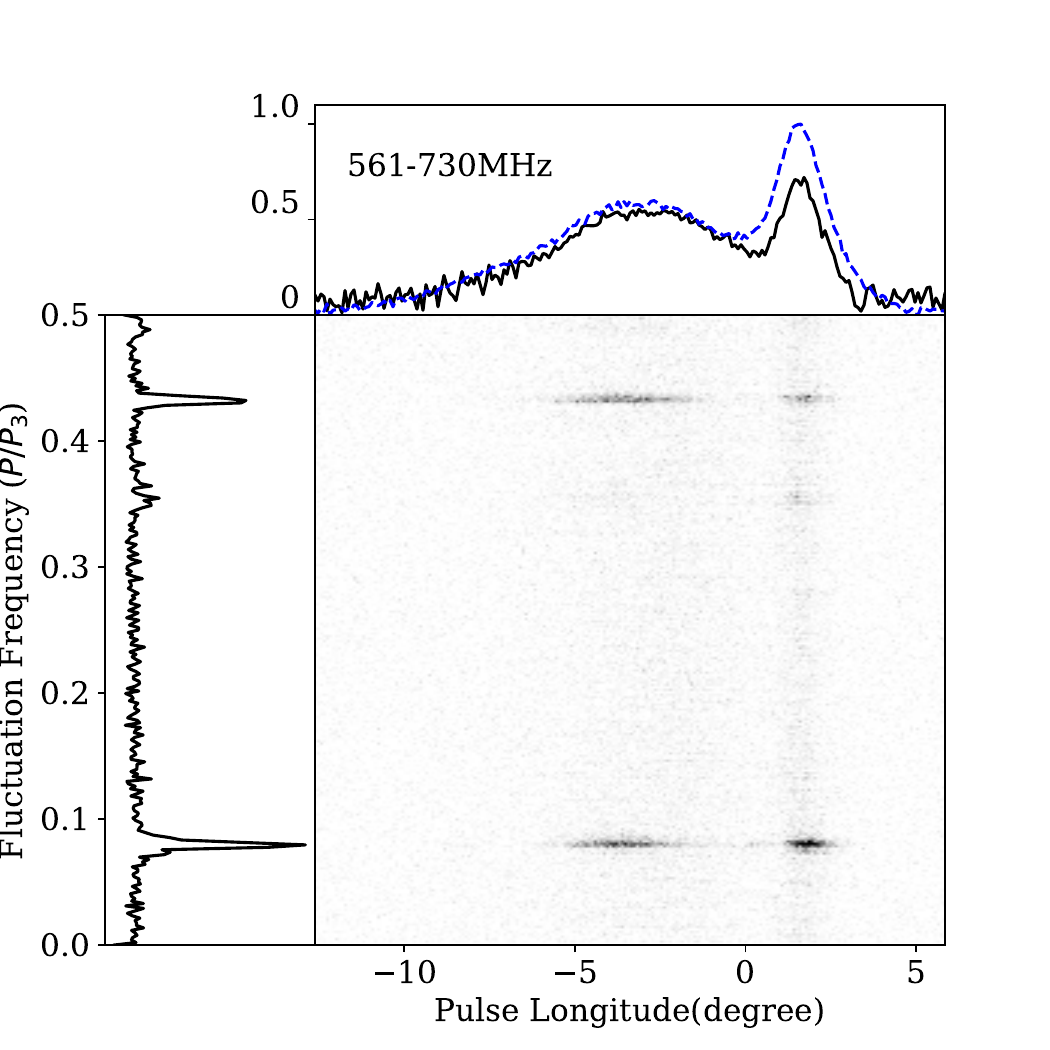}\\
    \end{tabular}
    \caption{Plots of PSR~J1803$-$3329 in three equally separated observing frequency bands with MK-UHF observation. The left column shows the $P_3$ fold plots at three frequency bands folded at $P_3 = 12.167P$. The smoothing factor is 4. The relative phases of the full pulse stack and two halves of the pulse stack are overlaid with the $P_3$ fold plots.
    The right column shows the LRFS. The FFT length used is 516. 
    The detailed description of the panels can be found in Fig.~\ref{fig:J1418-3921P3Fold} and Fig.~\ref{fig:J1418-3921TPAModesLRFS}. }
    \label{fig:J1803-3329UHF}
\end{figure*}

\subsection{PSR J1803--3329}
\label{sec:J1803}

PSR~J1803$-$3329 has a double-peaked profile, with a broad leading component and a brighter narrow trailing component.
There is a slow drift visible in the pulse stack, particularly in the leading component, as well as a more rapid intensity modulation showing as alternating bright pulses.
The LRFS shows the slower modulation has a fluctuation frequency around 0.082~cpp. 

Fig.~\ref{fig:J1803-3329UHF} shows more detailed LRFS and $P_3$ fold plots of the same pulses across three evenly separated frequency bands of the MK-UHF observation. 
The drift is associated with the fluctuation at 0.082~cpp seen in all three frequency bands, which is used to compute the $P_3$ fold plots in the left column of the figure.
An additional well-defined feature is visible at 0.43~cpp in the LRFS,
corresponding to a $P_3$ of about $2.33P$.
The relative intensity of the rapid modulation increases at lower observing frequency, and although these modulations are also visible in the MK-L, they are less pronounced.
Although the high-frequency modulation is close to five times the primary sub-pulse modulation frequency, it is not close enough to be a harmonic.
There are also weaker features above (around 0.49~cpp) and below (around 0.35~cpp) the high modulation frequency.
We attribute these to beat frequencies between the primary modulation and the faster modulation, noting that the beat at the higher modulation frequency is aliased.
Performing a $P_3$ fold at $2.33P$ does not show features with any drift, but rather a constant pulse phase across longitude.
Taken together, this suggests that this is a longitude-stationary modulation on top of the drifting pattern, and not directly related. This situation is comparable to what is reported in \cite{Ray2025} for PSR~J1514$-$4834.

The $P_3$ fold plots (left column of Fig.~\ref{fig:J1803-3329UHF}) with $P_3$ of $\sim 12.2P$ show clear drift characteristics.
As the three frequency bands are observed simultaneously, the folding phases are computed only for the central frequency band, and then identical folding is applied on the other two bands.
Overall, although the features are broadly similar, there seems to be a subtle evolution of the drift pattern with observing frequency.
In the higher frequency band, there is a negative drift across the broad `hump' of the profile, until around 0 degrees longitude, where the sub-pulse phases sharply change direction showing an overall positive drift, implying a bi-drifting feature.
There also seems to be an almost $\pi$ phase jump in phase between the leading shoulder and the central broad peak (at around longitude $-7$ degrees), which becomes clearer in the lower frequency bands. 
Similar phase jumps between components are seen in e.g. PSR~B0320+39 \citep{Ben2003}.

\begin{figure}
    \centering
    \includegraphics[width=0.45\textwidth]{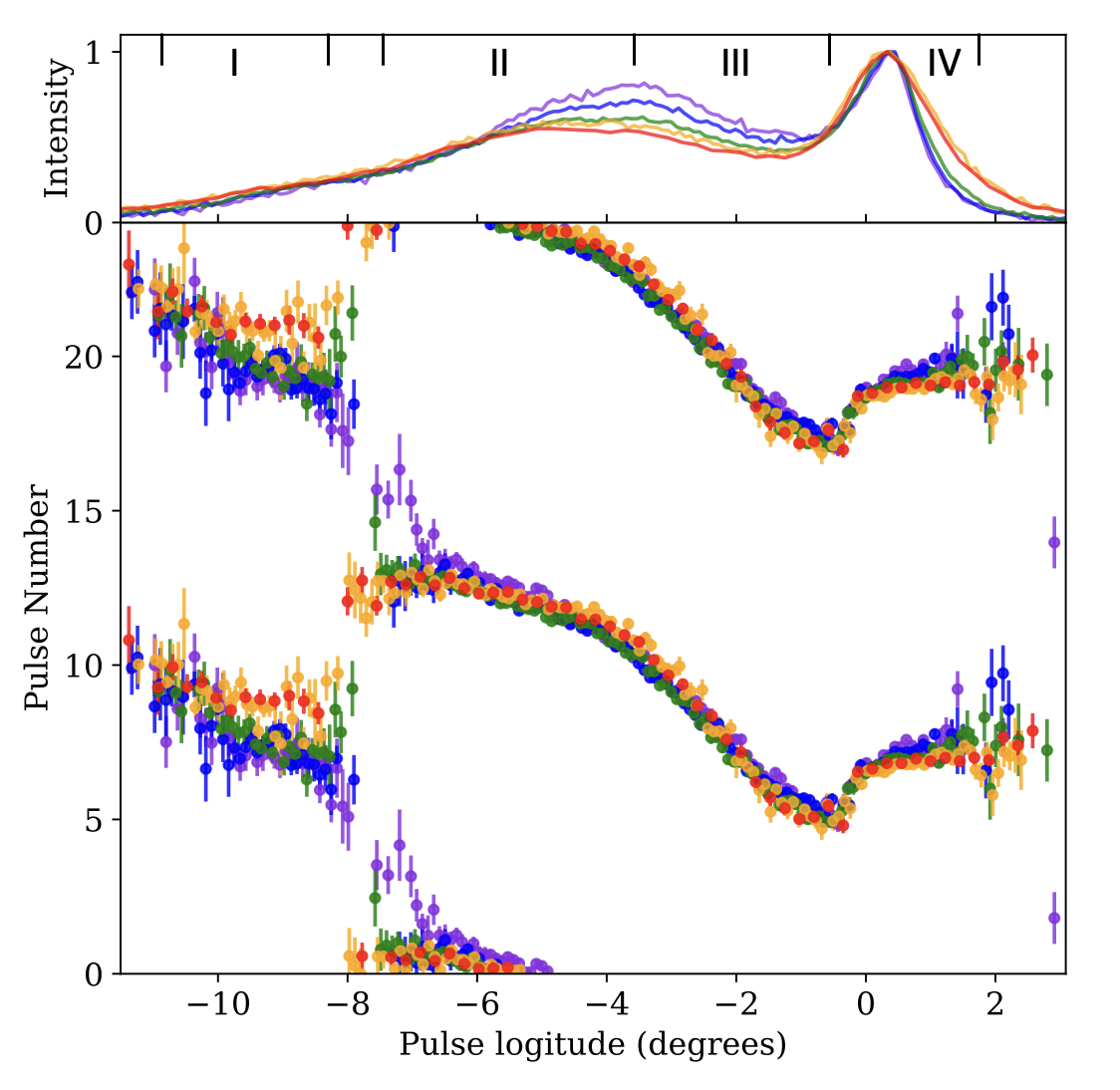}
    \caption{Plot of relative phases from the MK-L, three frequency bands of MK-UHF and GMRT observations of PSR~J1803$-$3329. For more detailed description of plotting criteria, see Fig.~\ref{fig:J1534-4428PhasesTogether}. 
    The black vertical lines at the top edge of the top panel mark the four component sections that will be discussed.}
    \label{fig:J1803-3329RelativePhasesALL}
\end{figure}

For comparison, the profiles and sub-pulse phase tracks from the three MK-UHF sub-bands as well as the MK-L and GMRT data are shown in Fig.~\ref{fig:J1803-3329RelativePhasesALL}.
The longitude axis of each profile has been rotated to align the trailing peak.
The sub-pulse phases of the MK-UHF and GMRT data have been adjusted to match the MK-L data using a method similar to that applied for PSR~J1534$-$4428 in Section\ref{sec:J1534}.
In order to interpret the phase tracks, we define four longitude sections of interest, I--IV, as labelled on the top of Fig.~\ref{fig:J1803-3329RelativePhasesALL}.
The profiles show that the trailing component in Section IV becomes wider and extends to the later longitude as the frequency gets lower.
Additionally, the relative amplitude of the leading peak across Sections II and III gets smaller as the frequency decreases. 

The relative phase of Section I evolves slightly toward later pulse numbers as the frequency gets lower, while the average phase of Sections II-IV remains much more constant. As a result, the phase jump between Sections I and II appears to be smaller at lower frequencies.
However, the gradient in Section I does not seem to change much with frequency. 
The phase track in Section II appears to have a flatter gradient at lower frequencies, while the gradient in Section III gets slightly steeper.
Consequently, the `curvature' of the central drift band (across Sections II and III) appears to change from a smooth sweep at high frequencies to more of a `kink' at lower frequencies. 
The later part (between about $-1.5$ degrees to about 0 degree) of Section III seems flatter at lower frequencies, resulting in a more `platform-like' section near the trailing component (Section IV). 
The phase gradient in Section IV flattens as the frequency decreases, although the relative phase difference between Sections III and IV remains constant.
It is worth noting that the phase track in Section IV in the GMRT observation shows a flatter gradient compared to the same section in the lowest MK-UHF frequency band.

\clearmaybe
\subsection{PSR J1834--1202}
\label{sec:J1834-1202}
\begin{figure}
    \centering
    \includegraphics[width=0.45\textwidth]{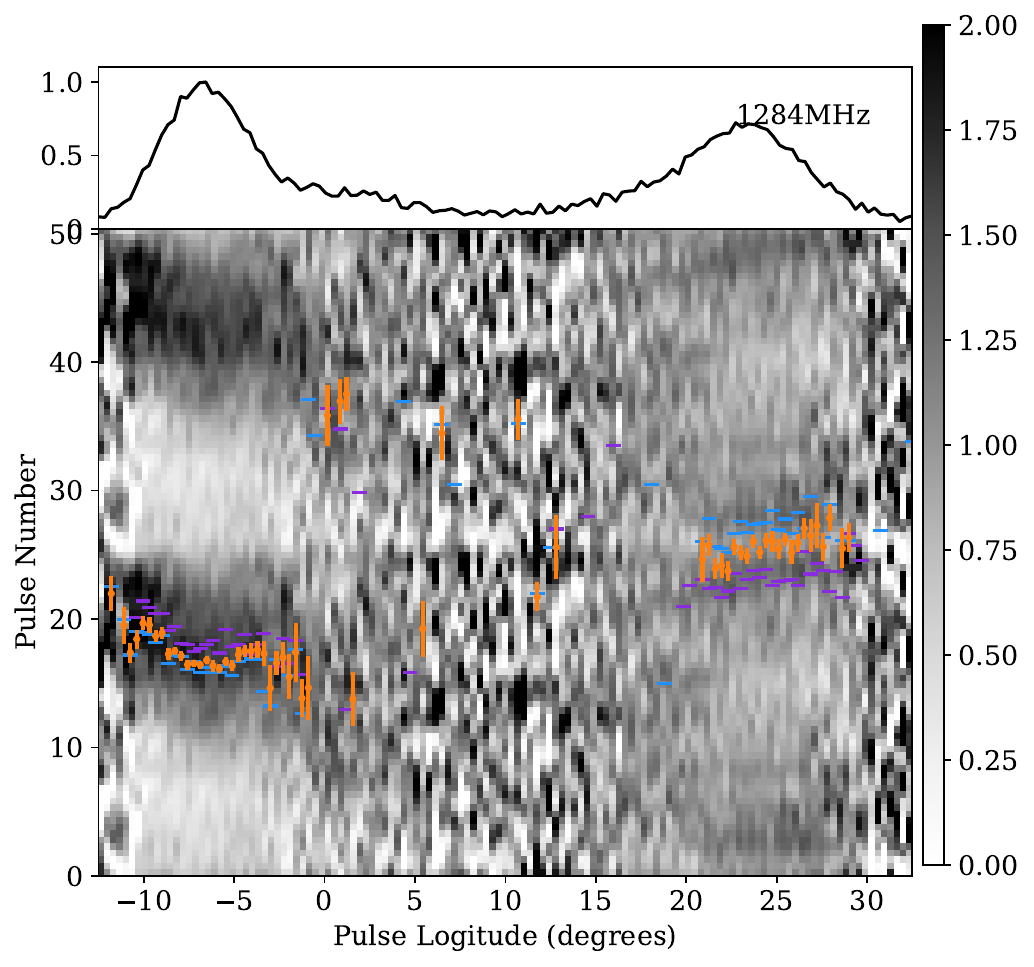}
    \caption{$P_3$ fold plot of the PSR~J1834$-$1202 MK-L observation at $P_3$ of 25.189$P$. The smoothing factor used is 2. For a detailed description of the plot, refer to Fig.~\ref{fig:J1418-3921P3Fold}.}
    \label{fig:J1834-1202TPA}
\end{figure}

\begin{figure}
    \centering
    \includegraphics[width=0.45\textwidth]{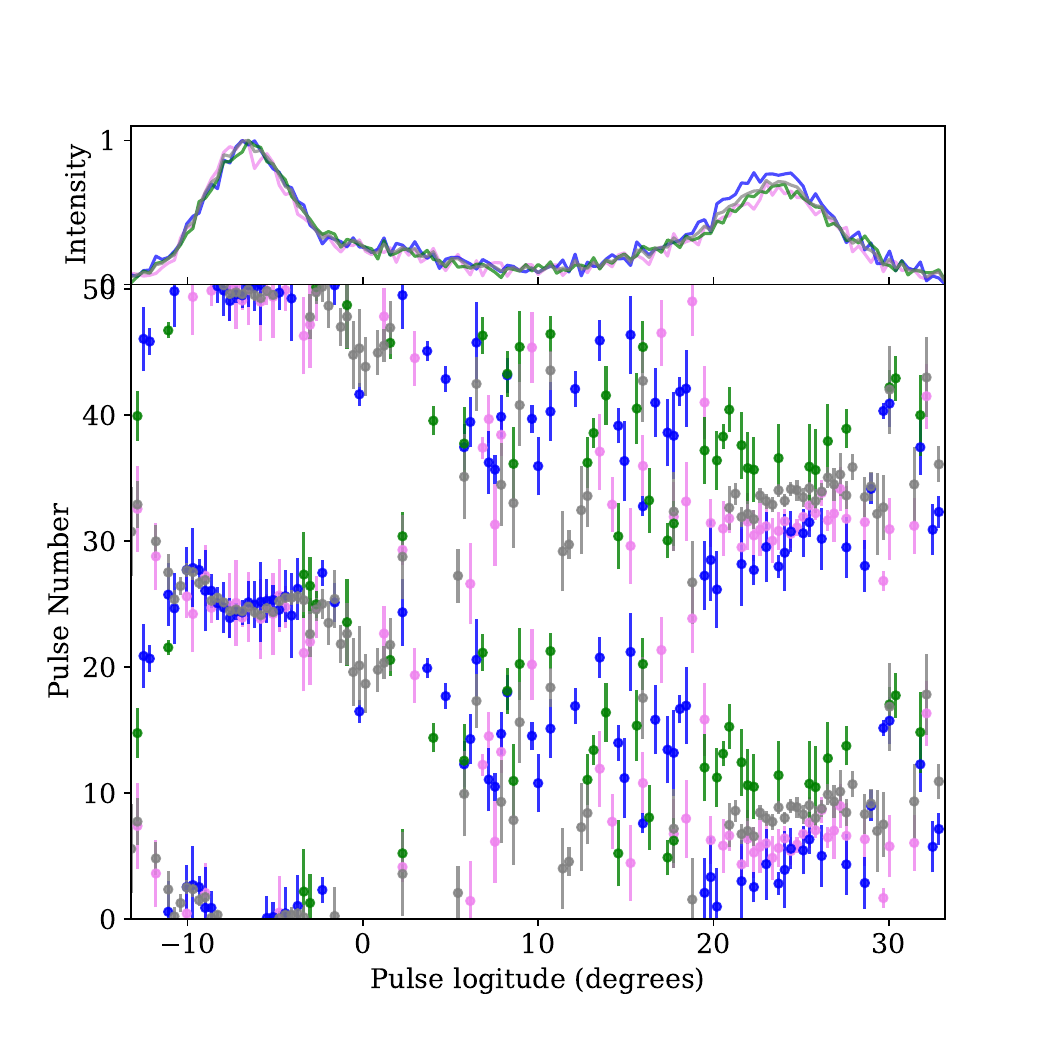}
    \caption{Plot of sub-pulse phases of PSR~J1834$-$1202, from three equal sections of the observation (346 pulses each). The phases of the leading component have been aligned. The phases of the first second, and third sections are represented in violet, blue, and green, respectively, while the phases of the full pulse stack are shown in gray for comparison. The plot only includes the phases with error bars smaller than 1/7 of $P_3$. The top panel shows the mean profiles of the sections in the corresponding colours. }
    \label{fig:J1834-1202ThreeSecPhases}
\end{figure}

PSR~J1834$-$1202 has a wide, symmetric, double-peaked pulse profile and both components show clear modulation with a fluctuation frequency of approximately 0.04~cpp.
From the LRFS (middle panel of Fig.~\ref{fig:PulseStacksTPA}) we can see that the trailing component has power across a broader range of fluctuation frequencies. 
Indeed, in the pulse stack we see that the trailing component has more irregular intensity variation than the leading component, both over relatively long timescales, combined with more rapid fluctuations.

The sub-pulse phase measurements (Fig.~\ref{fig:J1834-1202TPA}) show that the leading component has a subtle negative drift and the trailing component has a very shallow positive drift, and hence bi-drifting.
However, there is a degree of complexity in the relationship between the drift patterns of the two components. 
Over the course of the observation the relative phase of the two components seems to shift slightly.
This is more clearly seen in a cropped section of the pulse stack shown in Fig.~\ref{fig:J1834-1202TPASec}.

To explore this further, in Fig.~\ref{fig:J1834-1202ThreeSecPhases} we show the sub-pulse phase computed from three equal sections of the observation, aligned using the phase of the first component, as this seems more stable.
The phases of the trailing component in the first (violet) and second (blue) sections lie below those of the full pulse stack, whereas the phases of the third (green) section appear above.
However, it does not seem to be the case that the two profile components are entirely desynchronised, i.e., there does seem to be a consistent underlying $P_3$ across the profile.
This observation also challenges the classical carousel model, though perhaps not surprising, as some asymmetry is required to produce bi-drifting.
Small perturbations in this asymmetry could result in phase differences between components without producing large changes in the underlying $P_3$.

\clearmaybe
\subsection{PSR J1843--0211}
\label{sec:J1843}
\begin{figure}
    \centering
    \begin{tabular}{cc}
    \includegraphics[width=0.45\textwidth]{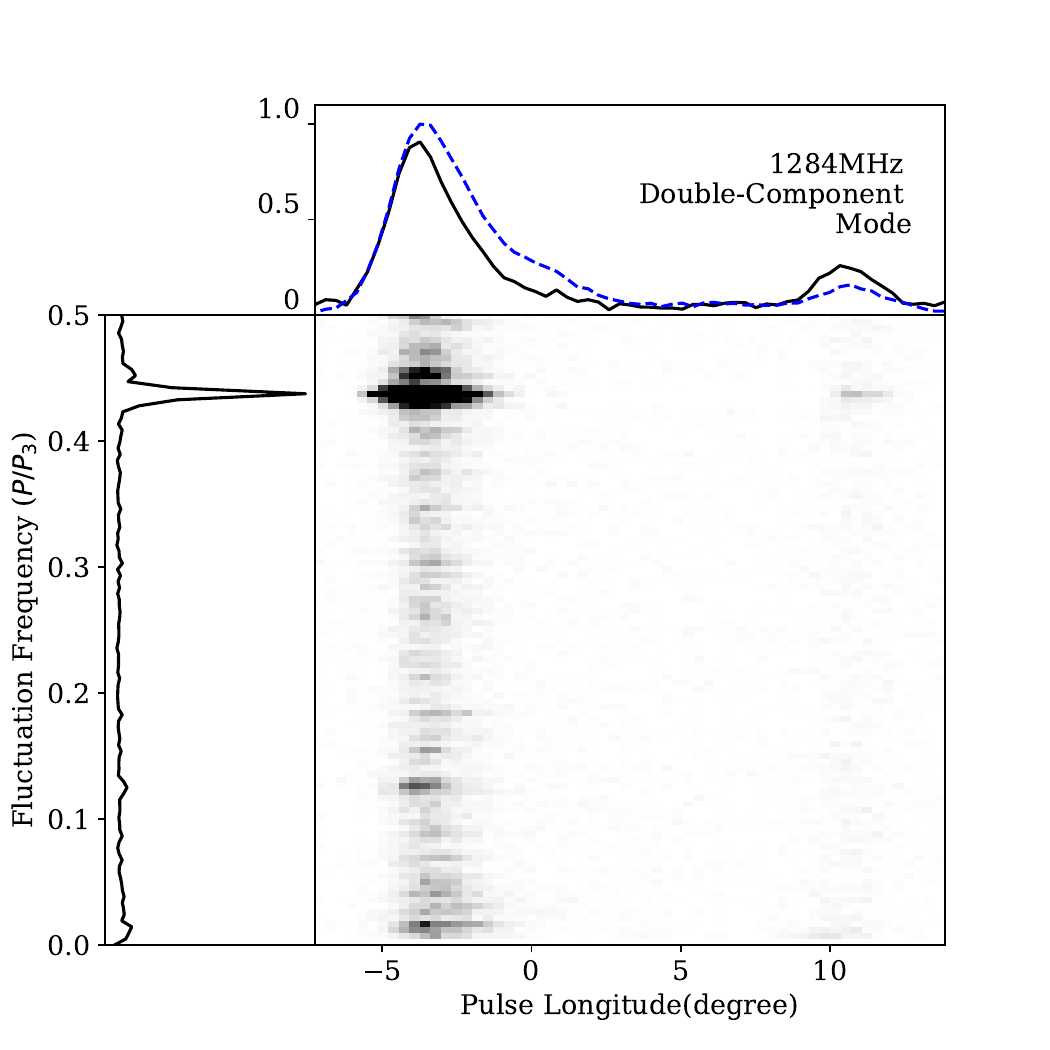}\\
    \includegraphics[width=0.45\textwidth]{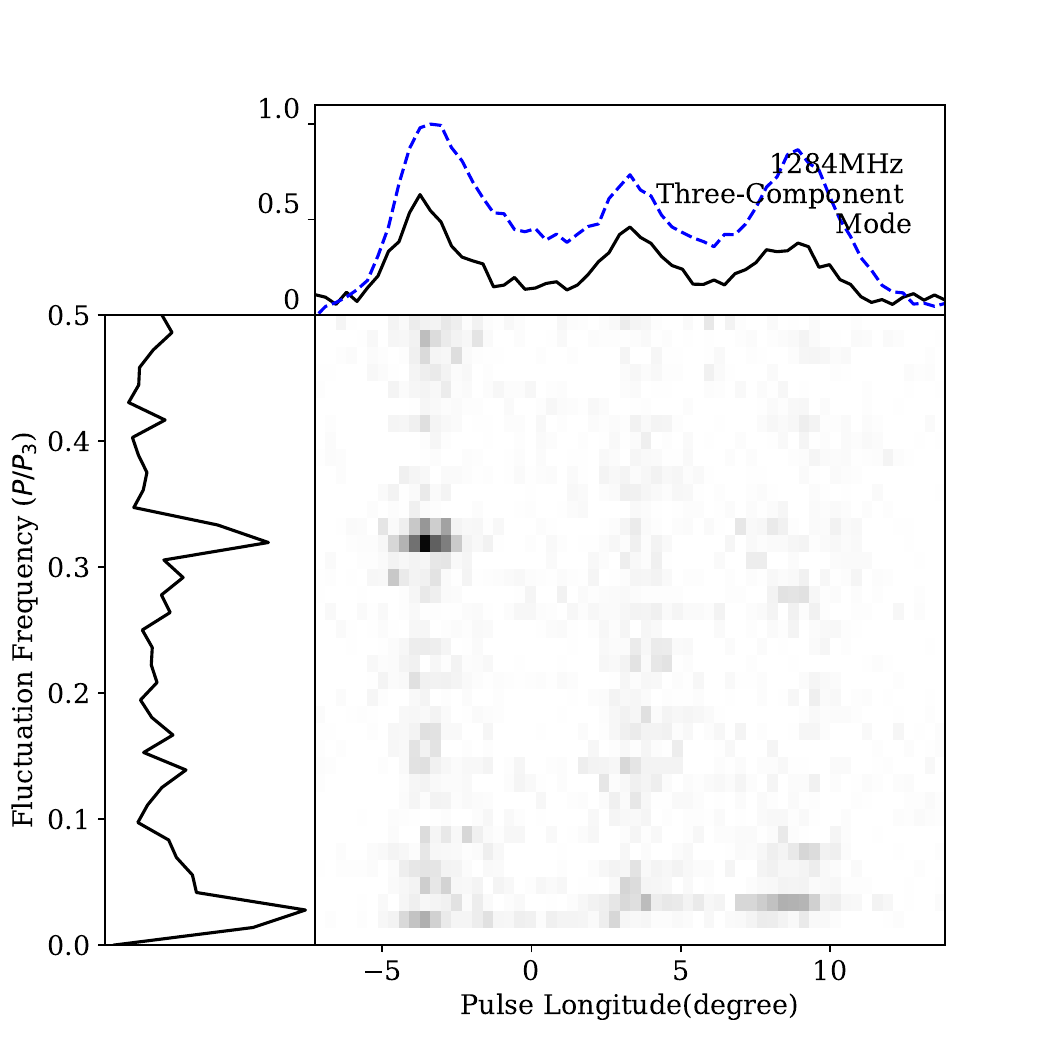}
    \end{tabular}
    \caption{Mode separated LRFS of PSR~J1843$-$0211 for the asymmetric double component mode (top), and the three component mode (bottom).
    The detailed description of the panels are in Fig.~\ref{fig:J1418-3921TPAModesLRFS}.
    The FFT length used for double component mode is 208, 
    and for the three component mode is 72.}
    \label{fig:J1843-0211CombinedMode1 and Mode2}
\end{figure}
\begin{figure}
    \centering
    \includegraphics[width=0.45\textwidth]{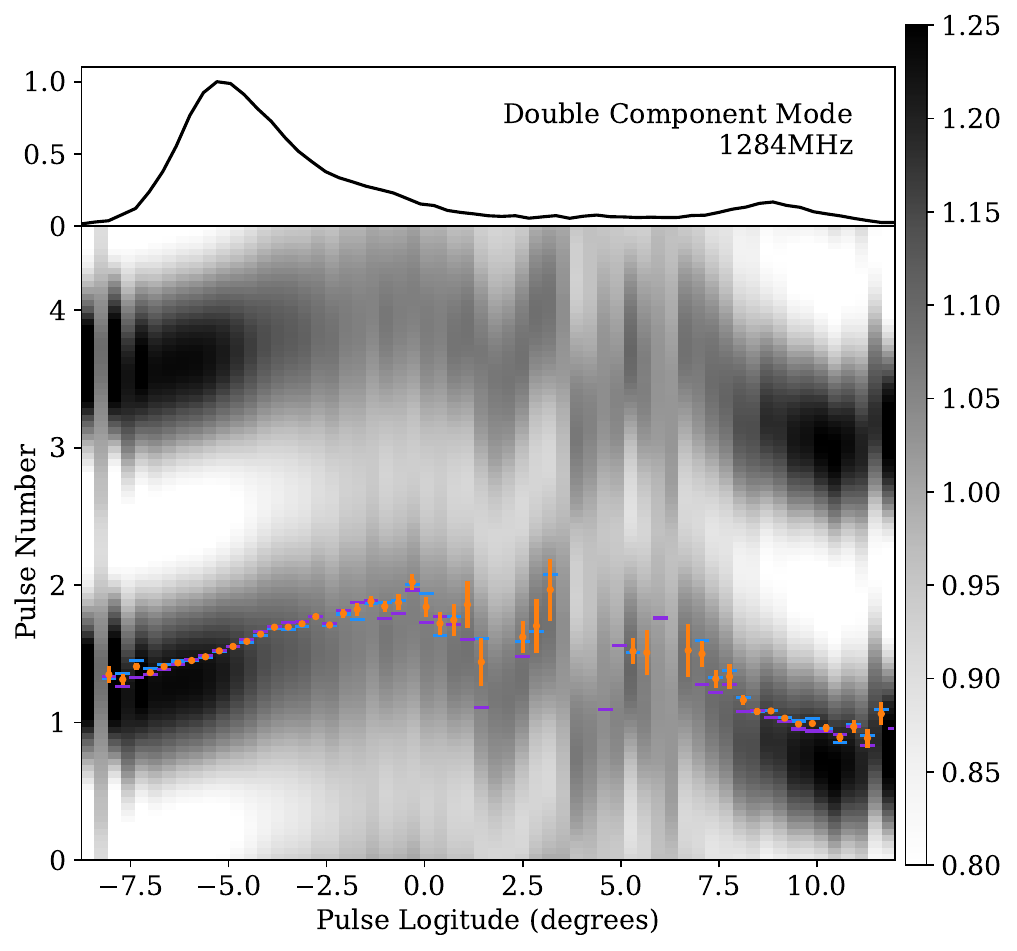}\\
    \includegraphics[width=0.45\textwidth]{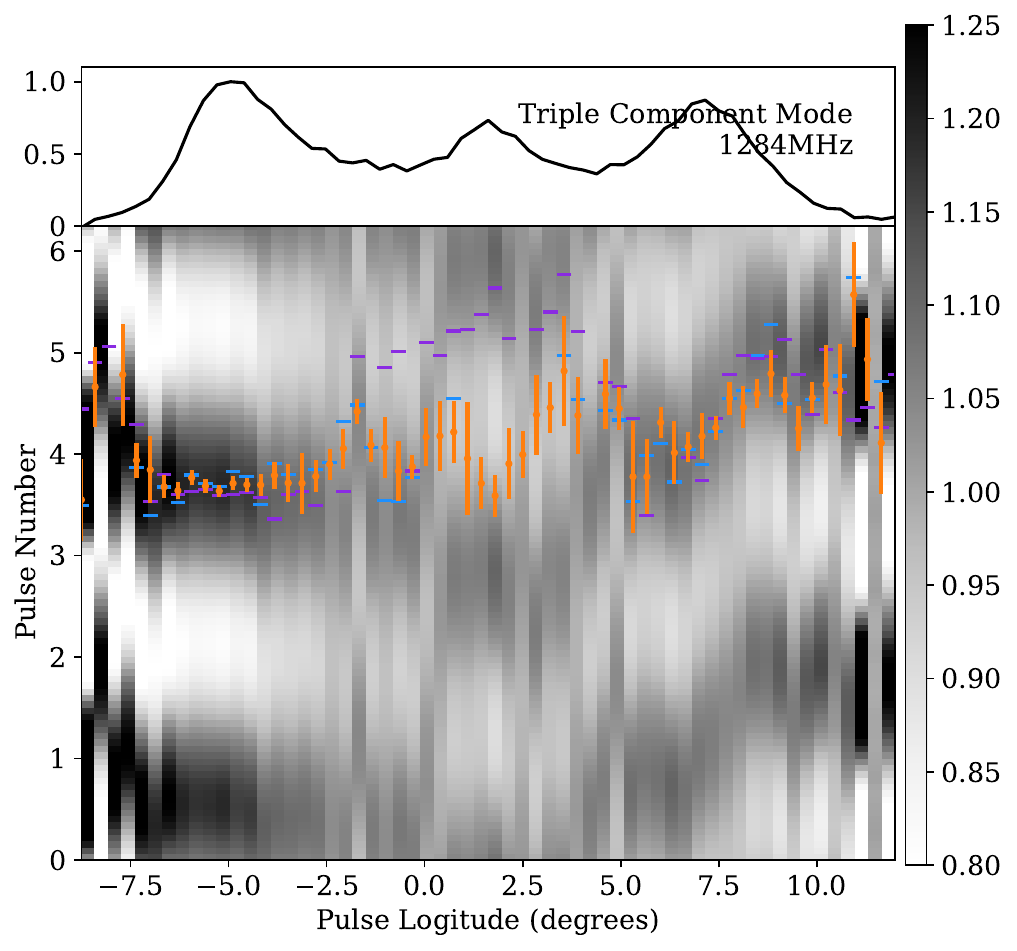}
    \caption{
    The $P_3$ fold of the asymmetric double component mode of PSR~J1843$-$0211 (top) and the three component mode (bottom), using the combined MK-L observations. The $P_3$ used for double component mode folding is 2.307$P$. The smoothing factor is 21. For the three component mode, the $P_3$ used is 3.125$P$, the smoothing factor is 16. A detailed description of the plot can be found in Fig.~\ref{fig:J1418-3921P3Fold}.
    }
    \label{fig:J1843-0211TPA}
\end{figure}

PSR~J1843$-$0211 shows at least two different modes (Fig.~\ref{fig:PulseStacksTPA}).
The most prominent mode is an asymmetric double-peaked profile with a low bridge of emission.
However, there are characteristic transition of profile shapes, first to a more symmetric double-peaked profile before a third component appears in the centre.
This mode changing behaviour is strikingly similar to that in PSR J1534$-$4428 (Section \ref{sec:J1534}).
The transition from the asymmetric double profile to the more symmetric profile is often associated with some nulling pulses, but the duration of the null is irregular. The overall null fraction is estimated to be $\sim$10\%.

For our analysis, we split the data into a `double component' mode (only the asymmetric double profile) and a `three component' mode (only the more symmetric profile with clear central component).
The modulation in the symmetric double component mode is similar to that in the three component mode.
The mode-separated LRFS of the full dataset is shown in Fig.~\ref{fig:J1843-0211CombinedMode1 and Mode2}.
The double component mode has strong modulation, especially in the leading component, at 0.43~cpp.
In the three-component mode, strong modulation occurs at fluctuation frequency of around 0.32~cpp, especially in the leading component. There is some modulation at 0.02~cpp, close to the typical timescale for the duration of this mode and likely to be an artifact of the mode separation, which is challenging for this mode.

The $P_3$ fold plot of the double component mode (Fig.~\ref{fig:J1843-0211TPA}) shows the two components with opposite drift directions, indicating bi-drifting.
However, with $P_3$ close to 2, the modulation is so fast that it is difficult to see the drift directly in the pulse stack, especially in the trailing component which shows weak and sporadic emission.

In the three-component mode, the leading component exhibits a relatively flat drift at $\sim$0.32~cpp. Although the error bars are relatively large, the trailing component shows positive drift, which is opposite to that in the double component mode. This behaviour of changing the drift direction in different modes is similar to what is observed in PSR~J1418$-$3921, although there are too few pulses to determine if this is a persistent behaviour, and longer observations are likely needed to fully investigate this mode.

\clearmaybe
\subsection{PSR J1921+1948} 
\label{sec:J1912+1948}
\begin{figure}
    \centering
    \includegraphics[width=0.45\textwidth]{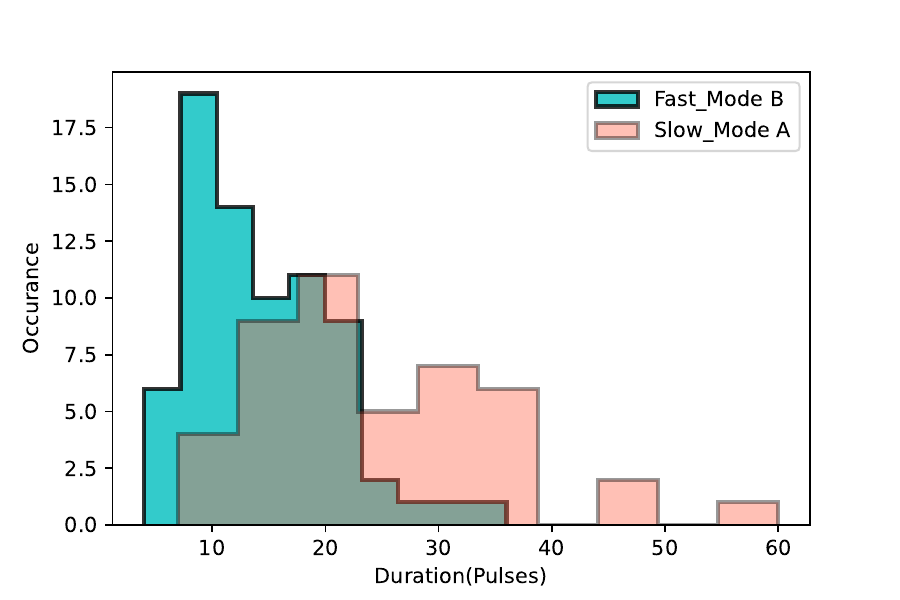}
    \caption{Occurrence of the fast and slow modes of PSR~J1921$+$1948. The histogram shows how often each mode persists for a given number of pulses (horizontal axis) before switching to another mode.}
    \label{fig:J1921+1948Occurrence}
\end{figure}
\begin{figure}
    \centering
    \begin{tabular}{c}
    \includegraphics[width=0.45\textwidth]{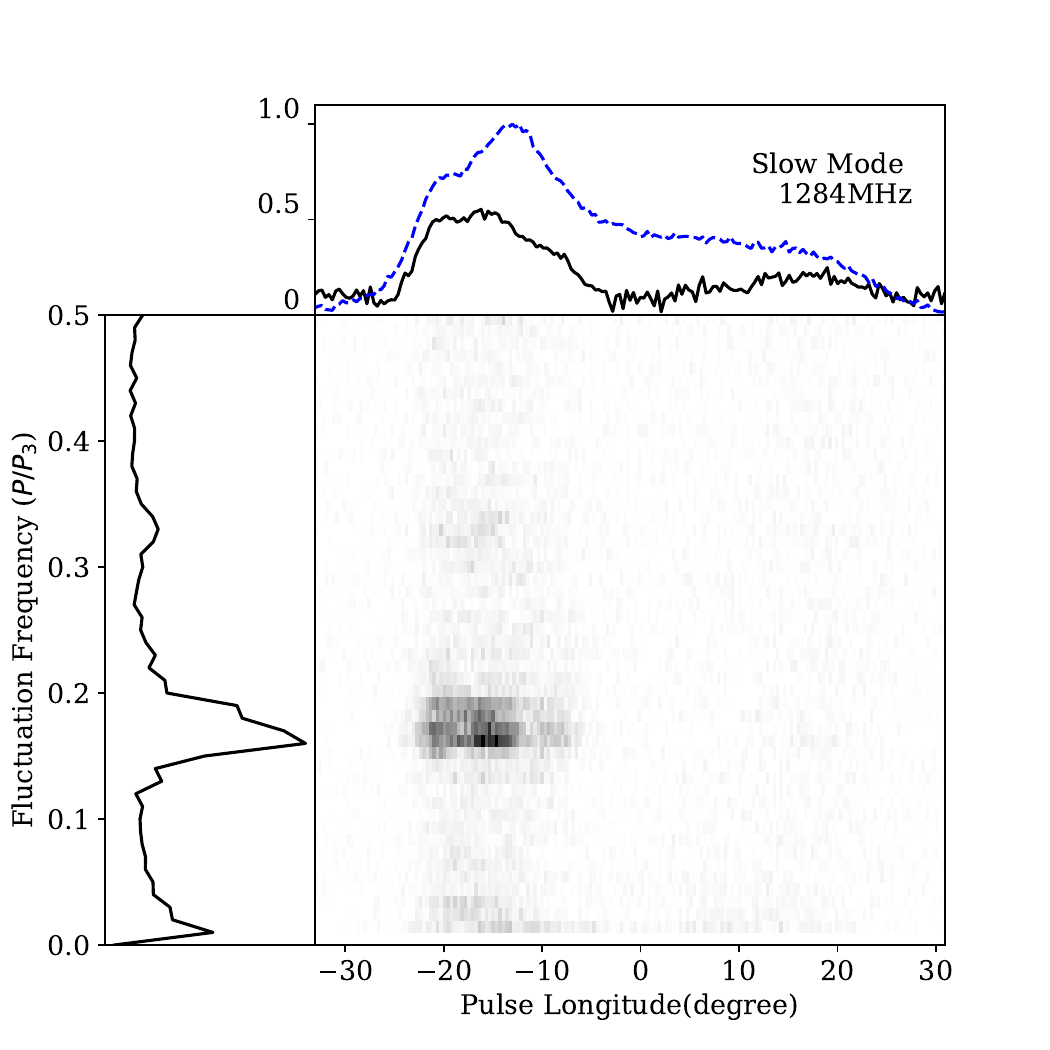}\\
    \includegraphics[width=0.45\textwidth]{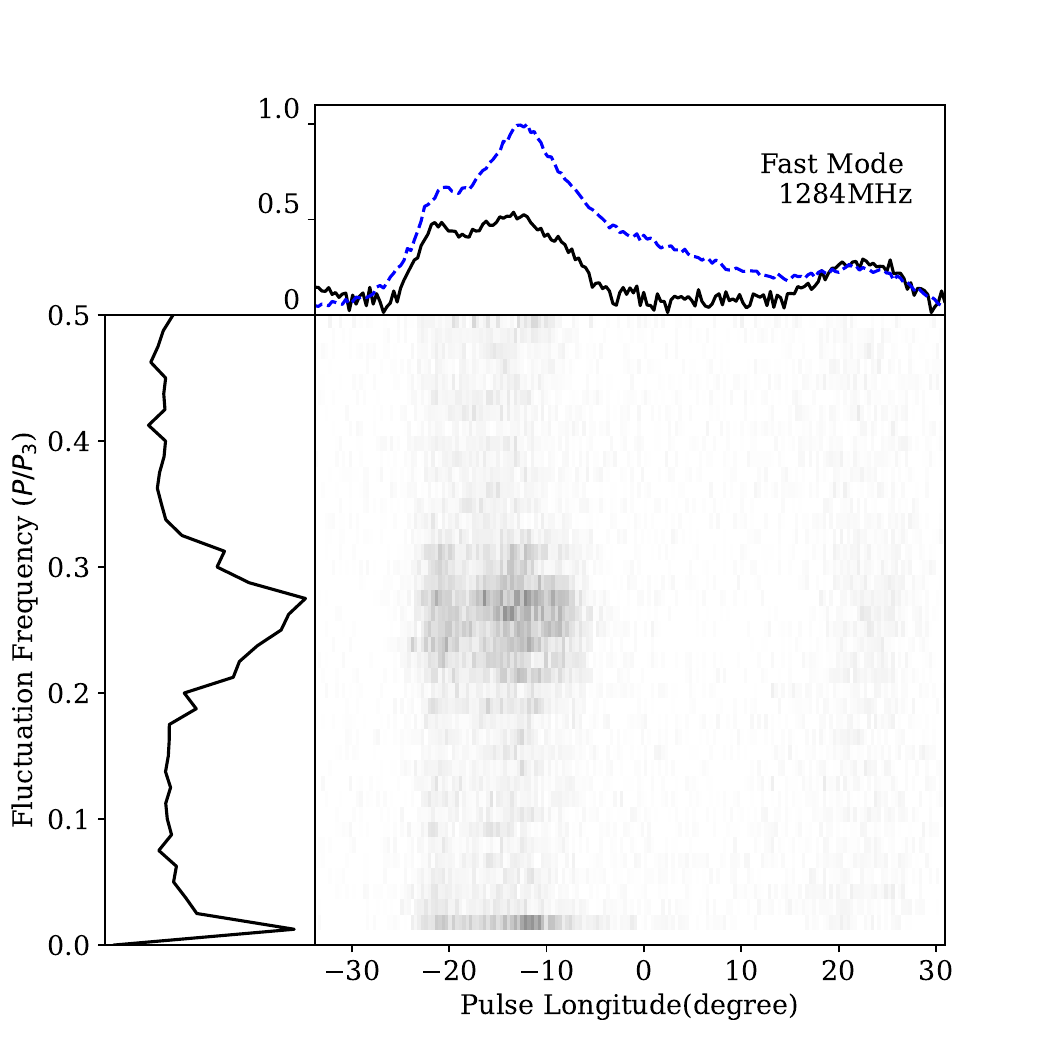}
    \end{tabular}
    \caption{Combined LRFS of PSR~J1921+1948 for slow (top) and fast (bottom) modes. The detailed description of the panels are in Fig.~\ref{fig:J1418-3921TPAModesLRFS}. The FFT length used for slow mode is 100 and for fast mode is 80. }
    \label{fig:J1921+1948LRFS}
\end{figure}

\begin{figure*}
    \centering
    \begin{tabular}{cc}
    \includegraphics[width=0.45\textwidth]{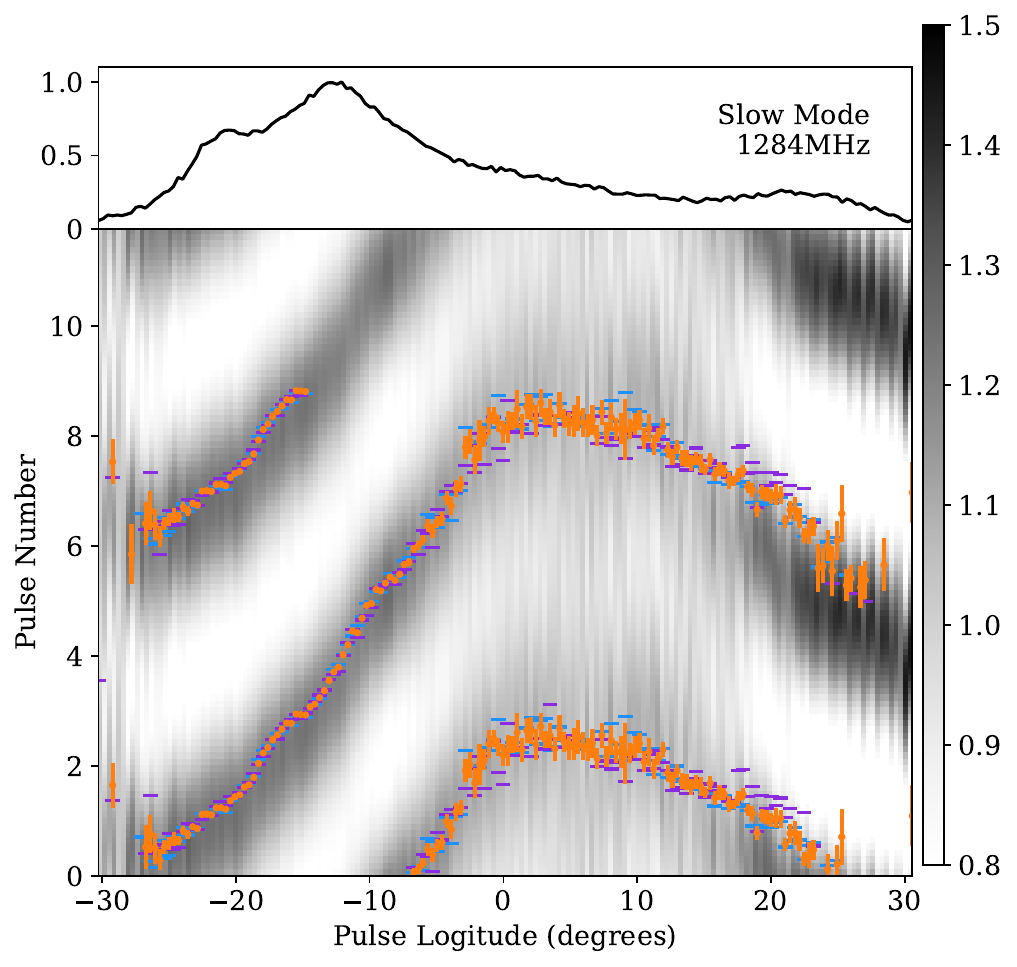}&
    \includegraphics[width=0.45\textwidth]{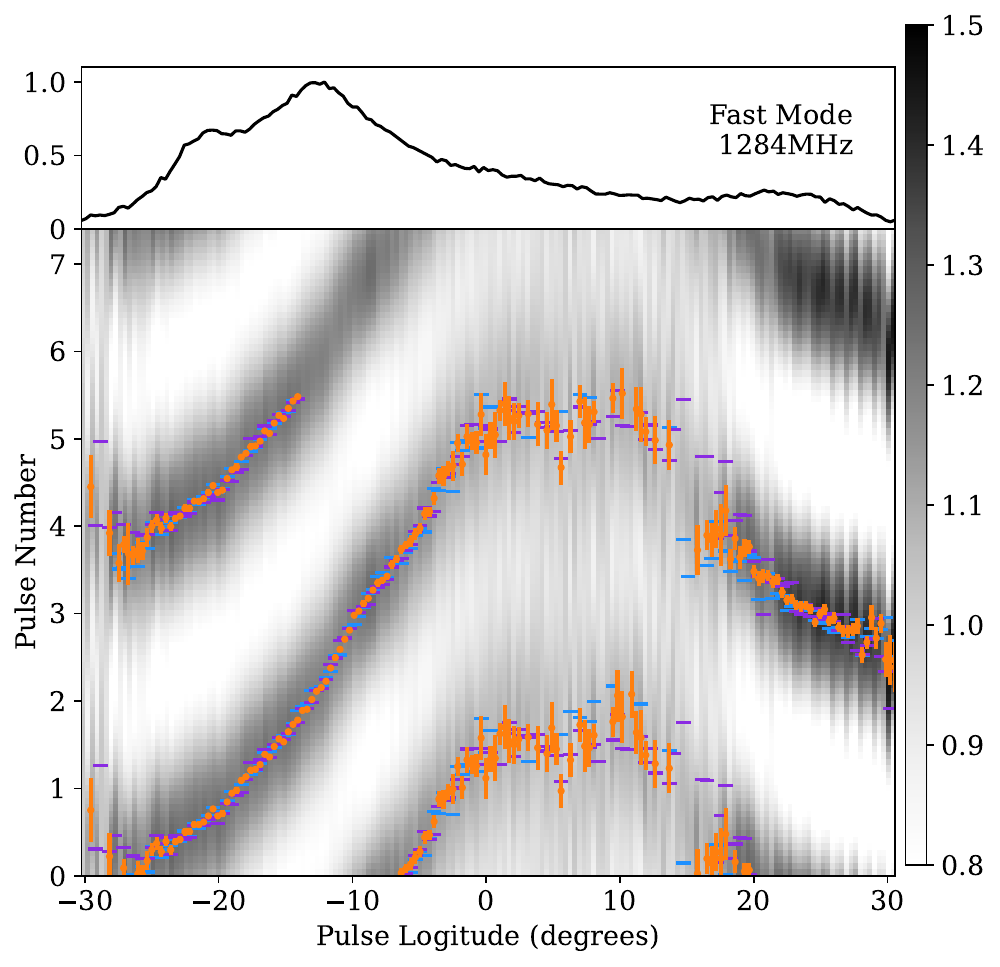}
    \end{tabular}
    \caption{The $P_3$ fold plots of PSR~J1921+1948 Slow (left) and Fast (right) modes of MK-L observations. The $P_3$ values used for folding are 5.88$P$ (Slow) and 3.7$P$(Fast), the smoothing factors are 8 and 13. The detailed description of the plot can be found in Fig.~\ref{fig:J1418-3921P3Fold}. }
    \label{fig:J1921+1948P3}
\end{figure*}

PSR~J1921+1948 shows multiple components over a total profile width about 60 degrees (Fig.~\ref{fig:PulseStacksTPA}).
The drift features and mode-changing behaviour of this pulsar have been discussed in \citet{Rankin2013} using data from the Arecibo Observatory at an observing frequency of 0.327\,GHz.  
Bi-drifting has also been reported in a recent analysis of Five-hundred-metre Aperture Spherical radio Telescope (FAST) data at 1250\,MHz observing frequency \citep{Shang2024}, which we also confirm with our MK-L data.

The four modes mentioned in \citet{Rankin2013} are also observed in the MK-L data.
In the MK-L observations, the `slow mode' (A in \citealt{Rankin2013}), which has the largest separation between the repeating drift bands, switches into the `fast mode' (B in \citealt{Rankin2013}) without intermediate features and is then followed by the fastest mode (C in \citealp{Rankin2013}). This process is also observed in \citet{Rankin2013}. 
For most of the MK-L pulse stack, the null-rich irregular mode (N in \citealp{Rankin2013}) is interposed between the A-B-C mode transforming `cycles'. Some pulses from mode C are mixed with mode N pulses between the `cycles'. These mode change behaviours are similar to what was found in \citet{Rankin2013}. However, the MK-L observations do not show extended sections of pulses in the fastest mode (C), as found in \citet{Rankin2013}.

In terms of the sub-pulse drift, in MK-L observations, the slow mode has very evident drifts in both leading and trailing components. Modes B and C have $P_3$ close to each other by inspection. The fastest mode C tends to have dimmer and flatter drift bands in the profile components. 

Notably, the pulse stack appears to be segmented into `chunks', with a trapezium shape, as seen in Fig.~\ref{fig:PulseStacksTPA}. This feature is a result of a combination of changes in the drift rate between different modes and bi-drifting. The trapezium in the pulse stack resembles the wave-like pattern in the pulse stack as observed in PSR~J1239$+$2453~\citep{WangZ2022}. This gives the impression of an apparent drift direction opposite to the actual drift bands within the modes.

The analysis is performed separately for the slow (A) and fast (B) modes.
Given the rapid fluctuations between modes, changes in $P_3$ within modes, and additional sporadic emission, it is not possible to firmly determine the mode boundaries.
Therefore, mode separation is performed by inspecting the pulse stack, generally discarding data where the mode is ambiguous.
The distribution of modes durations (e.g. \citealt{Ilie2020}) is shown in Fig.~\ref{fig:J1921+1948Occurrence}. The pulsar tends to stay in the fast mode for ten to twenty pulses, which include a few repeating drift bands. 
The distribution measurements indicate that while the fast mode may have more repeating drift bands, the slow mode tends to persist for more pulses before transitioning to other modes. Overall, there are generally more pulses in the slow mode than in the fast mode. This occurrence pattern is a bit different from the abundance of the B mode (with a $P_3$ estimated to be approximately $4P$) at 327\,MHz \citep{Rankin2013}, but it is similar to the findings in FAST observations \citep{Shang2024}, that the slow mode has the greatest number of pulses. 
Due to discontinuities in the resulting pulse stacks, the FFT length was selected to be three to four times the $P_3$ for Fourier analyses, because the pulsar does not stay in one mode for very long time. Separated modes with pulses shorter than the FFT length (a few mode segments contain just one $P_3$) were excluded from the analysis. 
Fig.~\ref{fig:J1921+1948LRFS} shows the combined LRFS plots of the separated modes. 

In the upper panels, the profiles have the greatest amplitude in the middle component. The fast mode also shows a distinct trailing peak. 
The slow mode, on the other hand, features a smoother trailing component. 
The slow mode has a fluctuation frequency about 0.16~cpp.
The fast mode has a fluctuation frequency about 0.27~cpp. 
The fastest mode has a modulation frequency about 0.37~cpp, but does not appear significantly in the fluctuation spectra of the full pulse stack (Fig.~\ref{fig:PulseStacksTPA}). Additionally, this mode is typically very short and its pulses are difficult to distinguish from the fast mode, so we do not include further analysis of the fastest mode.

A section of the pulse stack is shown in Fig.~\ref{fig:J1921+1948Sec}. It shows the slow mode at the bottom, followed by the fast mode, the fastest mode, and then the irregular mode with `incomplete' components. The bi-drifting features in the fast and slow modes are visible to the eye.

Fig.~\ref{fig:J1921+1948P3} shows the $P_3$ fold plots for both the slow and fast modes.
The plots show that the drift features in both modes are generally similar, with only subtle differences. The drift bands can be described with three sections.
The first section (from approximately $-28$ to $-5$ degrees in both slow mode and fast modes) exhibits a positive gradient, followed by a `bridge' section (about 10 degrees in longitude) with relatively weak modulation in the $P_3$ fold, and then the third section with a negative gradient.  
In both modes, the phase gradients of the first and third sections have opposite signs, indicating bi-drifting. 
In the slow mode, the third section appears closer to earlier longitudes and spans a wider longitude range, and appears to connect with the second section without phase discontinuity. Additionally, the drift band of the trailing component in the slow mode is more curved than in the fast mode. 
The fast mode shows modulations in the second section are not clearly resolved, and there is a phase discontinuity between the second and third sections.

\clearmaybe
\begin{figure}
    \centering
    \includegraphics[width=0.45\textwidth]{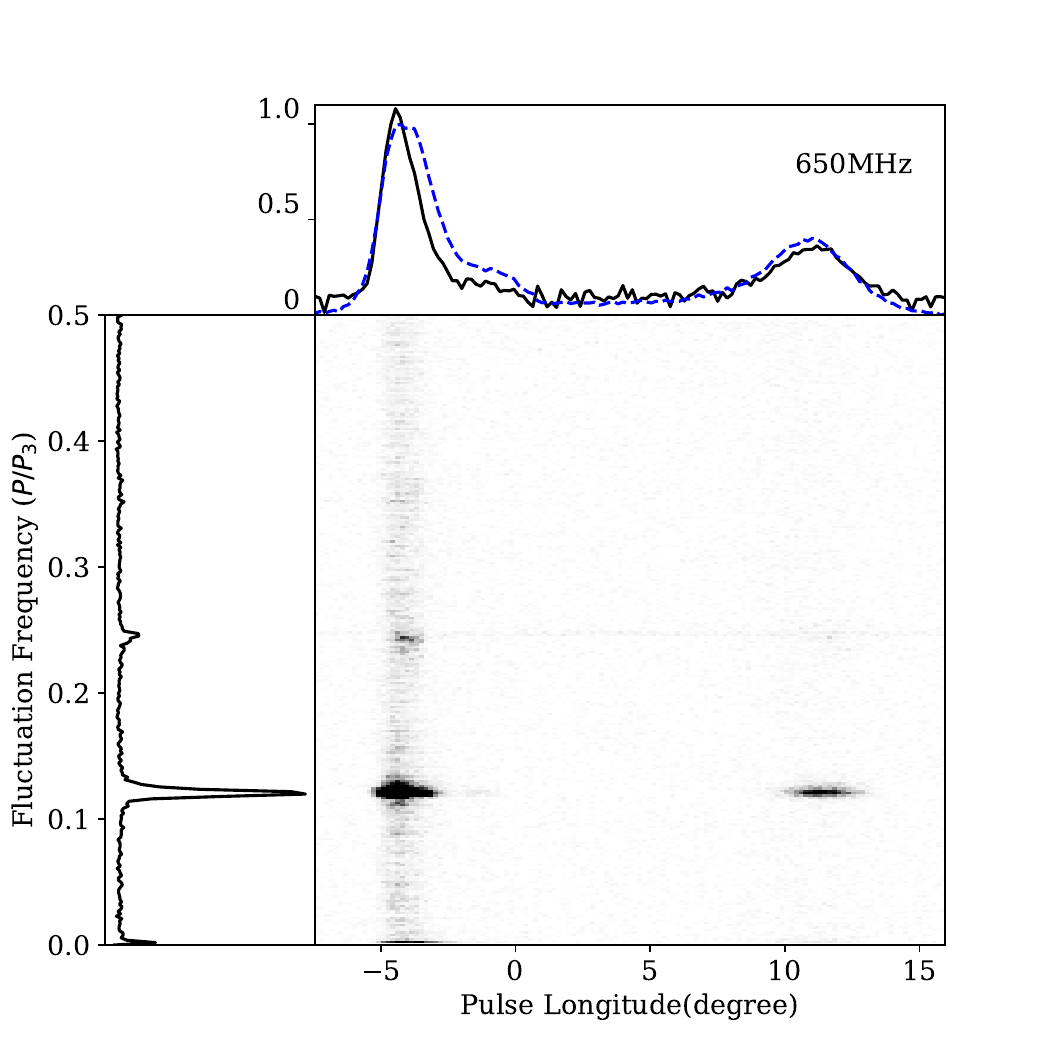}
    \caption{The LRFS of PSR~J1921$+$2003 GMRT observation. A detailed description of the panels can be found in Fig.~\ref{fig:J1418-3921TPAModesLRFS}. The features differ slightly from those in the MK-L observation in Fig.~\ref{fig:PulseStacksTPA}, with fluctuation frequencies appearing above and below the intensity modulation at around 0.12~cpp. The FFT length used is 526.}
    \label{fig:J1921+2003GMRTLRFS}
\end{figure}
\begin{figure*}
    \centering
    \begin{tabular}{cc}
    \includegraphics[width=0.45\textwidth]{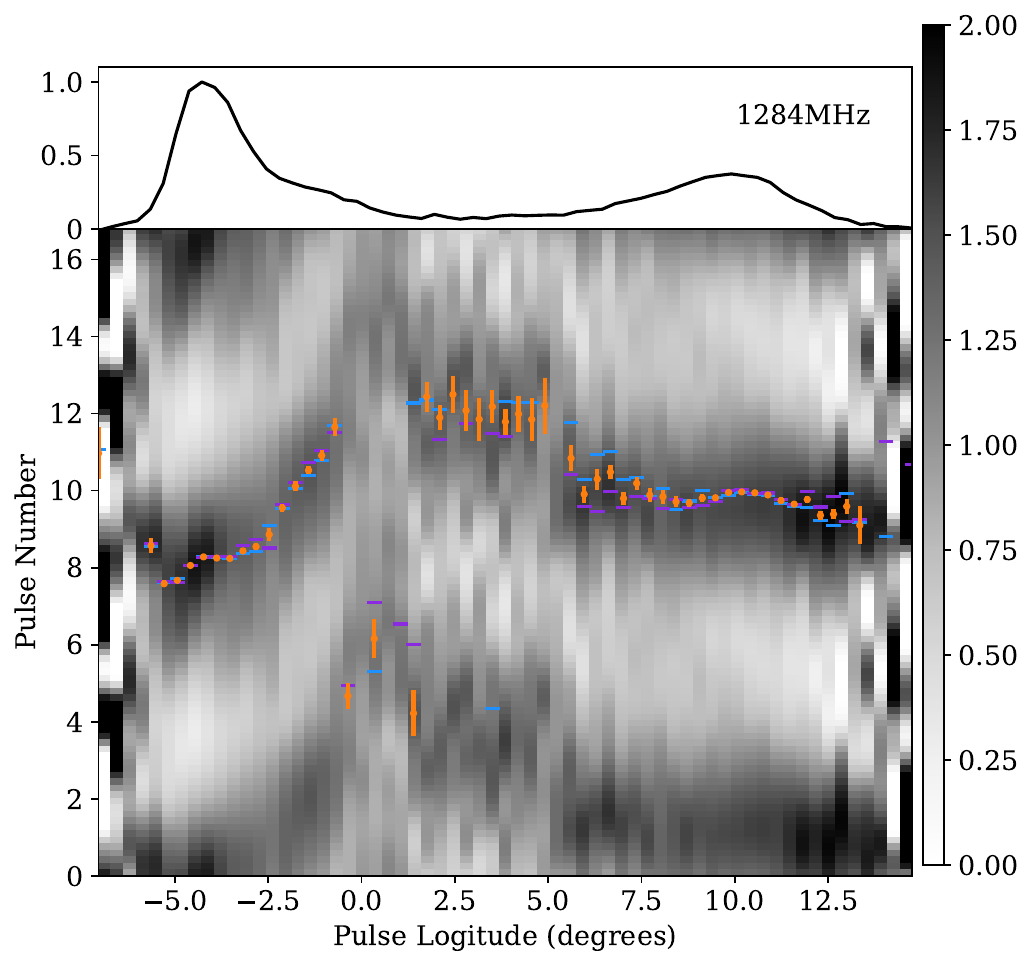}&
    \includegraphics[width=0.45\textwidth]{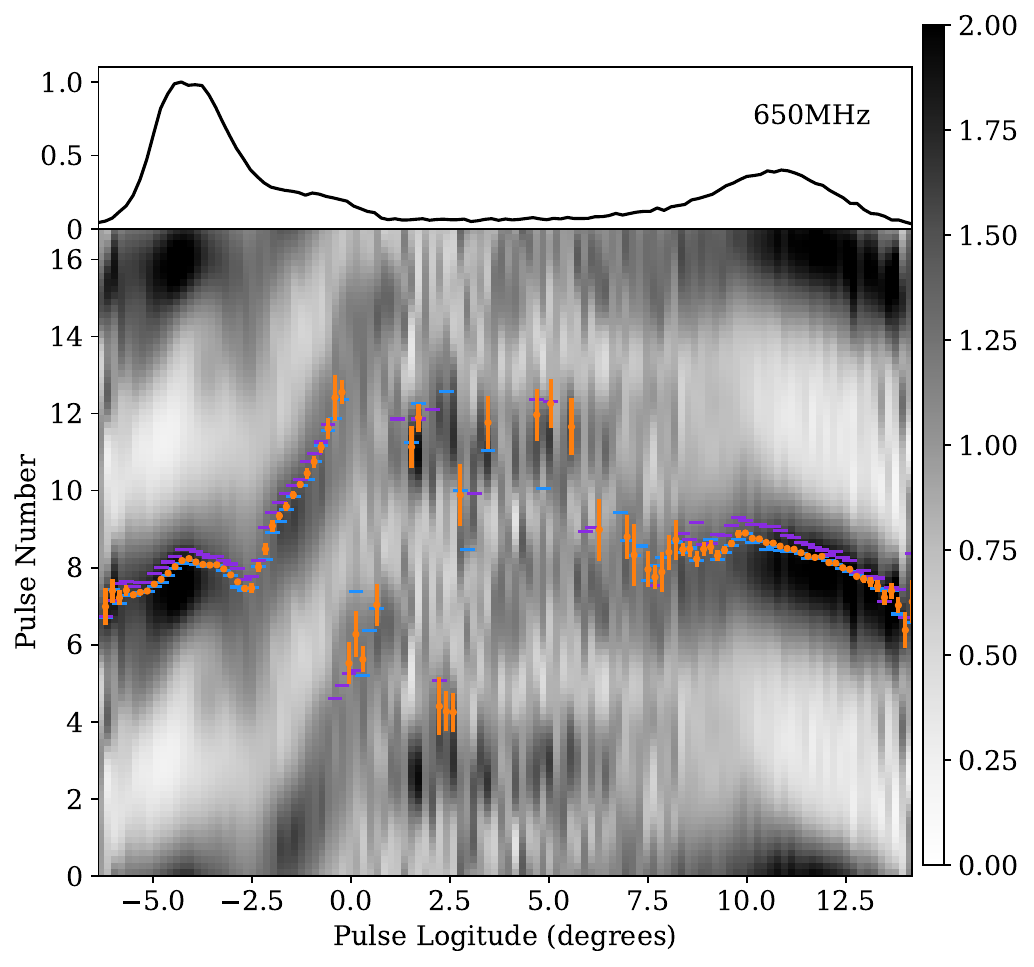}
    \end{tabular}
    \caption{The $P_3$ Fold plots of PSR~J1921+2003 from MK-L (left) and GMRT (right) observations. The $P_3$ used for folding is 8.397$P$ and the smoothing factor is 5. The detailed description of the plot can be found in Fig.~\ref{fig:J1418-3921P3Fold}. }
    \label{fig:J1921+2003P3Fold}
\end{figure*}
\subsection{PSR J1921+2003} 

We used both MK-L (Fig.~\ref{fig:PulseStacksTPA}) and GMRT (Fig.~\ref{fig:PulseStacksGMRT}) observations for the sub-pulse analysis of PSR~J1921+2003. 
Fig.~\ref{fig:PulseStacksTPA} shows two distinct profile components with a smooth, bump-like structure between about $-2$ and 1 degrees in longitude. This structure may be due to emissions from a different emission cone, resulting in a profile with multiple components. 
The LRFS from the MK-L observation (Fig.~\ref{fig:PulseStacksTPA} middle panel) shows a strong modulation around 0.12~cpp. However, in the GMRT observation (Fig.~\ref{fig:J1921+2003GMRTLRFS}), smaller fluctuation peaks also appear around 0.24, which is likely the harmonic of the primary modulation; and 0.01~cpp, especially in the leading component, which appears to be a long period intensity modulation. 

 Fig.~\ref{fig:J1921+2003P3Fold} shows the $P_3$ fold overlaid with sub-pulse phases. The drift features are similar in both observations. 
The bump-like component near the leading component (between $-$2.5 and 0 degrees) has the strongest subpulse drift.
 
The MK-L observation shows a smoother transition between the two leading components (between pulse longitude $-6$ and $-1$ degree), without significant change in the direction of the drift bands, compared to the GMRT observation. 
The trailing component (from about 10 to 12.5 degrees) shows a slight negative drift, which implies bi-drifting.  
In the GMRT observation, the trailing component shows a steeper negative drift compared to that in the MK-L observation.
In the leading component of the GMRT observation, there is a slight change in drift direction, transitioning from relatively flat to a positive drift around $-4$ degrees, resulting in a curved drift band between about $-6$ and $-2.5$ degrees. The drift then becomes positive at about $-2$ degree in the bump-like second profile component. The features in the middle section, between the leading and trailing components, are not well defined. 
In both observations, the sub-pulse drift is stable over time as the phases of the first and second halves of the pulse stack coincide with each other as well as with the phases of the full pulse stack.  

\clearmaybe
\section{Discussion}
\label{sec:Dis}
Section \ref{sec:results} demonstrated nine pulsars for which we identify different signs of the sub-pulse phase gradient at different longitudes.
We find that this change in slope is contemporaneous and persistent, suggesting that bi-drifting is present in all of these pulsars.

Unlike the previously published bi-drifting samples, which have wide profiles with clearly distinct drift bands, the bi-drifting in our sample is much harder to see directly in the pulse stacks.
However, it is perhaps no surprise that for every `obvious' bi-drifting pulsar there may be many more cases where the detection is marginal. 
The ability to discern bi-drifting requires at least two components with well-measured drift.
This can be hampered in pulsars with asymmetric profiles, where one component has a much lower signal-to-noise ratio than the other, or where one component has a weaker modulation.
Drift rates are also harder to establish for pulsars with short $P_3$ or narrow profile widths, where there is less longitude range over which to see the sub-pulse phase evolve.
The previously published bi-drifting pulsars typically have $P_3$ close to 10$P$ and pulse widths greater than 50 degrees.
Some of the pulsars in our sample have $P_3$ close to 2$P$, making the drift features harder to observe in the pulse stack, and most of the pulsars in our list have narrow profiles (smaller than 20 degrees).

Even with this more thorough search for bi-drifting, we find that it is a rare phenomenon.
Among the 1198 TPA pulsars, 418 show evidence of sub-pulse drift \citep{Song2023}, and of these we find only 11 exhibiting bi-drifting.
Two of these are
well studied bi-drifters: PSRs~J1842$-$0359 and J1034$-$3224 that also appear in the TPA analysis, but are not included in this work, as their properties are already established \citep{patrick2016,Szary2020,Basu2018}.

\begin{figure}
    \centering

     \includegraphics[width=0.45\textwidth]
     {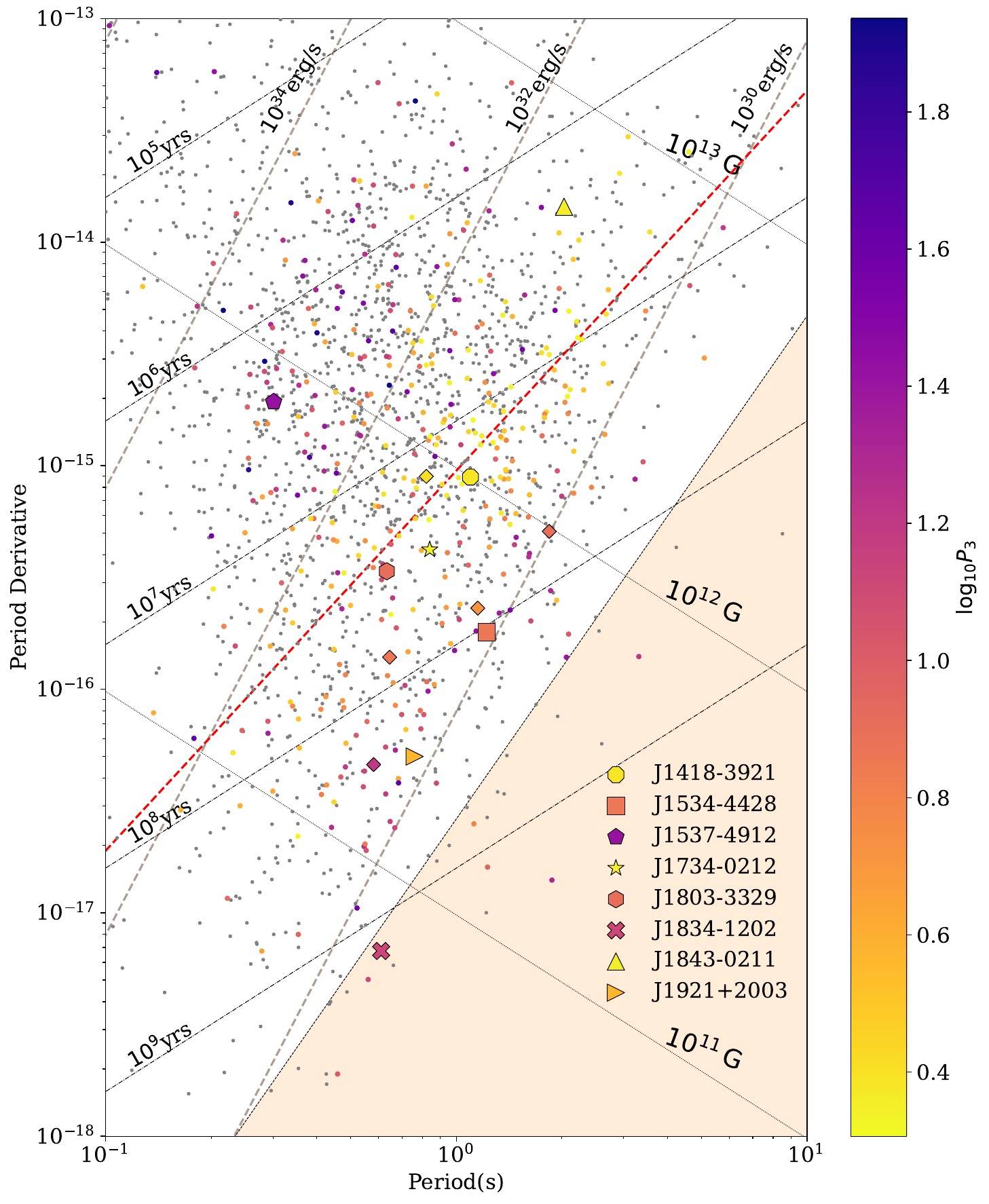}
  
    \caption{Plot of rotation period and period derivative. The grey dots are all radio pulsars in the ATNF pulsar catalogue (\citealp{Manchester2005}, \textsc{psrcat}, version 2.30, \url{https://www.atnf.csiro.au/research/pulsar/psrcat}). Note that not all pulsars in \textsc{psrcat} are shown due to the plot axis limits. The colours are the $P_3$s of the pulsars with evident drifting features, and the red dashed line lies where $P_3\sim2$ as identified by \citealt{Song2023}. For pulsars with multiple modes, the $P_3$ is selected to be the period of the strongest drift feature. The pulsars in our sample are marked in different symbols. The known bi-drifting pulsars are marked with diamonds. }
    \label{fig:ppdot}
\end{figure}

\subsection{Bi-drifting Pulsar Population}
In this sub-section we take a broader view of the 13 pulsars in the combined sample of the previously identified bi-drifters and the pulsars presented in this work.
Fig.~\ref{fig:ppdot} shows how our sample of pulsars lies in the $P$-$\dot{P}$ diagram, as well as the full sample of drifting pulsars from \citet{Song2023} and the previously published bi-drifting pulsars.
The $P_3$ values of the bi-drifting pulsars are consistent with the broader population of pulsars, implying that bi-drifting is not associated with any anomalous $P_3$.
It is notable that the bi-drifting pulsars predominantly occupy the lower right side of the pulsar population in the $P$-$\dot{P}$ diagram, even more so than the overall population of drifting pulsars.

In comparison to the broader population of drifting pulsars, 10 out of the 13 pulsars are near or below the $P_3\sim2$ valley (indicated by the red dashed line in Fig.~\
\ref{fig:ppdot}), where \citet{Song2023} suggest that the observed $P_3$ transitions from an aliased state to directly observing the underlying $P_3$.
Among the three pulsars located above the red dashed line,
PSR~J1843$-$0211 and the fastest mode of PSR~J1921$+$1948 have $P_3$s very close to 2, consistent with their proximity
to this aliasing boundary; and we have already noted that PSR~J1537$-$4912 shows very different properties to the rest of the sample and may not belong to the same family of bi-drifting pulsars. 
However, it is hard to see how bi-drifting could be affected by aliasing. Aliasing is only experienced in a frame that is not co-rotating with the pulsar, and hence it would only influence the emission processes if there were feedback from the environment onto the polar cap.

Additionally, 12 of the pulsars (all except PSR~J1537$-$4912) are in the `conal beam dominant' area of the $P-\dot{P}$ diagram, below $\dot{E}\sim10^{32.5} \textrm{erg\,s}^{-1}$, as defined by \citet{Rankin2022}. 
This is perhaps not surprising, as this region is strongly associated with conal drifting, and bi-drifting is defined by different perceived drift directions within the same emission cone.
To observe bi-drifting, some asymmetry in the emission region is required \citep{Geoff2017}.
Hence, it is possible that there is a link to bi-drifting as this conal emission is expected to arise from a wide range of emission heights, which may magnify any asymmetries in the emission beam.
  
Another possible explanation for the tendency of our sample to be in the lower right corner could be changes in the magnetic-field alignment as the pulsar evolves.
The magnetic inclination angle, $\alpha$, is thought to get smaller as the pulsar ages \citep{Tauris1998,Weltevrede2008,Johnston2017}, and hence the magnetic axis and rotation axis will be more aligned.
An elliptical beam model \citep{Geoff2017} suggests that bi-drifting would be more likely for pulsars with smaller values of $\alpha$. \citet{BasuMitra2020} also suggested that bi-drifting requires the path of the line-of-sight over the emission region to be highly curved, which can be achieved with small $\alpha$ values. However, unlike many of the previously published bi-drifting pulsars, many of the pulsars in our sample have narrow pulse widths. These are therefore unlikely to be highly aligned, and hence the bi-drifting cannot be easily explained by line-of-sight curvature.

\subsection{Mode Changing and Frequency Evolution}

Eight of the 13 pulsars with bi-drifting sub-pulses show clear mode changes on the typical observing timescale.
In our sample, PSRs~J1418$-$3921, J1534$-$4428, J1843$-$0211 and J1921$+$1948 show at least two modes with different profile shapes and drift properties (see, e.g., Figs. \ref{fig:PulseStacksTPA} and \ref{fig:PulseStacksGMRT}).
By performing mode-separated analysis we are confident that the bi-drifting is present within individual modes, i.e. the bi-drifting does not seem to be an artefact of the mode changes.
In most cases, bi-drifting is associated with one mode, while another mode may exhibit less clear or even undetectable drifting in some components. 
An exception is PSR~J1921$+$1948, which shows at least two distinct modes, each with well-defined $P_3$ periods, and both of these modes show bi-drifting features. 

In general, there does not appear to be a straightforward picture that indicates which mode would show bi-drifting, although it is notable that PSRs~J1534$-$4428 and J1843$-$0211 both show modes with similar profile shapes, similar transitions between modes and, similar sub-pulse drift properties in the two modes.
In particular, there is a two-component mode with stronger leading component and a three-component mode featuring components of relatively equal intensity, along with a slightly narrower overall profile width (see Sections~\ref{sec:J1534} and ~\ref{sec:J1843} for details). 
The two-component mode is associated with clear sub-pulse drift features, while the triple mode does not show a well-defined $P_3$.
However, this does not seem to be universal, for example PSR~J2006$-$0807 \citep{bpm19} also has mode changes involving the appearance of a central component, but this case shows clear bi-drifting in the mode with a central core component.

It should be noted that PSRs~J1418$-$3921 and J1843$-$0211 both exhibit components with drift in opposite directions in two distinct modes. 
Interestingly, and perhaps uniquely amongst such mode changes, in PSR~J1418$-$3921 the $P_3$ of both modes appears to be the same.

We do not see any strong evolution of the drifting or mode changing with observing frequency.
However, there are some subtle changes in the shape of the drift bands and profile shape with frequency, which may be attributed to changes to how the line-of-sight cuts through the emission beam.
It is hard to draw any conclusions from these relatively subtle changes.
Stronger evolution with frequency may be expected at lower observing frequencies ($\sim 100$~MHz) where profile shapes are known to evolve more strongly \citep{Pilia2016}.

\subsection{Implications of The Models}
\label{sec:Diss_implications}

\begin{figure}
    \centering
    \includegraphics[width=0.45\textwidth]{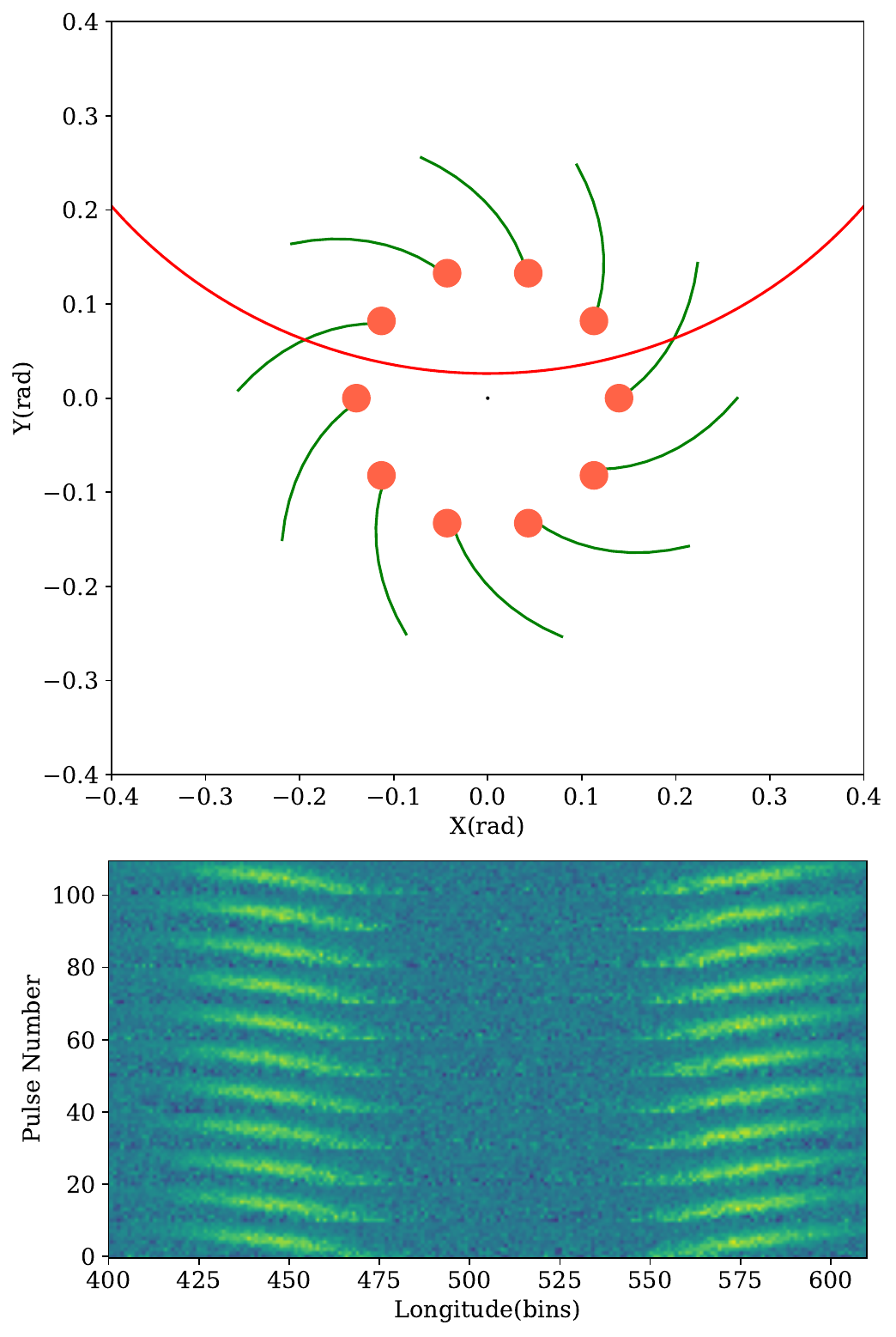}
\caption{Example of an spiral emission pattern. The upper panel shows a top-down view of the emission region containing a circular carousel with 10 sparks (red dots indicate their initial position). The sparks follow the green curves, defined by circulation combined with a linear expansion. In this simulation, after rotating by one spark separation,  the radius resets to the initial value. The resulting pulse stack, for an observer for which the line-of-sight intersects the emission region as shown by the red curve, is shown in the bottom panel. Here the line-of-sight  is defined by the angle between the rotation axis and the magnetic axis $\alpha=30\deg$ and the impact angle $\beta=-1.5\deg$. The black dot indicates the centre of the carousel.
    }
    \label{fig:spiralmodel}
\end{figure}
 
The models mentioned in Section~\ref{sec:Intro} 
have been shown to be able to explain the drifting subpulse patterns in some of the previously published cases of bi-drifting. 
However, some identifiable patterns seen in our extended sample have not yet been discussed in the context of existing models. 
Pulsars with bi-drifting sub-pulses can also have mode changes, sub-pulse phase jumps between the adjacent profile components, and additional intensity modulation. In addition, some pulsars in our sample have relatively narrow profile widths and/or small $P_3$ compared to the published bi-drifters. 
It would be interesting to explore these additional features in the context of models to explain bi-drift, as well as the changes in drift rate at different frequencies and drift modes.

The phenomenological model proposed by \cite{Qiao2004} in which a core gap coexists with an inner annular gap surrounded by an outer gap requires the pulsar to be a bare strange star, rather than a neutron star. Differences in the electric field direction in the different gaps allow for bi-drifting. Different spark numbers and rotation speeds for the inner and outer carousels can possibly produce bi-drifting in multiple profile components and might allow the phase jumps observed in PSRs~J1418$-$3921 and J1803$-$3329 in our sample. 
Similarly, the geometric model with tilted nested elliptical carousels proposed by \citet{Geoff2017} could also explain phase jumps associated with transitions between profile components.

However, in the model proposed by \citet{BasuMitra2020} sparks are not circulating about the magnetic axis, but are lagging behind co-rotation in the pulsar’s inner acceleration region. The spark trajectories and the polar cap geometry are further modified by a non-dipolar magnetic field structure. For a highly asymmetric non-dipolar polar cap, the spark tracks are curved and could lead to the change in drift direction in the form of bi-drifting \citep{Basu2023}. We find bi-drifting in pulsars with narrower profiles than those discussed in the literature. It would therefore be interesting to investigate whether the models for bi-drifting can account for these profiles, both in the context of the \citet{BasuMitra2020} model and in models that involve circulating sparks (e.g. \citealp{Geoff2017}).

To explain bi-drifting, it is not required that the sparks follow paths that are simple geometric shapes.
If the drift velocity of the plasma relative to the neutron star depends on the variation of the electric potential at the polar cap (\citealp{vanLeeuwen2012,Szary2017,Szary2020}), complex structures in the magnetic field result in complex paths of the spark motion. For some pulsars, the resulting deviation from a circular path could be similar to that described by~\citet{Geoff2017}.

We also encourage exploration of other models for emission geometry that may produce bi-drifting.
For example, bi-drifting would naturally occur if there is radial motion of the sparks, with or without circular drift around the magnetic axis.
In fact, such radial motion is perhaps to be expected given the curvature of the field lines (e.g.~\citealp{Cheng1980,Filippenko1982}). 
One can imagine a toy model where sparks travel in a spiral pattern, with sparks rotating in a circular motion as well as travelling radially outwards (or inwards). Indeed, there are examples of models that propose spiral emission regions in the literature \citep{Hankins1980,Dyks2017}, although these models were not developed to explain bi-drifting.
As an alternative to radial spark motion, a spiral pattern could arise from a simple circular carousel combined with a periodic variation in the emission height.
Fig.~\ref{fig:spiralmodel} shows a simple case where the sparks follow a fixed spiral pattern defined by a single periodicity, $P_3=10P$, which corresponds to both the normal $P_3$ associated with the circular motion (separation between two spars), as well as the period of the expansion.
Whilst this model is simplistic, it is easily able to produce bi-drifting sub-pulses.
Compared to the elliptical model, it is able to produce asymmetry in the drift patterns without requiring a highly asymmetric beam shape.
However, more work is needed to explore the range of possible phenomena that can be produced.
For example, if the spiral pattern itself is allowed to circulate, this can produce modulations at two frequencies and perhaps give rise to the beat frequencies we see in PSR~J1803$-$3329, as well as those described by \citet{Ray2025} for PSR~J1514$-$4834.

\clearmaybe

\section{Conclusion}
\label{sec:conclusion}

We analysed sub-pulse drifting patterns and mode changing of 9 bi-drifting candidates selected from the Thousand-Pulsar-Array single-pulse data. All of these we claim do show bi-drifting, although the evidence is primarily from statistical analysis, rather than directly from the pulse stack.
This sample of pulsars with complex drift patterns, phase jumps and mode changes will be important for challenging models of sub-pulse drifting, given that many drift features of our samples are not compatible with the classic carousel model and are difficult to be modelled with a circularly symmetric spark track.

Most of the pulsars in our sample have relatively narrow widths (<20\textdegree), which is also challenging for some proposed models for bi-drifting, which tend to favour wide profiles and low $\alpha$ (e.g. \citealp{Geoff2017,BasuMitra2020}).
Any model for drifting sub-pulses in pulsar emission needs to be able to explain both the phenomenon of bi-drifting, as well as complex mode changing: in our sample, PSRs~J1418$-$3921, J1534$-$4428, J1843$-$0211 and J1921$+$1948 show at least two modes, and the sub-pulse modulation features are generally different in different modes. 
In addition, the model also needs to explain the phase jumps between the adjacent profile components which are observed in e.g. PSRs~J1418$-$3921 and J1803$-$3329 in our sample. 

Although it is not obvious that the elliptical carousel model of \citet{Geoff2017} is sufficient to overcome all these issues, it can explain the bi-drifting features and the phase locking in some pulsars.
We intend to explore this model and the model with spiral spark tracks further in future work.
Clearly, pulsars are more complex than our models allow, and therefore we encourage the exploration of a wide range of models.

\clearmaybe
\section*{Acknowledgements}
The MeerKAT telescope is operated by the South African Radio Astronomy Observatory, which is a facility of the National Research Foundation, an agency of the Department of Science and Innovation.
The GMRT is run by the National Centre for Radio Astrophysics of the Tata Institute of Fundamental Research. 
Pulsar research at Jodrell Bank is supported by a consolidated grant (ST/T000414/1 and ST/X001229/1) from the UK Science and Technology Facilities Council (STFC).
GW thanks the University of Manchester for Visitor status. 
MeerTime data are housed and processed on the OzSTAR supercomputer at Swinburne University of Technology. 
The authors thank the reviewer for their insightful comments and suggestions, which helped us to improve the depth of the discussion.

\section*{Data Availability}

Data underlying this article will be shared upon reasonable request to the corresponding author.



\bibliographystyle{mnras}
\bibliography{BiDrift} 




\newpage
\appendix

\section{Table of Repeat MK-L Observations}

Sections~\ref{sec:obs} and \ref{sec:results} discussed combined observations with short
single-pulse MK-L observations for three of the pulsars to provide more pulses for analyses. The additional short observations are shown in Table~\ref{table: observation long}. 
For PSR~J1921$+$1948, relatively more recent observations from 2023 were included to provide more pulses for mode changes. 

\begin{table*}
\caption{\label{table: observation long}List of other (shorter) MK-L observations of possible bi-drifting pulsars that have been used for analysis.}
\centering
\begin{tabular}{lrrrr}\\
\hline
&J1537$-$4912 &  J1843$-$0211& J1921+1948 &J1921+1948\\
Total Nr of Pulses Used & 5624 & 1664&8356& \textit{cont...}\\
Duration(hr)&0.470&0.937&2.83&\\
OBS:&2019-06-21-19:24:09&2020-02-24-05:07:28&2019-10-18-14:45:19&2022-06-05-21:55:30\\
&2019-07-14-20:37:26&2020-03-29-05:36:54&2020-05-13-03:53:41&2022-06-14-23:59:34\\
&2019-10-30-17:09:28&2020-06-25-23:25:45&2020-07-03-22:55:03&2022-07-14-21:59:17\\
&2020-02-24-01:55:20&2020-09-19-16:16:10&2020-08-03-21:20:21&2022-08-05-17:45:49\\
&2020-04-27-22:48:41&2020-10-17-17:51:02&2020-08-29-19:35:27&2022-08-11-21:03:00\\
&2020-06-19-19:54:30&2020-12-12-08:51:55&2020-09-25-17:51:07&2022-09-05-15:57:32\\
&2020-09-13-12:15:39&2021-01-11-08:34:04&2020-10-24-15:51:07&2022-09-11-18:11:56\\
&2020-10-11-12:02:29&&2020-11-20-15:05:12&2022-10-08-13:31:58\\
&2020-11-02-12:20:32&&2020-12-24-13:18:32&2022-10-10-17:11:00\\
&2020-12-05-07:23:53&&2021-01-18-12:29:59&2022-11-12-11:08:57\\
&2020-12-29-08:59:05&&2021-02-12-10:22:38&2022-11-14-13:57:45\\
&2021-02-23-03:44:52&&2021-03-11-07:51:33&2022-12-16-11:51:42\\
&2021-05-03-21:54:23&&2021-04-07-01:55:03&2023-01-05-07:41:41\\
&2022-03-15-01:18:16&&2021-05-03-23:58:56&2023-01-22-09:26:54\\
&2022-03-29-01:50:46&&2021-06-03-22:00:32&2023-02-07-05:25:31\\
&2022-06-05-19:50:47&&2021-07-04-19:56:30&2023-02-16-06:15:32\\
&&&2021-08-02-18:31:55&2023-02-18-07:42:06\\
&&&2021-08-29-16:11:32&2023-03-12-03:18:42\\
&&&2021-09-30-15:06:13&2023-03-17-06:20:22\\
&&&2021-10-31-12:02:04&2023-04-12-01:13:52\\
&&&2021-12-02-10:03:20&2023-05-15-03:37:56\\
&&&2022-01-10-07:25:35&2023-06-13-23:26:34\\
&&&2022-01-16-10:54:43&2023-07-12-22:12:20\\
&&&2022-02-10-05:14:57&2023-08-12-20:27:16\\
&&&2022-02-19-08:09:43&2023-09-13-18:01:55\\
&&&2022-03-15-03:23:27&2023-10-10-17:12:58\\
&&&2022-03-21-05:39:08&2023-11-13-14:27:24\\
&&&2022-04-17-03:49:12&2023-12-14-12:07:47\\
&&&2022-05-15-23:00:26&2023-12-16-13:13:27\\
&&&2022-05-19-02:44:45&\\
\hline
\end{tabular}
\end{table*}

\newpage
\section{Plots of Pulse Stack Sections}

The pulse stacks of MK-L and GMRT are shown in Section~\ref{sec:results}, and the sub-pulse phase tracks are shown in e.g. Fig.~\ref{fig:J1418-3921P3Fold}. To compare the modulations from pulse to pulse and the overall sub-pulse phase modulation, we have plotted the sub-pulse phases on top of the smaller sections of pulse stacks. In general, the sub-pulse phase tracks describe phase modulations in the pulse stacks well.

\subsection{PSR J1418--3921}

\begin{figure*}
    \centering
    \begin{tabular}{cc}
    \includegraphics[width=0.42\textwidth]{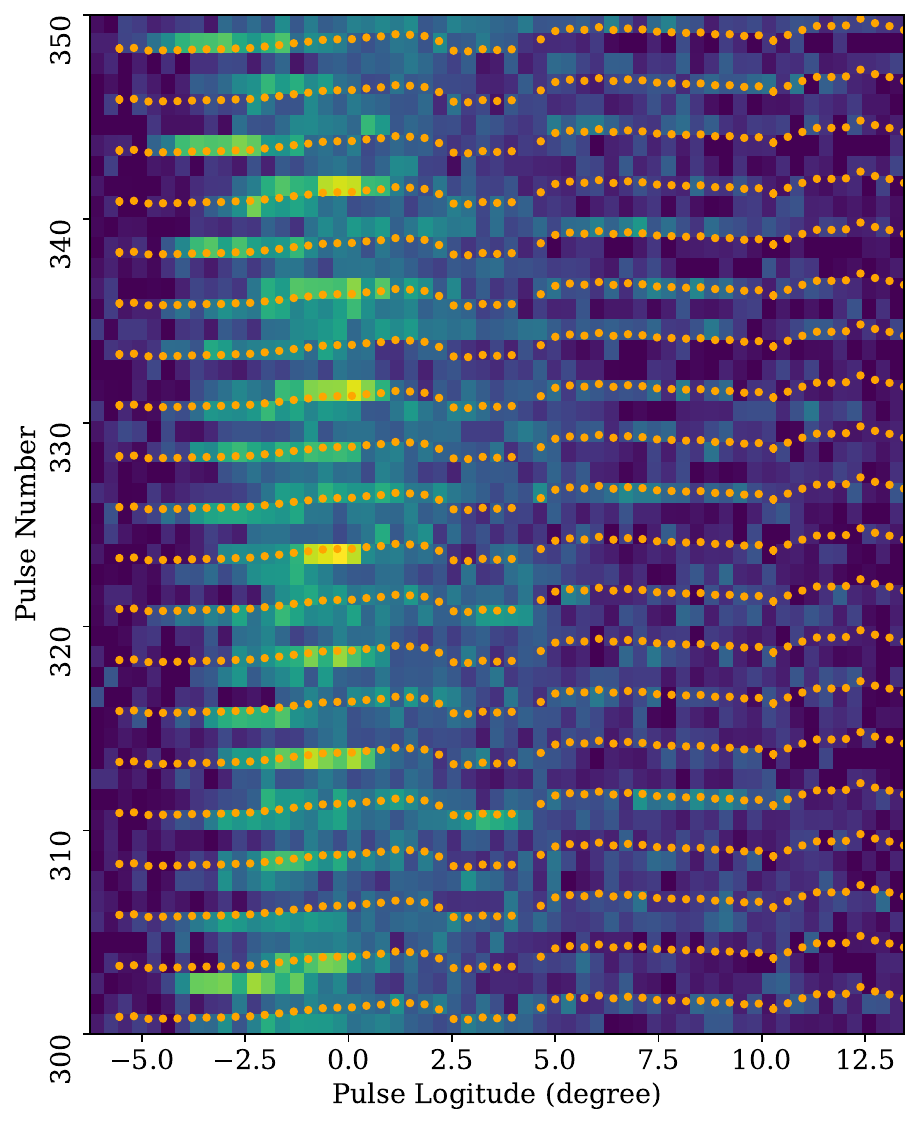}&
\includegraphics[width=0.42\textwidth]{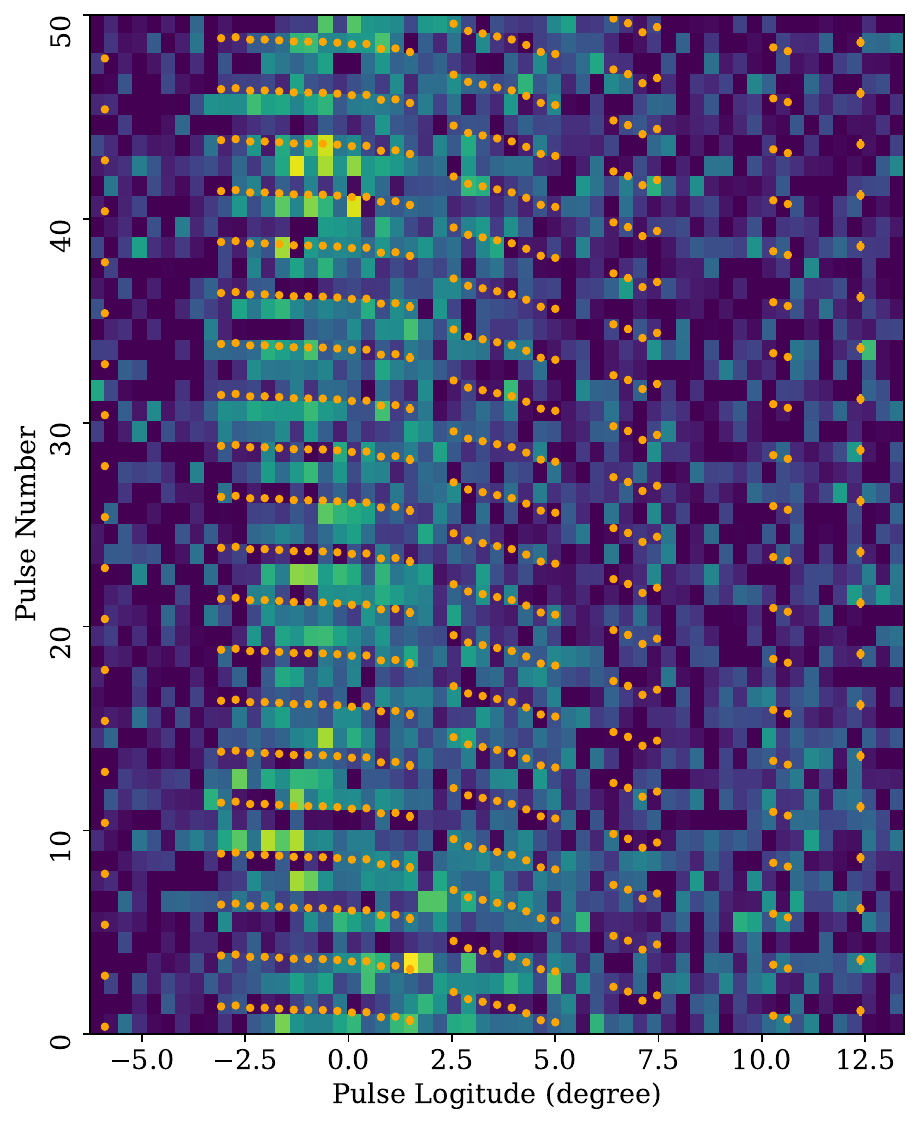}
    \end{tabular}
    \caption{Pulse stack sections (50 pulses each) of PSR~J1418$-$3921 from the MK-L observation. The left plot shows the bright mode and the right plot shows the weak mode. The plots are overlaid with sub-pulse phases (orange dots) with error bars, which show the relative sub-pulse phases derived from the LRFS. Only phases with errors within approximately 10\% of $P_3$ are shown. The phase tracks are repeated with the average $P_3$ separation. The longitude scale is the same as that in Fig.~\ref{fig:PulseStacksTPA}.}
    \label{fig:J1418-3921Sections}
\end{figure*}

The pulse stacks of 50 pulses and sub-pulse phases of the bright and weak modes of PSR~J1418$-$3921 are shown in Fig.~\ref{fig:J1418-3921Sections}. 
The phase jumps between the adjacent profile components, and the drift in every two pulses are clearly seen in the bright mode. On the other hand, it is harder to visualise the drift in the weak mode. There are additional modulations in the single pulses and some intensity modulation in both modes, but in general the single pulses follow the phase tracks. Evidence of bi-drifting is seen in the bright mode, while the leading component of the weak mode exhibits a drift direction opposite to that of the bright mode. 

\subsection{PSR J1834--1202}

\begin{figure}
    \centering  \includegraphics[width=0.42\textwidth]{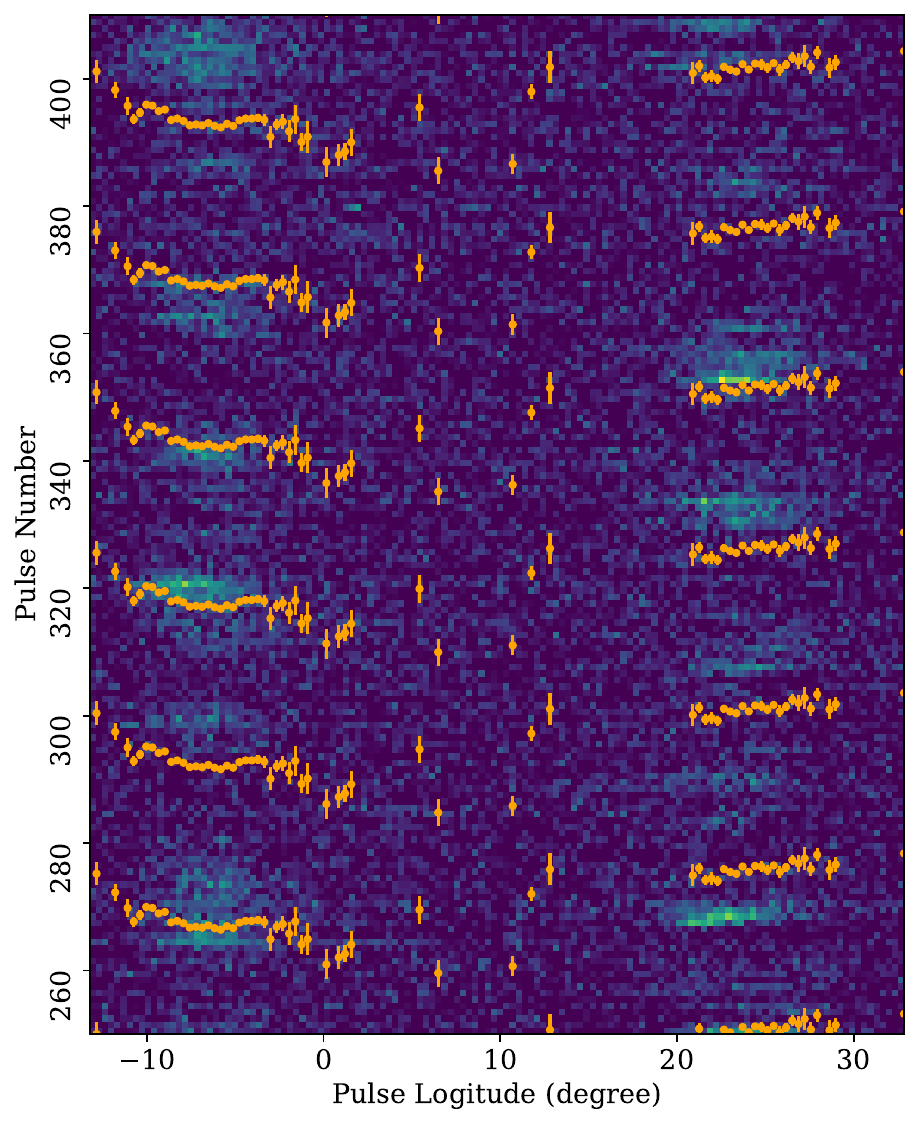}
    \caption{A section of the pulse stack (150 pulses) from the PSR~J1834$-$1202 MK-L observation. The plot is overlaid with relative sub-pulse phases, shown as orange dots with error bars. For a detailed description of the plot, refer to Fig.~\ref{fig:J1418-3921Sections}. }
    \label{fig:J1834-1202TPASec}
\end{figure}

Figs.~\ref{fig:J1834-1202TPA} and \ref{fig:J1834-1202ThreeSecPhases} show that the trailing component of this pulsar shows more irregular modulations compared to the leading component. This is more evident in Fig.~\ref{fig:J1834-1202TPASec}, where 150 pulses are overlaid with the sub-pulse phase tracks. The sub-pulse phases track the leading component relatively well, while the trailing component appears to be below the phase track at around, e.g. pulse number 270, and above at around, e.g. pulse number 330. 

\subsection{PSR J1921+1948}

\begin{figure}
    \centering
    \includegraphics[width=0.42\textwidth]{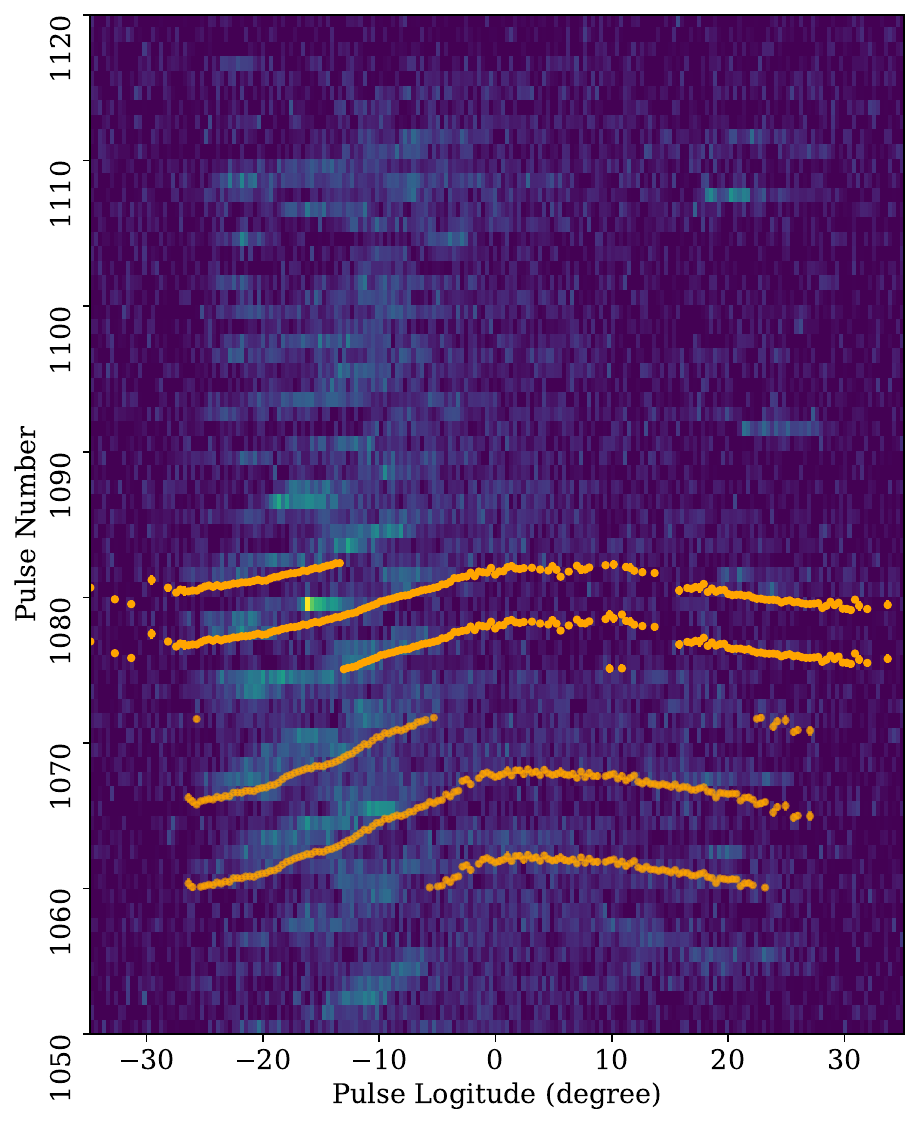}
    \caption{A section of 70 pulses from the PSR~J1921+1948 pulse stack, based on the longest MK-L observation. This section highlights the slow mode near pulse numbers 1060 to 1070; and fast mode around pulse number 1080. The plot is overlaid with relative sub-pulse phases.
    }
    \label{fig:J1921+1948Sec}
\end{figure}

Section~\ref{sec:J1912+1948} described the mode change of PSR~J1921+1948. Although the broad feature is shown in Fig.~\ref{fig:PulseStacksTPA}, the behaviour in each mode is not clear. Fig.~\ref{fig:J1921+1948Sec} shows a section of the pulse stack at which the drift mode changes from fast mode to slow mode and followed by a section of the irregular mode. The sub-pulse phases of the fast and slow modes track the pulses well. 

\clearpage


\newpage
\bsp	
\label{lastpage}
\end{document}